\newcount\mgnf\newcount\tipi\newcount\tipoformule\newcount\greco

\tipi=2          
\tipoformule=0   

\global\newcount\numsec\global\newcount\numfor
\global\newcount\numapp\global\newcount\numcap
\global\newcount\numfig\global\newcount\numpag
\global\newcount\numnf

\def\SIA #1,#2,#3 {\senondefinito{#1#2}%
\expandafter\xdef\csname #1#2\endcsname{#3}\else
\write16{???? ma #1,#2 e' gia' stato definito !!!!} \fi}

\def \FU(#1)#2{\SIA fu,#1,#2 }

\def\etichetta(#1){(\veroparagrafo.\veraformula)%
\SIA e,#1,(\veroparagrafo.\veraformula) %
\global\advance\numfor by 1%
\write15{\string\FU (#1){\equ(#1)}}%
\write16{ EQ #1 ==> \equ(#1)  }}
\def\etichettaa(#1){(A\veraappendice.\veraformula)
 \SIA e,#1,(A\veraappendice.\veraformula)
 \global\advance\numfor by 1
 \write15{\string\FU (#1){\equ(#1)}}
 \write16{ EQ #1 ==> \equ(#1) }}
\def\getichetta(#1){Fig. \verafigura
 \SIA g,#1,{\verafigura}
 \global\advance\numfig by 1
 \write15{\string\FU (#1){\graf(#1)}}
 \write16{ Fig. #1 ==> \graf(#1) }}
\def\retichetta(#1){\numpag=\pgn\SIA r,#1,{\verapagina}
 \write15{\string\FU (#1){\rif(#1)}}
 \write16{\rif(#1) ha simbolo  #1  }}
\def\etichettan(#1){(n\verocapitolo.\veranformula)
 \SIA e,#1,(n\verocapitolo.\veranformula)
 \global\advance\numnf by 1
\write16{\equ(#1) <= #1  }}

\newdimen\gwidth
\gdef\profonditastruttura{\dp\strutbox}
\def\senondefinito#1{\expandafter\ifx\csname#1\endcsname\relax}
\def\BOZZA{
\def\alato(##1){
 {\vtop to \profonditastruttura{\baselineskip
 \profonditastruttura\vss
 \rlap{\kern-\hsize\kern-1.2truecm{$\scriptstyle##1$}}}}}
\def\galato(##1){ \gwidth=\hsize \divide\gwidth by 2
 {\vtop to \profonditastruttura{\baselineskip
 \profonditastruttura\vss
 \rlap{\kern-\gwidth\kern-1.2truecm{$\scriptstyle##1$}}}}}
\def\verapagina{
{\romannumeral\number\numcap}.\number\numsec.\number\numpag}}

\def\alato(#1){}
\def\galato(#1){}
\def\veroparagrafo{\number\numsec}\def\veraformula{\number\numfor}
\def\veraappendice{\number\numapp}
\def\verapagina{\number\pageno}\def\veranformula{\number\numnf}
\def\verafigura{{\romannumeral\number\numcap}.\number\numfig}
\def\verocapitolo{\number\numcap}\def\veranformula{\number\numnf}
\def\Eqn(#1){\eqno{\etichettan(#1)\alato(#1)}}
\def\eqn(#1){\etichettan(#1)\alato(#1)}

\def\Eq(#1){\eqno{\etichetta(#1)\alato(#1)}}
\def\eq(#1){\etichetta(#1)\alato(#1)}
\def\Eqa(#1){\eqno{\etichettaa(#1)\alato(#1)}}
\def\eqa(#1){\etichettaa(#1)\alato(#1)}
\def\dgraf(#1){\getichetta(#1)\galato(#1)}
\def\drif(#1){\retichetta(#1)}

\def\eqv(#1){\senondefinito{fu#1}$\clubsuit$#1\else\csname fu#1\endcsname\fi}
\def\equ(#1){\senondefinito{e#1}\eqv(#1)\else\csname e#1\endcsname\fi}
\def\graf(#1){\senondefinito{g#1}\eqv(#1)\else\csname g#1\endcsname\fi}
\def\rif(#1){\senondefinito{r#1}\eqv(#1)\else\csname r#1\endcsname\fi}
\def\bib[#1]{[#1]\numpag=\pgn
\write13{\string[#1],\verapagina}}

\def\include#1{
\openin13=#1.aux \ifeof13 \relax \else
\input #1.aux \closein13 \fi}

\openin14=\jobname.aux \ifeof14 \relax \else
\input \jobname.aux \closein14 \fi
\openout15=\jobname.aux
\openout13=\jobname.bib


\ifnum\tipoformule=1\let\Eq=\eqno\def\eq{}\let\Eqa=\eqno\def\eqa{}
\def\equ{}\fi


{\count255=\time\divide\count255 by 60 \xdef\hourmin{\number\count255}
        \multiply\count255 by-60\advance\count255 by\time
   \xdef\hourmin{\hourmin:\ifnum\count255<10 0\fi\the\count255}}

\def\oramin{\hourmin }

\def\data{\number\day/\ifcase\month\or january \or february \or march \or
april \or may \or june \or july \or august \or september
\or october \or november \or december \fi/\number\year;\ \oramin}

\setbox200\hbox{$\scriptscriptstyle \data $}

\newcount\pgn \pgn=1
\def\foglio{\number\numsec:\number\pgn
\global\advance\pgn by 1}
\def\foglioa{A\number\numsec:\number\pgn
\global\advance\pgn by 1}

\footline={\hss\tenrm\folio\hss}

\def\TIPIO{
\font\setterm=amr7 
\def \settepunti{\def\rm{\fam0\setterm}
\textfont0=\setterm   
\normalbaselineskip=9pt\normalbaselines\rm
}\let\nota=\settepunti}

\def\TIPITOT{
\font\twelverm=cmr12
\font\twelvei=cmmi12
\font\twelvesy=cmsy10 scaled\magstep1
\font\twelveex=cmex10 scaled\magstep1
\font\twelveit=cmti12
\font\twelvett=cmtt12
\font\twelvebf=cmbx12
\font\twelvesl=cmsl12
\font\ninerm=cmr9
\font\ninesy=cmsy9
\font\eightrm=cmr8
\font\eighti=cmmi8
\font\eightsy=cmsy8
\font\eightbf=cmbx8
\font\eighttt=cmtt8
\font\eightsl=cmsl8
\font\eightit=cmti8
\font\sixrm=cmr6
\font\sixbf=cmbx6
\font\sixi=cmmi6
\font\sixsy=cmsy6
\font\twelvetruecmr=cmr10 scaled\magstep1
\font\twelvetruecmsy=cmsy10 scaled\magstep1
\font\tentruecmr=cmr10
\font\tentruecmsy=cmsy10
\font\eighttruecmr=cmr8
\font\eighttruecmsy=cmsy8
\font\seventruecmr=cmr7
\font\seventruecmsy=cmsy7
\font\sixtruecmr=cmr6
\font\sixtruecmsy=cmsy6
\font\fivetruecmr=cmr5
\font\fivetruecmsy=cmsy5
\textfont\truecmr=\tentruecmr
\scriptfont\truecmr=\seventruecmr
\scriptscriptfont\truecmr=\fivetruecmr
\textfont\truecmsy=\tentruecmsy
\scriptfont\truecmsy=\seventruecmsy
\scriptscriptfont\truecmr=\fivetruecmr
\scriptscriptfont\truecmsy=\fivetruecmsy
\def \eightpoint{\def\rm{\fam0\eightrm}
\textfont0=\eightrm \scriptfont0=\sixrm \scriptscriptfont0=\fiverm
\textfont1=\eighti \scriptfont1=\sixi   \scriptscriptfont1=\fivei
\textfont2=\eightsy \scriptfont2=\sixsy   \scriptscriptfont2=\fivesy
\textfont3=\tenex \scriptfont3=\tenex   \scriptscriptfont3=\tenex
\textfont\itfam=\eightit  \def\it{\fam\itfam\eightit}%
\textfont\slfam=\eightsl  \def\sl{\fam\slfam\eightsl}%
\textfont\ttfam=\eighttt  \def\tt{\fam\ttfam\eighttt}%
\textfont\bffam=\eightbf  \scriptfont\bffam=\sixbf
\scriptscriptfont\bffam=\fivebf  \def\bf{\fam\bffam\eightbf}%
\tt \ttglue=.5em plus.25em minus.15em
\setbox\strutbox=\hbox{\vrule height7pt depth2pt width0pt}%
\normalbaselineskip=9pt
\let\sc=\sixrm  \let\big=\eightbig  \normalbaselines\rm
\textfont\truecmr=\eighttruecmr
\scriptfont\truecmr=\sixtruecmr
\scriptscriptfont\truecmr=\fivetruecmr
\textfont\truecmsy=\eighttruecmsy
\scriptfont\truecmsy=\sixtruecmsy
}\let\nota=\eightpoint}

\newfam\msbfam   
\newfam\truecmr  
\newfam\truecmsy 
\newskip\ttglue
\ifnum\tipi=0\TIPIO \else\ifnum\tipi=1 \TIPI\else \TIPITOT\fi\fi

\global\newcount\numpunt

\magnification=\magstephalf
\baselineskip=16pt
\parskip=8pt

\def\a{\alpha}
\def\b{\beta}
\def\d{\delta}
\def\e{\epsilon}

\def\g{\gamma}

\def\l{\lambda}

\def\s{\sigma}
\def\t{\tau}

\def\z{\zeta}
\def\o{\omega}
\def\D{\Delta}
\def\L{\Lambda}
\def\G{\Gamma}
\def\O{\Omega}

\def\del #1{\frac{\partial^{#1}}{\partial\l^{#1}}}

\def\E{{I\kern-.25em{E}}}
\def\N{{I\kern-.22em{N}}}
\def\M{{I\kern-.22em{M}}}
\def\R{{I\kern-.22em{R}}}
\def\Z{{Z\kern-.5em{Z}}}
\def\1{{1\kern-.25em\hbox{\rm I}}}
\def\eu{{1\kern-.25em\hbox{\sm I}}}

\def\C{{C\kern-.75em{C}}}
\def\P{{I\kern-.25em{P}}}

\def\del{\partial}


\def\AA{{\cal A}}
\def\BB{{\cal B}}

\def\FF{{\cal F}}
\def\GG{{\cal G}}

\def\SS{{\cal S}}
\def\TT{{\cal T}}

\def\WW{{\cal W}}
\def\VV{{\cal V}}

\def\chap #1#2{\line{\ch #1\hfill}\numsec=#2\numfor=1}

\def\ba{{\backslash}}

\def\wt{\widetilde}

\def\sminn{\hbox{\ftn int}\,}

\def\jg{J_\g}


\newcount\foot
\foot=1
\def\note#1{\footnote{${}^{\number\foot}$}{\ftn #1}\advance\foot by 1}

\def\frac#1#2{{#1\over #2}}
\def\sfrac#1#2{{\textstyle{#1\over #2}}}
\def\text#1{\quad{\hbox{#1}}\quad}
\def\newpage{\vfill\eject}
\def\proposition #1{\noindent{\thbf Proposition #1:}}

\def\theo #1{\noindent{\thbf Theorem #1: }}
\def\lemma #1{\noindent{\thbf Lemma #1: }}

\def\corollary #1{\noindent{\thbf Corollary #1: }}
\def\proof{{\noindent\pr Proof: }}
\def\proofof #1{{\noindent\pr Proof of #1: }}
\def\endproof{$\diamondsuit$}
\def\remark{\noindent{\bf Remark: }}
\def\thanks{\noindent{\bf Aknowledgements: }}
\font\pr=cmbxsl10
\font\thbf=cmbxsl10 scaled\magstephalf

\font\ch=cmbx12
\font\ftn=cmr8

\font\it=cmti10
\font\bf=cmbx10
\font\sm=cmr7


\font\tit=cmbx12
\font\aut=cmbx12
\font\aff=cmsl12
\nopagenumbers
{$  $}
\vskip2truecm
\centerline{\tit DISTRIBUTION OF OVERLAP PROFILES}
\vskip.2truecm
\centerline{\tit IN THE ONE-DIMENSIONAL KAC-HOPFIELD MODEL \footnote{${}^\#$}{\ftn Work
partially supported by the Commission of the European Union
under contract No. CHRX-CT93-0411}}
\vskip1.5truecm
\centerline{\aut Anton Bovier
\footnote{${}^1$}{\ftn e-mail:
bovier@wias-berlin.de}}
\vskip.1truecm
\centerline{\aff Weierstra\char'31{}-Institut}
\centerline{\aff f\"ur Angewandte Analysis und Stochastik}
\centerline{\aff Mohrenstra\char'31 e 39, D-10117 Berlin, Germany}
\vskip.5truecm
\centerline{\aut  V\'eronique Gayrard\footnote{${}^2$}{\ftn
e-mail: gayrard@cpt.univ-mrs.fr}
and Pierre Picco\footnote{${}^3$}{\ftn
e-mail: picco@cpt.univ-mrs.fr}}
\vskip.1truecm
\centerline{\aff Centre de Physique Th\'eorique - CNRS}
\centerline{\aff Luminy, Case 907}
\centerline{\aff F-13288 Marseille Cedex 9, France}
\vskip1.5truecm\rm
\noindent {\bf Abstract:}
We study a one-dimensional version of the Hopfield model with 
long, but finite range interactions below the critical temperature.
In the thermodynamic limit we obtain large deviation estimates for the 
distribution of the ``local'' overlaps, the range of the interaction, 
$\gamma^{-1}$, being the large parameter. We show in particular that the 
local overlaps in a typical Gibbs configuration are constant and equal to one 
of the mean-field equilibrium values on a scale $o(\g^{-2})$. We also 
give estimates on the size of typical ``jumps''. i.e. the regions where 
transitions from one equilibrium value to another take place. Contrary to the 
situation in the ferromagnetic Kac-model, the structure of the 
profiles is found to be governed by the  quenched disorder rather than 
by entropy.
 


\vfill
$ {} $

\newpage
\count0=1
\footline={\hss\tenrm\folio\hss}

\chap{1.Introduction}1

 Models of statistical mechanics where particles (or spins) interact
through potentials $J_\g(r)\equiv \g^d J(\g r)$, $r\in \R^d$, with
$J$ some function that either has bounded support or is rapidly decreasing
were introduced by Kac et al. [KUH] in 1963 as links between
short-range, microscopic models and mean field theories such as the
van der Waals theory of the liquid-gas transition. The main success of these
models can be seen in that they explain, through the Lebowitz-Penrose theorem,
the origin of the Maxwell
rule that has to be invoked in an ad hoc way
to overcome the problem of the
non-convexity of the thermodynamic functions arising in mean-field theories.

Recently, there has been renewed interest in this model in the context of
attempting to obtain a precise description of equilibrium configurations [COP]
and their temporal evolution [DOPT] in magnetic systems at low temperatures.
In [COP] large deviation techniques were used to describe precisely the
profiles of local magnetization in a one dimensional Ising model with
Kac potential in infinite volume in the limit $\g\downarrow 0$. It turned out
that this apparently simple system exhibits a  surprisingly rich structure
when considered at appropriate scales and it appears that the Kac-type models
can still offer an interesting test ground for the study of low-temperature
phenomena. The purpose of the present paper is to extend such an analysis
to a class of models with {\it random interactions}.

Spin systems where spins at sites $i$ and $j$ interact through a
random coupling $J_{ij}$ whose mean value is zero (or close to
zero)
are commonly termed {\it spin glasses}.
The prototype models are the {\it Sherrington-Kirkpatrick
model }(SK-model) [SK] where the lattice is the completely
connected graph on $N$ vertices and the couplings $J_{ij}$ are
i.i.d. centered gaussian variables with variance $N^{-1/2}$, and
the {\it Edwards-Anderson model} [EA], defined on the lattice
$\Z^d$ and with $J_{ij}$  i.i.d. centered random variables
with variance $1$
if $i$ and $j$ are nearest neighbors in the lattice, whereas
$J_{ij}\equiv 0$ otherwise.

These systems
are notoriously difficult to analyse and little is known on a firm basis
about their low temperature properties. The situation is somewhat better
in the case of the {\it mean-field} SK-model, for which
 there is at least a rather elaborate
picture based on the so-called {\it replica-method} (for a review see
[MPV]) which is quite commonly accepted, although almost no results exist
that are mathematically rigorous. Exceptions concern the
high-temperature phase [ALR, FZ, CN, T1] and some  self-averaging
properties of the thermodynamic quantities [PS, BGP3]. For short-range models
(the Edwards-Anderson model [EA] the situation is much worse,
and there exist conflicting theories on such fundamental questions as the
upper and lower critical dimension and the number of low temperature
phases, all of which are more or less supported by heuristic arguments
(see e.g. [FH, BF, vE, NS]), and the
interpretation of numerical simulations on finite systems (for a recent
analysis and a critical assessment of the situation see [MPR]).

The difficulties with the SK-model have soon prompted the proposal of
simplified models for spin-glasses in which the statistics of the
random couplings was changed while some of the features
are conserved. The {\it Mattis-model} [Ma] where $J_{ij}\equiv \e_i\e_j$
with $\e_i$ independent symmetric Bernoulli variables was
realized to be trivially equivalent to a ferromagnet and lacking
the essential feature of {\it frustration}; Luttinger [Lu] amended
this by setting $J_{ij}\equiv \xi_i^1\xi_j^1+\xi_i^2\xi_j^2$ while
Figotin and Pastur [FP1, FP2] proposed and analysed a
generalization of this interaction with an arbitrary fixed number
of summands and more general distribution of the random variables
$\xi^\mu_i$. While these models could be solved exactly, they
lacked essential features expected for real spin glasses and thus
did not become very popular until they were again proposed in a
quite different context by Hopfield [Ho] as models for autoassociative
memory. Hopfield also considered the number of summands, $M$, to be
a function  of the size, $N$, of the graph (`network') and
observed numerically a drastic change of behaviour of the system
as the ratio $\a\equiv M/N$ exceeded a certain threshold.
This was confirmed by Amit et al. [AGS] through
a theoretical analysis using the replica trick. Indeed, the
Hopfield model can be seen as a family of models depending on the
different growth rate of $M(N)$ that mediates between simple
ferromagnets and the SK spin-glass.

The Hopfield model offers the advantage to be more amenable to a
mathematically rigorous analysis then the SK-model, at least as
long as $M(N)$ does not grow too fast with $N$. By now we have a
fairly complete understanding of the structure of the low
temperature Gibbs states [BGP1, BGP3, BG4] in the case where
$\lim_{N\uparrow \infty} M/N\leq \a_0$, for $\a_0$ sufficiently
small. It is thus interesting to take advantage of this situation
in order to get some insight into the relation between finite
dimensional  spin-glasses and the corresponding mean field models
by studying the finite dimensional version of the Hopfield model
with a Kac-type interaction. It should be noted that such a model
had already been considered by Figotin and Pastur [FP3] in 1982
in the case of bounded $M$. In a recent paper [BGP2] we have
proven the analogue of the classical Lebowitz-Penrose theorem for
this model, i.e. we have proven the convergence of the
thermodynamic functions to the convex hulls of those of the
mean-field model as $\g\downarrow 0$ under the condition that
$\lim_{\g\downarrow 0} M(\g)|\ln\g|/\g= 0$. In the present paper we
turn to the more detailed analysis of the Gibbs states of the
Kac-Hopfield model and consider, as a first step, the one
dimensional case along the lines of [COP].

Let us start by defining our model in a precise way and by fixing
our notations. Let $(\O,\FF,\P)$ be an
abstract probability space. Let $\xi\equiv
\{\xi_i^\mu\}_{i\in\Z,\mu\in\N}$
be a two-parameter family of independent, identically distributed random
variables on this space such that
$\P(\xi_i^\mu =1)=\P(\xi_i^\mu =-1)=\frac 12$.
(the precise form of the distribution of $\xi_i^\mu$ is not
really essential and far more general distributions can be
considered).
We denote by $\s$ a function $\s:\Z\rightarrow\{-1,1\}$ and call $\s_i$,
$i\in \Z$ the spin at site $i$. We denote by $\SS$ the space of
all such functions, equipped with the product topology of the
discrete topology in $\{-1,1\}$. We choose the function
$\jg(i-j)\equiv \g J\left(\g|i-j|\right)$, and
$$
J(x)=\cases{1,&if $|x|\leq 1/2$\cr 0,& otherwise}
\Eq(1.1)
$$
(Note that other choices for the function $J(x)$ are possible.
They must satisfy the conditions $J(x)\geq 0$, $\int dx J(x)=1$,
and must decay rapidly to zero on a scale of order unity. For
example, the original choice of Kac was $J(x)=e^{-|x|}$. For us,
the choice of the characteristic function is particularly
convenient).

The interaction between two spins at sites $i$ and $j$ will be
chosen for given $\o\in \O$, as
$$
-\frac
1{2}\sum_{\mu=1}^{M(\g)}
\xi_i^\mu[\o]\xi_j^\mu[\o] J_\g(i-j) \s_i\s_j
\Eq(1.2)
$$
and the {\it formal} Hamiltonian will be
$$
H_\g[\o](\s)=-\frac
1{2}\sum_{(i,j)\in\Z\times\Z}\sum_{\mu=1}^{M(\g)}
\xi_i^\mu[\o]\xi_j^\mu[\o] J_\g(i-j) \s_i\s_j
\Eq(1.3)
$$
As usual, to make mathematically meaningful statements, we have
to consider restrictions of this quantity to {\it finite
volumes}. We will do this in a particular way which requires some
prior discussion. Note that the parameter $\g$ introduces a
natural length scale $\g^{-1}$ into our model which is the
distance over which spins interact directly. We will be
interested later in the behaviour of the system on that and
larger scales and will refer to it as the {\it macroscopic
scale}, whereas the sites $i$ of the underlying lattice $\Z$ are
referred to as the microscopic scale. In the course of our
analysis we will have to introduce two more intermediate, {\it
mesoscopic scales}, as shall be explained later. We find it
convenient to measure distances and to define finite volumes in
the macroscopic rather than the microscopic scale, as this allows
to deal with volumes that actually do not change with $\g$.
Although this will require some slightly unconventional looking
definitions, we are convinced the reader will come to appreciate
the advantages of our conventions later on. Let thus
$\L=[\l_-,\l_+]\subset
\R$ be an interval on the real line. Thus for points $i\in \Z$
referring to sites on the microscopic scale we will write
$$
i\in\L\text{\it iff} \l_-\leq \g i\leq \l_+
\Eq(1.4)
$$
Note that we will stick very strictly to the convention that the
letters $i,j,k$ {\it always} refer to microscopic sites. The
Hamiltonian corresponding to a volume $\L$ (with free boundary
conditions) can then be written as
$$
H_{\g,\L}[\o](\s)=-\frac
1{2}\sum_{(i,j)\in\L\times\L}\sum_{\mu=1}^{M(\g)}
\xi_i^\mu[\o]\xi_j^\mu[\o] J_\g(i-j) \s_i\s_j
\Eq(1.5)
$$
We shall also write in the same spirit
$\SS_\L\equiv \times_{i\in\L} \{-1,1\}$ and
denote its elements by $\s_\L$. The {\it interaction} between the
spins in $\L$ and those outside $\L$ will be written as
$$
W_{\g,\L}[\o](\s_\L,\s_{\L^c})=-
\sum_{i\in\L}\sum_{j\in\L^c}\sum_{\mu=1}^{M(\g)}
\xi_i^\mu[\o]\xi_j^\mu[\o] J_\g(i-j) \s_i\s_j
\Eq(1.6)
$$
The finite volume Gibbs measure for such a volume $\L$ with fixed
external configuration $\s_{\L^c}$ (the `local specification')
is then defined by
assigning to each $\s_\L\in\SS_\L$ the mass
$$
\GG_{\b,\g,\L}^{\s_{\L^c}}[\o](\s_\L)\equiv
\frac 1{Z^{\s_{\L^c}}_{\b,\g,\L}[\o]}
e^{-\b \left[ H_{\g,\L}[\o](\s_\L)+W_{\g,\L}[\o](\s_\L,\s_{\L^c})\right]}
\Eq(1.7)
$$
where $Z^{\s_{\L^c}}_{\b,\g,\L}[\o]$ is a normalizing factor usually called 
{\it partition function}. 
We will also denote by 
$$
\GG_{\b,\g,\L}[\o](\s_\L)\equiv
\frac 1{Z_{\b,\g,\L}[\o]}
e^{-\b  H_{\g,\L}[\o](\s_\L)}
\Eq(1.7bis)
$$
the Gibbs measure with free boundary conditions.
It is crucial to keep in mind that we are always interested in
taking the infinite volume limit $\L\uparrow\R$ first for fixed
$\g$ and to study the asymptotic of the result as $\g\downarrow
0$ (this is sometimes referred  to as the `Lebowitz-Penrose limit').

In [BGP2] we have studied the distribution of the global `overlaps'
$m_\L^\mu(\s)\equiv\frac \g{|\L|}\sum_{i\in\L}\xi_i^\mu\s_i$
under the Gibbs measure (1.7). Here we are going into more detail 
in that we want to analyse the distribution of {\it local overlaps}. 
To do this we will actually have to introduce {\it two} intermediate
{\it mesoscopic} length scales, $1\ll \ell(\g)\ll L(\g)\ll \g^{-1}$. 
Note that both $\ell(\g)$ and $L(\g)$ will tend to infinity as $\g\downarrow 0$
while $\ell(\g)/L(\g)$ as well as $\g L(\g)$ tend to zero. 
We will assume that $\ell$, $L$ and $\g^{-1}$ are integer multiples
of each other.
Further conditions 
on this scales will be imposed later. To simplify notations, the dependence
on $\g$ of $\ell$ and $L$ will not be made explicit in the sequel.
 We now divide the real line into boxes of length
$\g\ell$ and $\g L$, respectively, with the first box, called $0$ being centered 
at the origin. The boxes of length $\g\ell$ will be called $x,y$, or $z$, 
and labelled by the integers. That is, the box $x$ is the interval of length
$\g\ell$ centered at the point $\g\ell x$. No confusion should arise from the 
fact that we use the symbol $x$ as denoting both the box {\it and}
its label, since again $x,y,z$ are used {\it exclusively} for 
this type of boxes. In the same way, the letters $r,s,t$ are reserved for 
the boxes of length $\g L$, centered at the points $ \g L \Z$, and finally
we reserve $u,v,w$ for boxes of length one centered at the integers. 
With these conventions, it makes sense to write e.g. $i\in x$ shorthand 
for $ \ell x-\ell/2\leq i\leq \ell x+ \ell/2$, 
etc.\note{On a technical level
we will in fact have to use even more auxiliary intermediate scales, but
as in [COP] we will try to keep this under the carpet as far as possible.} 
In this spirit we define the $M(\g)$ dimensional vector
$m_\ell(x,\s)$  and $m_L(r,\s)$ whose $\mu$-th components are
$$
m_\ell^\mu(x,\s)\equiv \frac 1{\ell} \sum_{i\in x}\xi^\mu_i\s_i
\Eq(1.8)
$$
and
$$
m_L^\mu(r,\s)\equiv \frac 1{ L} \sum_{i\in r}\xi^\mu_i\s_i
\Eq(1.9)
$$
respectively. 
Note that we have, for instance, that
$$
m_L^\mu(r,\s)=\frac {\ell}{L}\sum_{x\in r} m^\mu_\ell(x,\s)
\Eq(1.10)
$$
We will also have to be  able to indicate the box on some larger scale containing
a specified box on the smaller scale. Here we write simply, e.g.,
$r(x)$ for the unique box of length $L$ that contains the box $x$ of length 
$\ell$. Expressions like $x(i)$, $u(y)$ or $s(k)$ have corresponding meanings.
  
\remark It easy to connect from our notation to the continuum notation used in 
[COP]. For instance, \eqv(1.8) can be rewritten as
$$
m_\ell(x,u)=\frac 1{\g \ell}\g\sum_{i\in x}\xi^\mu_i\s_i
\Eq(1.11)
$$
where  $\g\sum_{i\in x}$ can be interpreted as a Riemann sum; the same occurs 
in  all other expressions. 
 
The r\^ole of the different scales will be the following. We will be 
interested  in the typical profiles of the overlaps on the scale $L$, 
i.e. the typical $m_L(r,\s)$ as a function of $r$; we will control 
these functions within volumes on the macroscopic scale $\g^{-1}$. The smaller 
mesoscopic scale $\ell$ enters only in an auxiliary way. Namely, we will use
a block-spin approximation of the Hamiltonian with blocks of that size. We 
will see that it is quite crucial to use a much smaller scale for that 
approximation than the scale on which we want to control the local overlaps.
This was noted already in [COP].    


We want to study the probability distribution induced by the Gibbs measure
on the functions $m_L(r)$ through the map defined by \eqv(1.9). The 
corresponding measure space is for fixed $\g$ simply the discrete space
$\{-1,-1+2/L,\dots,1-2/L,1\}^{M(\g)\times \Z}$, which should be equipped with 
the  product topology. Since this topology is quite non-uniform 
with respect to $\g$ (note that both $L$ and $M$ tend to infinity as
$\g\downarrow 0$), this is, however, not well adapted to 
take the limit $\g\downarrow 0$.  Thus we replace the discrete topology on 
$\{-1,-1+2/L,\dots,1-2/L,1\}^{M(\g)}$ by the Euclidean 
$\ell_2$-topology (which remains 
meaningful in the limit) and the product topology corresponding to $\Z$ 
is replaced by the weak local $L_2$ topology w.r.t. the measure 
$\g L\sum_{r\in \cdot}$; that is to say, a family of profiles
$m^n_L(r)$ converges to the profile $m_L(r)$, {\it iff} for all finite 
$R\in \R$, $\g L\sum_{r\in [-R,R]} 
\|m_L^n(r)-m_L(r)\|_2\downarrow 0$ as $n\uparrow\infty$. While for all finite
$\g$ this topology is completely equivalent to the product topology of the 
discrete topology, the point here is that it is meaningful to ask for 
{\it uniform} convergence with respect to the parameter $\g$.
We will denote this space by $\TT_\g$, or simply $\TT$ and call  it  the 
space of profiles (on scale $L$).

Before presenting our results, it may be useful to discuss
in a somewhat informal way the heuristic expectations based on the the work of 
[COP] and the results known from [BGP1, BGP3, BG4].  In [COP] it was shown that
the typical magnetization profiles are such that almost everywhere, $m_L(r,\s)$
is very close to one of the two equilibrium values of the mean field model,
$\pm a(\b)$; moreover, the profile is essentially constant over macroscopic
distances of the order $e^{\g^{-1}}$. The distances between 
jumps are actually independent exponentially distributed random variables.
Heuristically, this picture is not too difficult to understand. First, one 
approximates the Hamiltonian by a block-spin version by replacing the 
interaction potential by a function that is constant over blocks of length 
$L$. Ignoring the error term, the resulting model depends on $\s$ only
through the variables $m_L(r,\s)$. In fact,  at each block $r$
there is a little mean-field model and these mean field models interact 
through a ferromagnetic interaction of the form 
$J_{\g L}(r-s)(m_L(r)-m_L(s))^2$. This interaction can 
only bias a given block to choose between the two possible equilibrium 
values,  but never prevent it from taking on an equilibrium value over 
a longer interval. 
Moreover, it tends to align the blocks. To jump from one equilibrium into
the other costs in fact an energy of the order of $\g^{-1}$, so that the 
probability that this happens in a given unit interval is of the 
order $e^{-\g^{-1}}$. This explains why the entropy
can force this to happen only on distances of the order of the inverse of this 
value. Finally, the Markovian character of a one-dimensional model 
leaves only a Poisson-distribution as a candidate for the 
distribution of the jumps. The main difficulty in turning these
arguments  into rigorous proofs lies in the control of the error terms.

It is crucial for the above picture that there is a complete symmetry between
the two equilibrium states of the mean field model. As we have shown in [BGP2],
the Kac-Hopfield model can be approximated by a blocked model just the same, 
and in [BGP1] we have shown that the mean field Hopfield model
has its equilibrium states sharply concentrated at the $2M$ points 
$\pm a(\b) e^\mu$, where $e^\mu$ is the $\mu$-th standard unit vector. 
Thus we can again expect the overlap profiles to be over long distances
constant close to one of these values. What is different here, however, 
is that due to the disorder the different equilibrium positions are not 
entirely equivalent. We have shown in [BGP3] that the fluctuations are 
only of the order of the square root of the volume, but since they are 
independent from block to block, they can add up over a long distance 
and effectively enforce jumps to different equilibrium positions
at distances that are much shorter than those between entropic jumps. 
In fact, within the blocked approximation, it is not hard to estimate that
the typical distance over which the profiles remain constant should be of the
order $\g^{-1}$ on the macroscopic scale (i.e. $\g^{-2}$ on the microscopic 
scale). 
Using a  concentration of measure
estimates in a form  developed by M. Talagrand [T2], we extent these estimates 
to the full model. Our main results on the typical profiles can then be 
summarized (in a slightly informal way) as follows:

{\it Assume that 
$\lim_{\g\downarrow 0}\g M(\g)=0$. Then  there is a 
scale $L\ll\g^{-1}$  such that with $\P$-probability
tending to one (as $\g\downarrow 0$) the following holds:
\item{(i)} In any given macroscopic finite volume in any configuration 
that is ``typical'' with respect to the infinite volume 
Gibbs measure, for ``most'' blocks $r$, $m_L(r,\s)$ is very close to 
one of the values $\pm a(\b)e^\mu$ (we will say that $m_L(u,\s)$ 
is ``close to equilibrium'').
\item{(ii)}   In any macroscopic volume $\D$ that is small 
compared to $\g^{-1}$,
in a typical configuration, there  is at most one connected subset $J$ 
(called
a ``jump'')
with $|J|\sim \frac 1{\g L}$ on which $m_L$ is not close to equilibrium. 
Moreover, if such a jump occurs, 
then there exist $(s_1,\mu_1)$ and   $(s_2,\mu_2)$, such that 
for all $ u\in \D$ to the left of $J$, $m_L(u,\s)\sim s_1 a(\b)e^{\mu_1}$
and for all  $ u\in \D$ to the right  of $J$, 
$m_L(u,\s)\sim s_2 a(\b)e^{\mu_2}$
}

The precise statement of these facts will require more notation 
and is thus postponed to Section 6 where it will be stated as 
Theorem 6.15. That section contains also the 
large deviation estimates that are behind these results. 
We should mention that we have no result that would prove the existence 
of a ``jump'' in a sufficiently large region. We discuss this problem 
in Section 7 in some more detail.

We also remark that the condition $\lim_{\g\downarrow 0}\g M(\g)=0$ 
will be imposed thoughout the paper. It could be replaced with 
$\limsup_{\g\downarrow 0}\g M(\g)\leq \a_c(\b)$ 
for some strictly positive $ \a_c(\b)$ for
all $\b>1$. However, an actual estimate of this constant would be
outrageously tedious and does not really appear, in our view, to be worth 
the trouble.

The remainder of the paper is organized in the following way. The next two 
sections provide some technical tools that will be needed throughout.
Section 2 introduces the mesoscopic approximation of the Hamilitonian
and corresponding error estimates. Section 3 contains large deviation estimates
for the standard Hopfield model that are needed to analyse the mesoscopic 
approximation introduced before. Here we make use of some fundamental results
from [BGP2] and [BG3] but present them in a somewhat different form. 
In Section 4 we begin
the actual analysis of typical profiles. Here we show that for events that
are local, 
we can express
their probabilities in terms of a finite volume 
measure with random boundary conditions (see Corollary 4.2). 
In Section 5 we derive estimates on the random fluctuations of the 
free energies corresponding to these measures. 
In Section 6 we make use of these estimates to  show that
local events 
can be analysed 
using the mesoscopic approximation introduced in Section 2. This section is 
divided into three parts. Section 6.1 contains an analysis of measures 
with free boundary condition in macroscopic volumes of order $o\left(\g^{-1}
\right)$. 
It is shown that they are asymptotically concentrated on constant profiles
(see Theorem 6.1).
This result is already quite instructive, and technically rather easy.
In Sections 6.2 and 6.3 the measures with non-zero boundary conditions 
are studied. In Section 6.2 the case where the boundary conditions
are the same on both sides of the box. It is shown that here, too, 
the profiles are typically constant and take the value favored by the 
boundary conditions (see Theorem 6.9). In Section 6.3 the case with 
different boundary conditions is treated. Here we show that the typical profile
has exactly one ``jump'' and is constant otherwise (see Theorem 6.14).
The results of Sections 4 and 6 are then combined to yield Theorem 6.15
which gives a precise statement the result announced  above. 
In Section 7 we discuss some of the open points of our 
analysis. In particular we argue, that typical profiles are non-constant on a 
sufficiently large scale and that their precise form is entirely disorder 
determined (up to the global sign). We also formulate some conjectures 
for the model in dimensions greater than one. 
In Appendix A we give a proof 
of a technical
estimate on the minimal energy associated to profiles that 
contain ``jumps'' between different equilibrium positions that is needed in 
Section 6.
\newpage

\chap{2. Block-spin approximations}2

While mean-field models are characterized by the fact that the 
Hamiltonian is a function of global averages of the spin variables,
in Kac-models the Hamiltonian is ``close'', but not identical to a function 
of ``local''averages. In this section we make this statement precise
by introducing the block version of the Hamiltonian and deriving 
the necessary estimates  on the error terms. We define
$$
H_{\g,\L}(\s_{\L})=\g^{-1}E_{\g,\L}^\ell(m_\ell(\s)) + 
\D H_{\g,\L}^\ell(\s_{\L})
\Eq(2.1)
$$
and 
$$
W_{\g,\L}(\s_{\L},\s_{\L^c})=\g^{-1}E_{\g,\L}^{\ell,L}(m_\ell(\s),m_L(\s))
+\D W_{\g,\L}^{\ell,L}(\s_{\L},\s_{\L^c})
\Eq(2.2)
$$
where
$$
E_{\g,\L}^\ell(m)\equiv -\frac{1}{2}\g \ell\sum_{(x,y)\in\L\times\L} 
J_{\g \ell}(x-y)(m(x),m(y))
\Eq(2.3)
$$
and
$$
E_{\g,\L}^{\ell,L}(m, \tilde m)\equiv -\g \ell L\sum_{x\in\L}\sum_{r\in\L^c}
J_{\g}(\ell x-Lr)(m(x),\tilde m(r))
\Eq(2.4)
$$
For our purposes, we only need to consider volumes $\L$ of the form
$\L=[\l^-,\l^+]$ with $|\L|>1$. For such volumes we set
$\partial\L\equiv\partial^-\L\cup\partial^+\L$,
$\partial^-\L\equiv[\l^--\frac{1}{2},\l^-)$, and 
$\partial^+\L\equiv(\l^+,\l^++\frac{1}{2}]$. Thus, obviously,
$W_{\g,\L}(\s_{\L},\s_{\L^c})=W_{\g,\L}(\s_{\L},\s_{\partial\L})$
and $\D W_{\g,\L}^{\ell,L}(\s_{\L},\s_{\L^c})=
\D W_{\g,\L}^{\ell ,L}(\s_{\L},\s_{\partial\L})$.

\lemma {2.1} {\it For all $\d>0$
\item{i)}
$$
\P\left[\sup_{\s\in \SS_\L}
\frac{\g}{|\L|}|\D H_\L(\s)|\geq \g \ell (\g)8\sqrt 2(\log2 +\d)
+2\sqrt 2\g M(\g)
\right]\leq
16e^{-\d\frac{|\L|}{\g}}
\Eq(2.9)
$$
\item{ii)} 
$$
\P\left[\sup_{\s\in \SS_{\L\cup\partial\L}} 
\g |\D W_{\g,\L}^{\ell ,L}(\s_{\L},\s_{\partial\L})|>
(4\g L(\g)(\log2+\d) +\g M(\g))\left(1+\frac{\ell }{L}\right)^{\frac{1}{2}}
\right]\leq 8e^{-\frac{\d}{\g}}
\Eq(2.10)
$$
}



\proof We will give the proof of (ii) only; the proof of (i) is 
similar and can be found in [BGP2].
Since $|\L|>1$, the spins inside $\partial^-\L$ do not interact with those
inside $\partial^+\L$ and $\D W_{\g,\L}^{\ell ,L}(\s_{\L},\s_{\partial\L})$
can be written as
$$
\D W_{\g,\L}^{\ell ,L}(\s_{\L},\s_{\partial\L})=
\D W_{\g,\L}^{\ell ,L}(\s_{\L},\s_{\partial^-\L})+
\D W_{\g,\L}^{\ell ,L}(\s_{\L},\s_{\partial^+\L})
\Eq(2.11)
$$
where
$$
\D W_{\g,\L}^{\ell ,L}(\s_{\L},\s_{\partial^{\pm}\L})=-
\sum_{x\in\L}\sum_{r\in\partial^{\pm}\L}\sum_{i\in x}\sum_{j\in r}
[J_{\g}(i-j)-J_{\g}(\ell x-Lr)](\xi_i,\xi_j)\s_i\s_j
\Eq(2.12)
$$
Both terms \eqv(2.11) being treated similarly, we will only consider
$\D W_{\g,\L}^{\ell ,L}(\s_{\L},\s_{\partial^+\L})$. First notice that
since
$$
J_{\g}(i-j)-J_{\g}(\ell x-Lr)
=\g\left[
\1_{\{|i-j|\leq (2\g)^{-1}\}}\1_{\{|\ell x-Lr|>(2\g)^{-1}\}}
-\1_{\{|i-j|>(2\g)^{-1}\}}\1_{\{|\ell x-Lr|\leq(2\g)^{-1}\}}
\right]
\Eq(2.13)
$$
we can write $\D W_{\g,\L}^{\ell ,L}(\s_{\L},\s_{\partial^+\L})
=\g\left[\D^1 W_{\g,\L}^{\ell ,L}(\s_{\L},\s_{\partial^+\L}) 
-\D^2 W_{\g,\L}^{\ell ,L}(\s_{\L},\s_{\partial^+\L})\right]$
with
$$
\D^1 W_{\g,\L}^{\ell ,L}(\s_{\L},\s_{\partial^+\L})=-
\sum_{x\in\L}\sum_{r\in\partial^+\L}\sum_{i\in x}\sum_{j\in r}
\1_{\{|i-j|\leq (2\g)^{-1}\}}\1_{\{|\ell x-Lr|>(2\g)^{-1}\}}
(\xi_i,\xi_j)\s_i\s_j
\Eq(2.14)
$$
and
$$
\D^2 W_{\g,\L}^{\ell ,L}(\s_{\L},\s_{\partial^+\L})=-
\sum_{x\in\L}\sum_{r\in\partial^+\L}\sum_{i\in x}\sum_{j\in r}
\1_{\{|i-j|>(2\g)^{-1}\}}\1_{\{|\ell x-Lr|\leq(2\g)^{-1}\}}
(\xi_i,\xi_j)\s_i\s_j
\Eq(2.15)
$$
Again, both terms $\D^1 W_{\g,\L}^{\ell ,L}(\s_{\L},\s_{\partial^+\L})$ and 
$\D^2 W_{\g,\L}^{\ell ,L}(\s_{\L},\s_{\partial^+\L})$ can be treated in the 
same way so that we only present an estimate of the former. 
Using the identity
$$
\1_{\{|i-j|\leq (2\g)^{-1}\}}\1_{\{|\ell x-Lr|>(2\g)^{-1}\}}=
\1_{\{|i-j|\leq (2\g)^{-1}\}}\1_{\{(2\g)^{-1}<|\ell x-Lr|\leq(2\g)^{-1}
+(\ell +L)/2\}}
\Eq(2.16)
$$
and setting
$$
g_{\g,\L}^{\mu}(r)=
\sum_{{x\in\L : }\atop{(2\g)^{-1}<|\ell x-Lr|\leq(2\g)^{-1}+(\ell +L)/2}}
\sum_{i\in x}\sum_{j\in r}
\1_{\{|i-j|\leq (2\g)^{-1}\}}
\xi_i^{\mu}\xi_j^{\mu}\s_i\s_j
\Eq(2.17)
$$
we have
$$
\D^1 W_{\g,\L}^{\ell ,L}(\s_{\L},\s_{\partial^+\L})=-
\sum_{\mu=1}^M\sum_{r\in\partial^+\L} g_{\g,\L}^{\mu}(r)
\Eq(2.18)
$$
Note that the right hand side of \eqv(2.18) is a sum of independent random 
variables since for any two distinct $r_1\,,\,r_2$ in $\partial^+\L$, 
the sets $\{x\in\L:{(2\g)^{-1}<|\ell x-Lr_1|\leq(2\g)^{-1}+(\ell +L)/2}\}$ and 
$\{x\in\L:{(2\g)^{-1}<|\ell x-Lr_2|\leq(2\g)^{-1}+(\ell +L)/2}\}$ are disjoint.
Therefore,
$$
\P\left[\sup_{\s\in \SS_{\L\cup\partial^+\L}} {\g}^2 
|\D^1 W_{\g,\L}^{\ell ,L}(\s_{\L},\s_{\partial^+})|>\frac{\e}{4}\right]
\leq 2^{({\g}^{-1}+1)}
\P\left[\sum_{\mu=1}^M\sum_{r\in\partial^+\L} g_{\g,\L}^{\mu}(r)
>{\g}^{-2}\frac{\e}{4}\right]
\Eq(2.19)
$$
where the probability in the right hand side is independent of the 
chosen spin configuration $\s_{\L\cup\partial^+\L}$. For convenience we 
will choose the configuration whose spins are all one's. Using the
exponential Markov inequality together with the independence, we
get 
$$
\P\left[\sup_{\s\in \SS_{\L\cup\partial^+\L}} {\g}^2 
|\D^1 W_{\g,\L}^{\ell ,L}(\s_{\L},\s_{\partial^+\L})|>\frac{\e}{4}\right]
\leq 2^{({\g}^{-1}+1)}
\inf_{t\geq 0} e^{-t{\g}^{-2}\frac{\e}{4}}
\left[\prod_{r\in\partial^+\L}\E e^{t g^1_{\g,\L}(r)}\right]^{M}
\Eq(2.20)
$$
Thus we have to estimate the Laplace-transform of $g^1_{\g,\L}(r)$
for any $r\in\partial^+\L$. We write
$$
\E e^{tg^1_{\g,\L}(r)}=\E \exp\left\{t
\sum_{j\in r}\xi_j^{1}
\sum_{{x\in\L : }\atop{(2\g)^{-1}<|\ell x-Lr|\leq(2\g)^{-1}+(\ell +L)/2}}
\sum_{i\in x}\1_{\{|i-j|\leq (2\g)^{-1}\}}\xi_i^{1}
\right\}
\Eq(2.21)
$$
Note that all the $\xi^1_j$ with $j\in r$ are independent of
the $\xi_i^1$ with $i\in x$ for $x$ satisfying
$(2\g)^{-1}<|\ell x-Lr|\leq(2\g)^{-1}+(\ell +L)/2$, and that,
conditioned on these latter variables, the variables
$\xi_j^{1}\sum_{x\in\L}\1_{\{(2\g)^{-1}<|\ell x-Lr|\leq(2\g)^{-1}+(\ell +L)/2\}}
\1_{\{|i-j|\leq (2\g)^{-1}\}}\xi_i^{1}$ are independent. 
If we denote by $\E_j$ the expectation w.r.t. $\xi^1_j$, this allows us 
to write
$$
\eqalign{
\E e^{tg^1_{\g,\L}(r)}&=\E \prod_{j\in r}\E_j
\exp\left\{t\xi_j^{1}
\sum_{{x\in\L : }\atop{(2\g)^{-1}<|\ell x-Lr|\leq(2\g)^{-1}+(\ell +L)/2}}
\sum_{i\in x}\1_{\{|i-j|\leq (2\g)^{-1}\}}\xi_i^{1}
\right\}\cr
&\leq
\E \prod_{j\in r}
\exp\left\{\frac{t^2}{2}\left(
\sum_{{x\in\L : }\atop{(2\g)^{-1}<|\ell x-Lr|\leq(2\g)^{-1}+(\ell +L)/2}}
\sum_{i\in x}\1_{\{|i-j|\leq (2\g)^{-1}\}}\xi_i^{1}\right)^2
\right\}
\cr
}
\Eq(2.22)
$$
where we have used that $\ln\cosh x\leq  \frac 12 x^2$. Using the
H\"older-inequality on the last line, we arrive at
$$
\E e^{tg^1_{\g,\L}(r)}\leq
\prod_{j\in r}\left[
\E \exp\left\{\frac{Lt^2}{2}\left(
\sum_{{x\in\L : }\atop{(2\g)^{-1}<|\ell x-Lr|\leq(2\g)^{-1}+(\ell +L)/2}}
\sum_{i\in x}\1_{\{|i-j|\leq (2\g)^{-1}\}}\xi_i^{1}\right)^2
\right\}\right]^{\frac{1}{L}}
\Eq(2.23)
$$
Now
$$
\eqalign{
&\E \exp\left\{\frac{Lt^2}{2}\left(
\sum_{{x\in\L : }\atop{(2\g)^{-1}<|\ell x-Lr|\leq(2\g)^{-1}+(\ell +L)/2}}
\sum_{i\in x}\1_{\{|i-j|\leq (2\g)^{-1}\}}\xi_i^{1}
\right)^2\right\}\cr
\leq &
\E \exp\left\{\frac{Lt^2}{2}\left(
\sum_{{x\in\L : }\atop{(2\g)^{-1}<|\ell x-Lr|\leq(2\g)^{-1}+(\ell +L)/2}}
\sum_{i\in x}\xi_i^{1}\right)^2
\right\}\cr
\leq &
\frac 1{\sqrt{1-t^2 L(L+\ell )/2}}\cr
}
\Eq(2.24)
$$
where we have used the Khintchine inequality and the fact
that, for all $r\in\partial^+\L$,  
$$
\sum_{x\in\L}\sum_{i\in x}
\1_{\{(2\g)^{-1}<|\ell x-Lr|\leq(2\g)^{-1}+(\ell +L)/2\}}\leq \frac{L+\ell }{2}
\Eq(2.241)
$$ 
Since for $0\leq x\leq 1/2$, $1/\sqrt{1-x}\leq e^{x}$,  
for $t^2\leq \frac 1{\ell (L+\ell )}$,
we finally get, collecting \eqv(2.22)-\eqv(2.24),
$$
\E e^{tg^1_{\g,\L}(r)}\leq e^{t^2 \frac{L(L+\ell )}{2}}
\Eq(2.25)
$$ 
Therefore, since 
$\sharp\{r\in\partial^+\L\}\leq (2\g L)^{-1}$, choosing 
$t=\frac{1}{\sqrt{L(L+\ell )}}$ in \eqv(2.25) yields
$$
\P\left[\sup_{\s\in \SS_{\L\cup\partial^+\L}} {\g}^2 
|\D^1 W_{\g,\L}^{\ell ,L}(\s_{\L},\s_{\partial^+\L})|>\frac{\e}{4}\right]
\leq 2^{2/\g+1} \exp\left\{-\frac{1}{\g}
\left[\frac{\e}{4\g\sqrt{L(L+\ell )}}
\right]\right\} \exp\left\{\frac{M}{4\g L}\right\}
\Eq(2.26)
$$
Choosing $\e$ in 2.6  as 
$\e=4\g\sqrt{L(L+\ell )}\left(\log 2 +\frac {M(\g)}{4\ell (\g)}+\d\right)$
for some $\d>0$, gives \eqv(2.10).
\endproof

\newpage

\chap{3. Some large deviation estimates for the Hopfield model}3

In the preceeding chapter we have introduced the block-approximation
for the Hamiltonian of the Kac-Hopfield model. To make use of these,
we need some large deviation results for the standard Hopfield model.
They are essentially contained in [BGP1] and [BGP2],
but we present them here in a
slightly different way that focuses on our actual needs.
We set $\frac MN\equiv \a$ throughout this section.

Recall that we have to consider the quantities
$$
Z_{N,\b,\rho}( m) \equiv 2^{-N} \sum_{\s\in\SS_N}
e^{\frac {\b N}2 \|m_N[\o](\s)\|_2^2}
\1_{\left\{\|m_N(\s)-m\|_2\leq \rho\right\}}
\Eq(B.1)
$$
We set $f_{N,\b,\rho}(m)\equiv -\frac 1{\b N}\ln
Z_{N,\b,\rho}( m)$.
In this paper we are mostly interested in the localization of the minima
of the functions $f_{N,\b,\rho}(m)$.
Thus we will only need the following estimates:

\lemma {3.1} {\it Define the random function
$$
\Phi_{N,\b}(m)\equiv \frac 12\|m\|_2^2-\frac 1{\b N}\sum_{i=1}^N
\ln\cosh\left(\b(\xi_i,m)\right)
\Eq(B.01)
$$
Then
$$
f_{N,\b,\rho}(m)\geq
\Phi_{N,\b}(m)
-\sfrac 12 \rho^2
\Eq(B.012)
$$
and for $\rho\geq \sqrt{2\a}$, if $m^*$ is a critical point of
$\Phi_{N,\b}(m)$,
$$
f_{N,\b,\rho}(m^*)\leq
\Phi_{N,\b}(m^*)
+\sfrac {\ln 2}{\b N}
\Eq(B.013)
$$
}

\proof To prove Lemma 3.1, we define
probability measures $\tilde \P$ on $\{-1,1\}^N$
through their expectation $\tilde\E_\s$, given by
$$
\tilde \E_\s\bigl(\cdot\bigr)\equiv\frac{\E_\s e^{\b N\left(m,m_N(\s)\right)}
\bigl(\cdot\bigr)}{\E_\s e^{\b N\left(m,m_N(\s)\right)}}
\Eq(B.7)
$$
We have obviously that
$$
\eqalign{
Z_{N,\b,\rho}(m)&= \tilde\E_\s e^{\frac {\b N}2\|m_N(\s)\|_2^2-
\b N\left(m,m_N(\s)\right)}
\1_{\left\{\| m_N(\s)-m\|_2\leq \rho\right\}}
\E_\s e^{\b N\left(m,m_N(\s)\right)}
\cr
&=
e^{-\frac {\b N}2 \|m\|_2^2}
 \tilde\E_\s e^{\frac {\b N}2\left\|m_N(\s)-m\right\|_2^2}
\1_{\left\{\| m_N(\s)-m\|_2\leq \rho\right\}}
\E_\s e^{\b N\left(m,m_N(\s)\right)}\cr
&=e^{\b N\left(-\frac 12\|m\|_2^2+\frac 1{\b N}
\sum_{i=1}^N\ln\cosh\b(\xi_i,m)\right)}
\tilde\E_\s e^{\frac {\b N}2\left\|m_N(\s)-m\right\|_2^2}
\1_{\left\{\| m_N(\s)-m\|_2\leq \rho\right\}}
}
\Eq(B.08)
$$
But
$$
\1_{\left\{\| m_N(\s)-m\|_2\leq \rho\right\}}\leq
e^{\frac {\b N}2\left\|m_N(\s)-m\right\|_2^2}
\1_{\left\{\| m_N(\s)-m\|_2\leq \rho\right\}}
\leq
e^{\frac {\b N}2 \rho^2}\1_{\left\{\| m_N(\s)-m\|_2\leq \rho\right\}}
\Eq(B.09)
$$
so that we get on the one hand
$$
Z_{N,\b\rho}(m)\leq e^{-\b N \left[\Phi_{N,\b}(m)-\frac 12\rho^2\right]}
\Eq(B.0008)
$$
which yields \eqv(B.012), and on the other hand
$$
Z_{N,\b\rho}(m)\geq
e^{-\b N\Phi_{N,\b}(m)}
\tilde \P\left[\| m_N(\s)-m\|_2\leq \rho\right]
\Eq(B.8)
$$
But, using Chebychev's inequality, we have that
$$
\tilde\P\left[\|m_N(\s)-m\|_2\leq \rho\right]
\geq 1-\frac 1{\rho^2} \tilde\E_\s \|m_N(\s)-m\|_2^2
\Eq(B.9)
$$
and
$$
\eqalign{
&\tilde\E \|m_N(\s)-m\|_2^2=\cr
&\frac{\E_\s \prod_{i=1}^Ne^{\b (m,
\xi_i\s_i)}\sum_{\nu}\left(
N^{-2}\sum_{j,k}\xi_j^\nu\xi_k^\nu\s_j\s_k -2m^\nu N^{-1}\sum_j
\mu_j^\nu\s_j+(m^\nu)^2\right)}{\prod_{i=1}^N\cosh\b (\xi_i,m)}\cr
&=\frac 1{N^2}\sum_\nu\sum_j 1+\frac 1{N^2}
\sum_{\nu}\sum_{j\neq k}\tanh(\b (m,\xi_j))
\tanh(\b (m,\xi_k))\xi_j^\nu\xi_k^\nu\cr
&
-\frac 2N\sum_{j}\sum_{\nu} m^\nu\tanh(\b (m,\xi_j))\xi_j^\nu
+\sum_\nu (m^\nu)^2\cr
&=\frac MN
-\sum_{\nu}\frac 1N\sum_i\tanh^2(\b(m,\xi_i))+
\sum_\nu\left(\frac 1N \sum_i \xi_i^\nu\tanh(\b(m,\xi_i))-m^\nu\right)^2
}
\Eq(B.22)
$$
IF  $m^*$ is a critical point  of $\Phi$, 
$$
m^*=\frac 1N\sum_i \xi_i\tanh(\b( m^*,\xi_i))
\Eq(B.23)
$$
and so the last terms in \eqv(B.22) vanish and we remain with
$$
\tilde\E \|m_N(\s)-m\|_2^2\leq \frac MN\left(1-\frac 1N\sum_i\tanh^2(\b
(\xi_i,m))\right)\leq \a
\Eq(B.24)
$$
from which \eqv(B.013) follows immediately.\endproof

Given the upper and lower bounds in terms of $\Phi$, it remains to show that
this function takes its absolute minima near the points
$m^{(\mu,s)}\equiv s a(\b)e^\mu$ only. This
was done in [BGP1] and, with sharper estimates in [BG3].
We recall this result
in a form suitable for our purposes. 
We denote by $a(\b)$ the positive solution
of the equation $a=\tanh(\b a)$. 

\proposition {3.2} {\it Assume that $\sqrt\a /a(\b)^2$ is sufficiently 
small.
Then there exists a set $\O_4(N)\subset\O$ with
$\P(\O_4(N))\geq 1-e^{-c M}$
such that for all $\o\in\O_4$, for all $m\in\R^M$
$$
\Phi_{N,\b}[\o](m)-\Phi_{N,\b}[\o](m^{(\mu,s)})\geq \e(m)
\Eq(B.27)
$$
where  
$\e$ is a non random function that satisfies 
$$
\e(m)=\cases { 0,&if $\,\,\inf_{\mu,s}\|m-m^{(\mu,s)}\|_2\leq c_1 
               \sqrt\a/a(\b)$\cr
              c a(\b)^2 \inf_{\mu,s}\|m-m^{(\mu,s)}\|_2^2,&
              if $\,\, c_1 \sqrt\a/a(\b)\leq\inf_{\mu,s}\|m-m^{(\mu,s)}\|_2\leq
              c_2 a(\b)$\cr
              c c_2 a(\b)^4,& if $\,\,\inf_{\mu,s}\|m-m^{(\mu,s)}\|_2\geq
              c_2 a(\b)$
}
\Eq(B.28)
$$
where $c,c_1,c_2$ are finite positive constants.
}

\proof By some trivial
changes of notations  this follows from the estimates in Section 
3 of [BG3], in particular Theorem  3.1 and 
Lemma 3.9. \endproof

\newpage
\def\hrho{\hat\rho}
\overfullrule=0pt
\chap{4. Local effective measures}4

In Section 2 we have seen that the Kac-Hopfield Hamiltonian 
can be approximated by a block-spin Hamiltonian up to errors that
are essentially proportional to $\g\ell$ times the volume. This means of
course that we cannot use this approximation throughout the 
entire volume $\L$ if we are interested in controlling local observables, 
as the  errors would grow without bounds in the thermodynamic limit. 
A clever way to solve this difficulty was given in [COP] for the   
ferromagnetic Kac-model. The crucial point is that 
if one is interested in local observables in a box $V$, it is possible to 
show that with large probability
(w.r.t. the Gibbs measure) not far away from this volume, 
there are intervals of 
macroscopic length $1$ where the mesoscopic magnetization profiles are 
very close to one of the ``equilibrium'' values of the mean-field model.
This knowledge allows to 
effectively decouple the system inside and outside this region, 
with the outside acting only as a ``boundary condition''. 
Due to the randomness of the interaction, an additional difficulty presents 
itself in terms of the randomness of the effective boundary conditions. 
This makes it necessary to perform this analysis on two separate length
scales: in this section we consider a rather large volume 
(which we will see later can be chosen of order $o(\g^{-1})$
(on the macroscopic scale); 
in Section 6 these measures will be further
analyzed by localizing them to much smaller boxes.

To begin, we imitate [COP] by defining variables $\eta$ that serve as a 
decomposition of the configuration space through
$$
\eta(u,\s)\equiv\eta_{\z,L}(u,\s)=\cases{
s e^{\mu} &if $\,\forall_{r\in u}\, 
\|m^{(\mu,s)}-m_L(r,\s)\|_2\leq\z$\cr
0 &if $\,\forall_{\mu,s}\,\exists_{r\in u} : 
\|m^{(\mu,s)}-m_L(r,\s)\|_2>\z$ \cr
}
\Eq(4.1)
$$
(This definition is unequivocal if $\z$ is chosen small enough i.e.
$\z< \sqrt 2 a(\b)$).
For a given configuration $\s$, $\eta$ determines whether a unit interval is 
close to equilibrium on the scale $L$.
For a given volume $V\equiv[v_-,v_+]\subset \L$, with $|V|>1$, we set
$$
\tau^+=\cases{
\inf\{u\geq v_+ : \eta(u,\s)\neq 0\}   &\cr
\infty \enskip\hbox{if no such}\enskip u\enskip\hbox{ exists} &\cr
}
\Eq(4.2)
$$
and
$$
\tau^-=\cases{
\sup\{u\leq v_- : \eta(u,\s)\neq 0\}   &\cr
-\infty \enskip\hbox{if no such}\enskip u\enskip\hbox{ exists} &\cr
}
\Eq(4.3)
$$
for a given configuration $\s$, $\t^\pm$ indicates the position of the 
first unit interval to the right, respec. the left,
of $V$ where the configurations $\s$ is close to equilibrium.

Let us introduce the indices $\mu^+, \mu^-, s^+, s^-, w_+, w_-$ where
$\mu^{\pm}\in\{1,\dots,M(\g)\}$, $s^{\pm}\in\{-1,1\}$
and $w_+\in [v_+,\infty)$, $w_-\in (-\infty, v_-]$. In the sequel, if not 
otherwise specified, all sums and unions over these indices run over
the above sets. With these notations we define a partition
of the configuration space $\SS$ 
whose atoms are given by
$$
\AA(\mu^{\pm}, s^{\pm}, w_{\pm})\equiv
\left\{
\s\in\SS : \tau^{\pm}=w_{\pm} , \eta(\tau^{\pm},\s)=s^{\pm}e^{\mu^{\pm}} 
\right\}
\Eq(4.4)
$$
and we denote by
$$
\SS_R=\bigcup_{{\mu^{\pm}, s^{\pm}, w_{\pm}}\atop 
{ 0\leq \pm (w_\pm-v_\pm)\leq R}}
\AA(\mu^{\pm}, s^{\pm}, w_{\pm})
\Eq(4.5)
$$
Notice that 
$$
\SS_R^c = A^+(R) \cup A^-(R)
\Eq(4.5bis)
$$
where
$$
 A^+(R)\equiv \left\{
\s\in\SS : \tau^+ > v_+ + R \right\}= \left\{
\s\in\SS : \forall_{v_+ \leq w \leq  v_+ + R }\,\, \eta(w,\s)=0\right\}
\Eq(4.5ter) 
$$
and 
$$
 A^-(R)\equiv \left\{
\s\in\SS : \tau^- < v_- - R \right\}= \left\{
\s\in\SS : \forall_{v_- -R\leq w \leq  v_- }\,\, \eta(w,\s)=0\right\}
\Eq(4.5quater) 
$$

%
Before stating the main results of this chapter we need some further
notations. For given indices $\mu^{\pm}, s^{\pm}, w_{\pm}$ we write 
$\D\equiv[w_-+\frac{1}{2},w_+-\frac{1}{2}]$ and we set
$$
\widehat\AA(\mu^{\pm}, s^{\pm}, w_{\pm})\equiv
\left\{\s\in\SS : \eta(w_{\pm},\s)=s^{\pm}e^{\mu^{\pm}}\right\}
\Eq(4.700)
$$ 
We define the Gibbs measure on $\D$ with 
{\it mesoscopic boundary conditions} $m^{(\mu^{\pm}, s^{\pm})}$ as
the measure that assigns, to each $\s_{\D}\in\SS_{\D}$, the mass,
$$
\GG_{\b,\g,\D}^{\mu^{\pm},s^{\pm}}[\o](\s_{\D})=
\frac{1}{Z_{\b,\g,\D}^{\mu^{\pm},s^{\pm}}[\o]}
e^{-\b \left\{H_{\g,\D}[\o](\s_{\D})
+W_{\g,\D}[\o](\s_{\D},m^{(\mu^{\pm}, s^{\pm})})\right\}}
\Eq(4.7)
$$
where $Z_{\b,\g,\D}^{\mu^{\pm},s^{\pm}}[\o]$ is the corresponding
normalization factor and
$$
W_{\g,\D}[\o](\s_{\D},m^{(\mu^{\pm}, s^{\pm})})\equiv
-\sum_{i\in\D}s^-a(\b)\xi^{\mu^-}_i\s_i\sum_{j\in\partial^-\D}J_\g(i-j)
-\sum_{i\in\D}s^+a(\b)\xi^{\mu^+}_i\s_i\sum_{j\in\partial^+\D}J_\g(i-j)
\Eq(4.8)
$$

\proposition{4.1} {\it Let $F$ be a cylinder event with base contained in 
$[v_-,v_+]$. Then
\item{i)} 
There exists a positive constant $c$ such that, for all integer $R$, 
there exists $\O_R$ with $\P(\O_R)\geq 1-Re^{-c \g^{-1}}$
such that
for all $ \mu^\pm, s^\pm, w_\pm,
v_+\leq w_+ \leq v_+ +R, v_-- R \leq w \leq v_- $ and $\o \in \O_R$ 
for all $\L\supset [v_--R,v_++R]$
$$
\GG_{\b,\g,\L}[\o]\Bigl(F\cap \AA(\mu^{\pm}, s^{\pm}, w_{\pm})\Bigr)
\leq
\GG_{\b,\g,\D}^{\mu^{\pm},s^{\pm}}[\o]
\left(F\right)
\GG_{\b,\g,\L}[\o]\left(\widehat\AA(\mu^{\pm}, s^{\pm}, w_{\pm})\right)
e^{8\b\g^{-1}(\zeta+2\g L)}
\Eq(4.9)
$$
and for any $u_+\geq v_+$, $u_-\leq v_-$,
$$
\eqalign{
&\GG_{\b,\g,\L}[\o]\Bigl(F\cap \widehat\AA(\mu^{\pm}, s^{\pm}, u_{\pm})\Bigr)
\geq
\GG_{\b,\g,[u_-,u_+]}^{\mu^{\pm},s^{\pm}}[\o]
\left(F\right)
\GG_{\b,\g,\L}[\o]\left(\widehat\AA(\mu^{\pm}, s^{\pm}, u_{\pm})\right)
e^{- 8\b\g^{-1}(\zeta+2\g L)}
}\Eq(4.9bis)
$$

\item{ii)} There exist a positive constant $c'$ such that for all integer $R$,
there exists $\O_R$ with $\P(\O_R)\geq 1-\g^{-1}Re^{-c'M}$ and there exist 
finite positive 
constants $c_1$ and $c_2$ such that if
$\z\e(\z)\g L>2 c_1 \sqrt {\frac M\ell} $, then
for all  $\o \in \O_R$ and  $\L\supset [v_--R,v_++R]$
$$
\GG_{\b,\g,\L}[\o](F\cap S_R^c)
\leq \exp\left(-\b  LR c_2\z\e(\z)\right)
\Eq(4.10)
$$
}

\corollary{4.2}{\it Let $F$ be a cylinder event with base contained in 
$[v_-,v_+]$. 
Then there exist a positive constant $c'$ such that for all 
integer $R$,
there exists $\O_R$ with $\P(\O_R)\geq 1-\g^{-1}Re^{-c'M}$ and there exist 
finite positive 
constants $c_1$ and $c_2$ such that if
$\z\e(\z)\g L>2 c_1 \sqrt {\frac M\ell} $, then
 for all  $\o \in \O_R$ and  $\L\supset [v_--R,v_++R]$
$$
\eqalign{
\GG_{\b,\g,\L}[\o](F) \leq &
\sum_{{{\mu^{\pm}, s^{\pm}}\atop{-R<w_-\leq v_-}}\atop{v_+\leq w_+<R}}
\GG_{\b,\g,\D}^{\mu^{\pm},s^{\pm}}[\o]
\left(F\right)
\GG_{\b,\g,\L}[\o]\left(\widehat\AA(\mu^{\pm}, s^{\pm}, w_{\pm})\right)
e^{8\b\g^{-1}(\zeta+2\g L)}\cr
&+ \exp\left(-\b  LR c_2\z\e(\z)\right)
}
\Eq(4.10bis)
$$
and there exist $u_\pm $ with $\pm (u_\pm-v_\pm)\leq R$
 such that for all  $\L\supset [v_--R,v_++R]$
$$
\GG_{\b,\g,\L}[\o](F) \geq
\sum_{\mu^\pm,s^\pm}
\GG_{\b,\g,[u_-,u_+]}^{\mu^{\pm},s^{\pm}}[\o]
\left(F\right)
\GG_{\b,\g,\L}[\o]\left(\widehat\AA(\mu^{\pm}, s^{\pm}, u_{\pm})\right)
e^{-8\b\g^{-1}(\zeta+2\g L)}
\Eq(4.10ter)
$$
and there exists $(\mu^\pm,s^\pm)$ such that 
$$
\GG_{\b,\g,\L}[\o]\left(\widehat\AA(\mu^{\pm}, s^{\pm}, u_{\pm})\right)
\geq \frac 1{8R^2M^2}
\Eq(4.10terbis)
$$
}

\remark Corollary 4.2 tells us that in order to estimate the probability of
 some local event in $V$ with respect to the infinite volume Gibbs measure
we only need to control finite volume Gibbs measures in volumes $|\D|$
with all possible boundary conditions corresponding to 
one of the mean field equilibrium states. This analysis will be performed in 
Section 6. On the other hand, it appears quite hopeless to get a more precise 
information than \eqv(4.17) on the terms 
$\GG_{\b,\g,\L}[\o]\left(\widehat\AA(\mu^{\pm}, s^{\pm}, u_{\pm})\right)$
appearing in both bounds. This is, after some thought, not surprising, but 
reflects the fact that the exact shape of typical profiles depends strongly 
on the disorder and only some of their properties on relatively short 
scales can be effectively controlled. In particular, it is clear 
that we cannot hope to get something like a full large deviation 
principle (in the sense of the results of [COP] in the 
deterministic case) for the infinite volume Gibbs measures.

\proof The first assertion of Corollary 4.2 is obvious from \eqv (4.9) and 
\eqv(4.10). 
To prove the second, we need to show that 
$$
\sup_{\mu^\pm, s^\pm}\sup_{\pm(u_\pm-v_\pm)\leq R}
\GG_{\b,\g,\L}[\o]\left(\widehat\AA(\mu^{\pm}, s^{\pm}, u_{\pm})\right)
\geq \frac 1{8R^2M^2}
\Eq(4.10quater)
$$
But from \eqv(4.10) we see that
$$
\eqalign{
\frac 12&\leq 1- \exp\left(-\b  LR c_2\z\e(\z)\right)
\leq 1-\GG_{\b,\g,\L}[\o](S_R^c)\cr
&\leq \GG_{\b,\g,\L}[\o]\left(\t_+\leq v_++R,\t_-\geq v_--R\right)\cr
&\leq \sum_{\pm(u_\pm-v_\pm)\leq R} \GG_{\b,\g,\L}[\o]\left(\t_-=u_-,\t_+=u_+
\right)\cr
&\leq \sum_{\pm(u_\pm-v_\pm)\leq R} \GG_{\b,\g,\L}[\o]\left(
\eta(u_-,\s)\neq 0,\eta(u_+,\s)\neq 0\right)\cr
&\leq  \sum_{\pm(u_\pm-v_\pm)\leq R}\sum_{\mu^\pm,s^\pm}
\GG_{\b,\g,\L}[\o]\left(
\eta(u_-,s)=s^- e^{\mu^-},\eta(u_+,s)=s^+ e^{\mu^+}\right)\cr
&\leq 4R^2 M^2 \sup_{\pm(u_\pm-v_\pm)\leq R}\sup_{\mu^\pm,s^\pm}
 \GG_{\b,\g,\L}[\o]\left(\hat\AA(\mu^\pm,s^\pm, u_\pm)\right)
}
\Eq(4.10cinq)
$$
which gives \eqv(4.10quater). 
\endproof

In order to prove Proposition 4.1, we need  the following lemmata.

\lemma{4.3}{\it For any finite subset $\G\subset \Z$
we denote by $A(\G)$  the $M\times M$-matrix with elements
$$
A_{\mu,\nu}(\G)=\frac{1}{|\G|}\sum_{i\in \G}\xi_i^{\mu}\xi_i^{\nu}
\Eq(4.11)
$$
and let $B$ be the $N\times N$-matrix with entries
$$
B_{i,j}=\frac{1}{N}\sum_{\mu=1}^M\xi_i^{\mu}\xi_j^{\mu}
\Eq(4.11bis)
$$
Set $N=|\G|$ and assume that $M\geq N^{1/6}$.
Then,
\item{(i)} 
$$
\E\|A(\G)-\1\| \leq \sqrt{\sfrac M{N}}\left(2+ \sqrt{\sfrac M{N}}\right)
+C\sfrac {\ln N}{N^{1/6}}
\Eq(4.12)
$$
and 
\item {(ii)} There exists a universal constant $K<\infty$ such that for all
$0\leq \d\leq 1$.
$$
\P\left[\bigl|\|B(\G)\|- \E\|B(\G)\|\bigr|>\d\right]
\leq K\exp\left(-N\frac {\d^2}{2K}\right)
\Eq(4.13)
$$
In particular,
$$
\P\left[\|A(\G)\|\geq \left(1+\sqrt{\sfrac M{N}}\right)^2(1+\d)
\right] \leq K\exp\left(-N\frac {\d^2}{2K}\right)
\Eq(4.14)
$$
} 

\proof For the proof of this Lemma, see [BG3], Section 2.
Somewhat weaker estimates were previously obtained in [Ge,ST,BG1,BGP1].
\endproof

\lemma{4.4}{\it Let $\{X_i(n), i\geq 1\}$ be independent random variables
with $X_i(n)\geq 0$, satisfying, for any $z\geq 0$,
$$
\P\left[X_i(n)\geq (1+z)a_n\right]\leq c_n e^{-z b_n}
\Eq(4.15)
$$
where $a_n, b_n, c_n$ are strictly positive parameters satisfying
$b_n\uparrow\infty$ and $(\ln c_n)/b_n\downarrow 0$ as 
$n\uparrow\infty$. Then,
$$
\E(X_i(n))\leq a_n\left(1+\frac{\ln c_n}{b_n}\right)
\Eq(4.16)
$$
and, for all $\e>0$ and $n$ sufficiently large,
$$
\P\left[\frac{1}{K}\sum_{i=1}^K X_i(n)\geq (1+z+\e)a_n\right]
\leq e^{-z b_n(1-\eta)K}
\Eq(4.17)
$$
where $\eta\equiv\eta(\e,b_n,c_n)\downarrow 0$ as $n\uparrow\infty$.
}

\proof Setting $Y_i(n)\equiv X_i(n)/a_n$, 
we have, 
$$
\E(Y_i(n))=\E\int_0^{\infty}\1_{\{y\leq Y_i(n)\}}dy 
= \int_0^{\infty}\P(Y_i(n)\geq y)dy
\Eq(4.20)
$$
Thus, for any $x\geq 0$,
$$
\E(Y_i(n))\leq 1+x + \int_{1+x}^{\infty}\P(Y_i(n)\geq y)dy
\Eq(4.21)
$$
Performing the change of variable $y=1+z$ and making use of (4.15)
yields
$$
\E(Y_i(n))\leq 1+x + c_n\int_{x}^{\infty}e^{-b_n z}dz 
= 1+x+\frac{c_n}{b_n}e^{-x b_n}
\Eq(4.22)
$$
Now, choosing $x=(\ln c_n)/b_n$ 
minimizes the r.h.s. of \eqv(4.22) and gives
\eqv(4.16). To prove \eqv(4.14) we first use that, by the exponential
Markov inequality, for any $t>0$,
$$
\P\left[\frac{1}{K}\sum_{i=1}^K Y_i(n)\geq 1+z+\e\right]
\leq e^{-Kt(1+z+\e)}\prod_{i=1}^K\E e^{tY_i(n)}
\Eq(4.23)
$$
To estimate the Laplace transform of $Y_i(n)$, we write that,
$$
\E e^{tY_i(n)}=\E (1+ \int_0^{\infty} t e^{ty}\1_{\{y\leq Y_i(n)\}}dy)
= 1 + \int_0^{\infty} t e^{ty}\P(Y_i(n)\geq y)dy
\Eq(4.24)
$$
and, for any $x\geq 0$,
$$
\eqalign{
\E e^{tY_i(n)} &=1 + \int_0^{1+x} t e^{ty}\P(Y_i(n)\geq y)dy
+\int_{1+x}^{\infty} t e^{ty}\P(Y_i(n)\geq y)dy\cr
&\leq e^{t(1+x)} +
\int_{1+x}^{\infty} t e^{ty}\P(Y_i(n)\geq y)dy\cr
&\leq e^{t(1+x)} +
c_n te^{t}\int_{x}^{\infty} e^{-z(b_n-t)}dz\cr
}
\Eq(4.25)
$$
where we used \eqv(4.15) in the last line after having performed the 
change of variable $y=1+z$. Choosing $t=b_n(1-\eta)$ for some 
$0<\eta\leq 1$, we get
$$
\eqalign{
\E e^{tY_i(n)}&\leq e^{b_n(1-\eta)(1+x)}
\left[1+c_n \frac{1-\eta}{\eta}e^{-x b_n}\right]\cr
&\leq
\exp\left(b_n(1-\eta)(1+x)+c_n\frac{1-\eta}{\eta}e^{-x b_n}\right)
\cr
}
\Eq(4.26)
$$
and finally, inserting \eqv(4.26) in \eqv(4.23) yields
$$
\P\left[\frac{1}{K}\sum_{i=1}^K Y_i(n)\geq 1+z+\e\right]
\leq e^{-z b_n(1-\eta)K} \exp\left(-(1-\eta)K\left[
b_n(\e-x)-\frac{c_n}{\eta}e^{-x b_n}\right]\right)
\Eq(4.27)
$$
For $n$ large enough, choosing $x =\e/2$, one can always choose
$\eta\equiv\eta(\e,b_n,c_n)$ such that the last exponential in 
\eqv(4.27) is less than 1 and $\eta(\e,b_n,c_n)\downarrow 0$ as
$n\uparrow\infty$. 
\endproof  

\lemma{4.5}{\it  
There exists a positive constant $c$ such that, for all integer $R$, 
there exists $\O_R$ with $\P(\O_R)\geq 1-R\g^{-1} e^{-c \g^{-1}}$
such that
for all $ \mu^\pm, s^\pm, w_\pm,
v_+\leq w_+ \leq v_+ +R, v_-- R \leq w_- \leq v_- $ and $\o \in \O_R$
\item{(i)}
$$
\sup_{\s:\eta(w_\pm,\s)=s^\pm e^{\mu^\pm}}\Bigl|\g^{-1}E_{\g,\D}^{1,L}[\o](\s_{\D}, m_L(\s_{\partial\D}))-
W_{\g,\D}[\o](\s_{\D},m^{(\mu^{\pm}, s^{\pm})})\Bigr|
\leq  \z\g^{-1}(1+\sqrt{2\g M(\g)})\sqrt 2
\Eq(4.18)
$$
and
\item{(ii)}  
$$
\sup_\s\left |W_{\g,\D}[\o](\s_{\D}, \s_{\partial\D})\right |\leq
\g^{-1}4(1+\sqrt{M/\ell})^2
\Eq(4.19)
$$
where $\D=[w_-+\sfrac 12,w_+-\sfrac 12]$.
}

\proof 
We first prove (i).
We set
$$
W_{\g,\D}[\o](\s_{\D},m^{(\mu^{\pm}, s^{\pm})})=
W_{\g,\D}^+[\o](\s_{\D},m^{(\mu^+, s^+)})+
W_{\g,\D}^-[\o](\s_{\D},m^{(\mu^-, s^-)})
\Eq(4.27bis)
$$
where
$$
W_{\g,\D}^-[\o](\s_{\D},m^{(\mu^{-}, s^{-})})\equiv
-L\sum_{i\in\D}s^-a(\b)\xi^{\mu^-}_i
\s_i\sum_{r\in\partial^-\D}J_\g(i-Lr)
\Eq(4.27ter)
$$
and
$$
W_{\g,\D}^+[\o](\s_{\D},m^{(\mu^{+}, s^{+})})
\equiv
-L\sum_{i\in\D}s^+a(\b)\xi^{\mu^+}_i
\s_i\sum_{r\in\partial^+\D}J_\g(i-Lr)
\Eq(4.28)
$$
We will consider only the terms corresponding to the interaction with the 
right part of $\D$, the other ones being similar. We have
$$
\eqalign{
&\left|\g^{-1}E_{\g,\D}^{1,L}[\o](\s_{\D}, m_L(\s_{\partial^+\D}))-
W_{\g,\D}^+[\o](\s_{\D},m^{(\mu^{+}, s^{+})})\right|
\1_{\{\s\in \widehat\AA(\mu^{\pm}, s^{\pm}, w_{\pm})\}} \cr
&\leq 
L\left|
\sum_{i\in\D}\sum_{r\in\partial^+\D}J_\g(i-Lr)
\s_i\left(\xi_i, \left[m_L(r,\s_{\partial^+\D})
-m^{(\mu^+,s^+)}\right]\right)
\right|
\1_{\{\s\in \widehat\AA(\mu^{\pm}, s^{\pm}, w_{\pm})\}}
\cr
&\leq 
L\sum_{r\in\partial^+\D}
\left\|\sum_{i\in\D}J_\g(i-Lr)\xi_i\s_i\right\|_2\,
\left\|m_L(r,\s_{\partial^+\D})-m^{(\mu^+,s^+)}\right\|_2
\1_{\{\s\in \widehat\AA(\mu^{\pm}, s^{\pm}, w_{\pm})\}}
\cr
&\leq
\zeta L
\sum_{r\in\partial^+\D}
\left\|\sum_{i\in\D}J_\g(i-Lr)\xi_i\s_i\right\|_2
\equiv T^+(\s)
\cr
}
\Eq(4.29)
$$
$T^-(\s)$ is defined in an analogous way.
Recalling the definition \eqv(4.11bis) we have
$$
\eqalign{
T^+(\s)=&\zeta L\sum_{r\in\partial^+\D}
\left(\sum_{i\in [w_+-1, w_+-\frac{1}{2}]}
\sum_{j\in [w_+-1, w_+-\frac{1}{2}]}(\xi_i,\xi_j)\s_i\s_j
J_\g(i-Lr)J_\g(j-Lr)\right)^{\frac{1}{2}}\cr
\leq&\zeta L\sum_{r\in\partial^+\D}\left(\g^{-1}\|B\|
\sum_{i\in [w_+-1, w_+-\frac{1}{2}]}
(\s_iJ_\g(i-Lr))^2\right)^{\frac{1}{2}}\cr
\leq&\zeta L\sum_{r\in\partial^+\D}\|B\|^{\frac{1}{2}}\cr
\leq&\zeta (2\g)^{-1}\|B\|^{\frac{1}{2}}\cr
}
\Eq(4.30)
$$
where we have used in the last inequality that
$\#\{r\in\partial^+\D\}=(2\g L)^{-1}$. Thus, by Lemma 4.3, for
all $\e>0$,
$$
\P\left[\sup_{\s\in\SS}
T^+(\s)\geq \zeta (2\g)^{-1}(1+\sqrt{2\g M})\sqrt{1+\e}
\right]\leq 2K\g^{-1}\exp\left(-\frac{\e}{2K\g}\right)
\Eq(4.31)
$$
from which (i) follows. 

We turn to the  proof of (ii). 
Using \eqv(2.2) we have, for all $\e>0$,
$$
\eqalign{
&\P\left[\sup_{\s\in\SS}
\left |W_{\g,\D}[\o](\s_{\D}, \s_{\partial\D})\right |
\geq 4\e^2
\right]
\cr
\leq
&\P\left[\sup_{\s\in\SS}
\left|\g^{-1}E_{\g,\D}^{\ell,\ell}
(m_{\ell}(\s_{\D}),m_{\ell}(\s_{\partial\D}))\right|
\geq 2\e^2
\right]
+
\P\left[\sup_{\s\in\SS}
\left|\D W_{\g,\D}^{\ell,\ell}(\s_{\D},\s_{\partial\D})\right|
\geq 2\e^2
\right]
\cr
}
\Eq(4.32)
$$
Let us consider the first probability in the r.h.s. of \eqv(4.32).
By definition,
$$
E_{\g,\D}^{\ell,\ell}(m_{\ell}(\s_{\D}),m_{\ell}(\s_{\partial\D}))
=\g \ell \sum_{x\in\D}\sum_{y\in\partial\D}J_{\g \ell}(x-y)
(m_{\ell}(x,\s_{\D}),m_{\ell}(y,\s_{\partial\D}))
\Eq(4.33)
$$
Now
$$
(m_{\ell}(x,\s_{\D}),m_{\ell}(y,\s_{\partial\D}))\leq
\|m_{\ell}(x,\s_{\D})\|_2\,\|m_{\ell}(y,\s_{\partial\D})\|_2
\leq \|B(x)\|^{\frac{1}{2}}\|B(y)\|^{\frac{1}{2}}
\Eq(4.34)
$$
where $B(x)$ is the $\ell\times \ell$-matrix 
$B(x)=\left\{B(x)_{i,j}\right\}_{i\in x, j\in x}$
with $B(x)_{i,j}=\frac{1}{\ell}\sum_{\mu=1}^{M}\xi_i^{\mu}\xi_j^{\mu}$.
Thus
$$
\eqalign{
\left|E_{\g,\D}^{\ell,\ell}(m_{\ell}(\s_{\D}),m_{\ell}
(\s_{\partial\D}))\right|
&\leq (\g \ell)^2
\sum_{x\in\D}\sum_{y\in\partial\D}\1_{\{|\ell x-\ell y|\leq (2\g)^{-1}\}}
\|B(x)\|^{\frac{1}{2}}\|B(y)\|^{\frac{1}{2}}\cr
&\leq \left(\g \ell\sum_{x\in [w_+-1,w_+-\frac{1}{2}]} 
\|B(x)\|^{\frac{1}{2}}\right)
\,\left(\g \ell\sum_{y\in [w_+-\frac{1}{2},w_++1]} 
\|B(y)\|^{\frac{1}{2}}\right)\cr
&+\left(\g \ell\sum_{x\in [w_-+\frac{1}{2},w_-+1]} 
\|B(x)\|^{\frac{1}{2}}\right)
\,\left(\g \ell\sum_{y\in [w_-,w_-+\frac{1}{2}]} 
\|B(y)\|^{\frac{1}{2}}\right)\cr
&\equiv T_1 T_2+T_3 T_4
}
\Eq(4.35)
$$
and,
$$
\P\left[\sup_{\s\in\SS}
\left|E_{\g,\D}^{\ell,\ell}(m_{\ell}(\s_{\D}),m_{\ell}
(\s_{\partial\D}))\right|
\geq 2\e^2\right]\leq\sum_{k=1}^4\P(T_k\geq \e)
\Eq(4.36)
$$
where the last equality in \eqv(4.35) defines the quantities $T_k$.
All four probabilities on the right hand side of \eqv(4.36) will be 
bounded
in the same way. Let us consider $\P(T_1\geq \e)$. Note that
$\left\{\|B(x)\|^{\frac{1}{2}}\right\}_{x\in [w_+-1,w_+-\frac{1}{2}]}$
are independent random variables. It follows from
Lemma 4.3 that, for all $\tilde \e>0$,
$$
\P\left[\|B(x)\|^{\frac{1}{2}}>\left(1+\sqrt{M/\ell}\right)
(1+\tilde\e)\right]\leq 4K\ell\exp\left(-\frac{\tilde\e \ell}K\right)
\Eq(4.37)
$$
and by Lemma 4.4, we get that for large enough $\ell$,
$$
\P\left[T_1\geq \frac{1}{2}(1+\sqrt{M/\ell})(1+\tilde\e)\right]
\leq
K\exp\left(-\frac{\tilde\e}{2K\g}\right)
\Eq(4.38)
$$
Therefore, choosing $\e\equiv\frac{1}{2}(1+\sqrt{M/\ell})(1+\tilde\e)$
in \eqv(4.32), \eqv(4.36) yields
$$
\P\left[\sup_{\s\in\SS}
\left|\g^{-1}E_{\g,\D}^{\ell,\ell}(m_{\ell}(\s_{\D}),m_{\ell}
(\s_{\partial\D}))\right|
\geq (2\g)^{-1}(1+\sqrt{M/\ell})^2(1+\tilde\e)^2
\right]\leq
4K\exp\left(-\frac{\tilde\e}{2K\g}\right)
\Eq(4.39)
$$
Choosing  $\tilde\e=1$ and using   Lemma 2.1 to bound the second term in 
\eqv(4.32) we get \eqv(4.19) which concludes the proof of Lemma 4.5.
\endproof

We are now ready to prove Proposition 4.1.

\proofof{Proposition 4.1 part i)} 
Setting $\D^c\equiv\L\setminus\D$, some simple manipulations allow us
to write
$$
\eqalign{
&\GG_{\b,\g,\L}[\o](F\cap \AA(\mu^{\pm}, s^{\pm}, w_{\pm}))\leq
\GG_{\b,\g,\L}[\o](F\cap \widehat\AA(\mu^{\pm}, s^{\pm}, w_{\pm}))
\cr
&=
\frac{1}{ Z_{\b,\g,\L}[\o]}
\E_{\s_{\D}}
\Biggl[
e^{-\b\left[H_{\g,\D}[\o](\s_{\D})
+W_{\g,\D}[\o](\s_{\D},m^{(\mu^{\pm}, s^{\pm})})\right]}\cr
&\times\E_{\s_{\D^c}}
e^{-\b\left[H_{\g,\D^c}[\o](\s_{\D^c})
+\left[W_{\g,\D}[\o](\s_{\D},\s_{\D^c})
-W_{\g,\D}[\o](\s_{\D},m^{(\mu^{\pm}, s^{\pm})})\right]\right]}
\1_{\{\s\in F\cap \widehat\AA(\mu^{\pm}, s^{\pm}, w_{\pm})\}}
\Biggr]
\cr
&=\E_{\s_{\D}}
\Biggl[
\frac{1}{Z_{\b,\g,\D}^{\mu^{\pm},s^{\pm}}[\o]}
e^{-\b\left[H_{\g,\D}[\o](\s_{\D})
+W_{\g,\D}[\o](\s_{\D},m^{(\mu^{\pm}, s^{\pm})})\right]}\cr
&\times
\E_{\s_{\D^c}}\E_{\tilde\s_{\D}}\frac{1}{ Z_{\b,\g,\L}[\o]}
e^{-\b\left[H_{\g,\D^c}[\o](\s_{\D^c})+
H_{\g,\D}[\o](\tilde\s_{\D})
+W_{\g,\D}[\o](\tilde\s_{\D},\s_{\D^c})
\right]}\1_{\{\s\in F\cap \widehat\AA(\mu^{\pm}, s^{\pm}, w_{\pm})\}}\cr
&\quad\times
e^{-\b\left[
\left[W_{\g,\D}[\o](\s_{\D},\s_{\D^c})
-W_{\g,\D}[\o](\s_{\D},m^{(\mu^{\pm}, s^{\pm})})\right]
+\left[W_{\g,\D}[\o](\tilde\s_{\D},m^{(\mu^{\pm}, s^{\pm})})
-W_{\g,\D}[\o](\tilde\s_{\D},\s_{\D^c})\right]
\right]}
\Biggr]
\cr
&=\E_{\s_{\D}}
\Biggl[\GG_{\b,\g,\D}^{\mu^{\pm}, s^{\pm}}[\o](\s_{\D})
\1_{\{\s\in F\}}
\E_{\bar\s_{\L}}
\GG_{\b,\g,\L}[\o](\bar\s_{\L})
\1_{\{\bar\s\in \widehat\AA(\mu^{\pm}, s^{\pm}, w_{\pm})\}}\cr
&\quad\times e^{-\b\left[
\left[W_{\g,\D}[\o](\s_{\D},\bar\s_{\D^c})
-W_{\g,\D}[\o](\s_{\D},m^{(\mu^{\pm}, s^{\pm})})\right]
+\left[W_{\g,\D}[\o](\bar\s_{\D},m^{(\mu^{\pm}, s^{\pm})})
-W_{\g,\D}[\o](\bar\s_{\D},\bar\s_{\D^c})\right]
\right]}
\Biggr]
}
\Eq(4.40)
$$
Now, if 
$\bar\s\in\widehat\AA(\mu^{\pm}, s^{\pm}, w_{\pm})$
$$
\eqalign{
&\left|
\left[W_{\g,\D}[\o](\s_{\D},\bar\s_{\D^c})
-W_{\g,\D}[\o](\s_{\D},m^{(\mu^{\pm}, s^{\pm})})\right]
+\left[W_{\g,\D}[\o](\bar\s_{\D},m^{(\mu^{\pm}, s^{\pm})})
-W_{\g,\D}[\o](\bar\s_{\D},\bar\s_{\D^c})\right]
\right|\cr
\leq&
2\sup_{\bar\s\in\widehat\AA(\mu^{\pm}, s^{\pm}, w_{\pm})}
\left|
W_{\g,\D}[\o](\bar\s_{\D},\bar\s_{\D^c})-
W_{\g,\D}[\o](\bar\s_{\D},m^{(\mu^{\pm}, s^{\pm})})
\right|\cr
\leq&
2\sup_{\bar\s\in\widehat\AA(\mu^{\pm}, s^{\pm}, w_{\pm})}
\left|\g^{-1}E_{\g,\D}^{1,L}[\o](\bar\s_{\D}, m_L(\bar\s_{\partial\D}))-
W_{\g,\D}[\o](\bar\s_{\D},m^{(\mu^{\pm}, s^{\pm})})\right|
\cr
&+
2\sup_{\bar\s\in\SS}
\left |\D W_{\g,\D}^{1,L}[\o](\bar\s_{\D}, \bar\s_{\partial\D})\right |
\cr
}
\Eq(4.41)
$$
Finally, by Lemma 4.5 and Lemma  2.1, the 
supremum over $\mu^{\pm}, s^{\pm}$ and $w_{\pm}$,
$v_+\leq w_+\leq v_+ +R$ $v_- -R \leq w_- \leq v_-$, of the
last line of \eqv(4.41) 
is bounded from above by $8\g^{-1}(\zeta+2\g L)$ with a 
$\P_{\xi}$-probability, greater than 
$1-4\g^{-1}R\exp(-c\g^{-1})$  for some positive constant $c$. Thus from 
\eqv(4.40) and \eqv(4.41) follow both \eqv(4.9) and \eqv(4.9bis).
\endproof

\noindent {\bf Proof of Proposition 4.1 part ii):} 
Using \eqv(4.5bis) the l.h.s. of \eqv(4.10) is bounded from above  by 
$\GG_{\b,\g,\L}[\o](A^+(R)) +\GG_{\b,\g,\L}[\o](A^-(R))$. We estimate the 
first term, the second one being similar.  Since
the spin configuration are away from the equilibria on a length $R$,
 we can  decouple the interaction
between this part and the rest of the volume $\L$, by making a rough
estimate of those interaction  terms. The fact that we are out of equilibrium
will give  terms proportional to $R$ that will be dominant if $R$ is 
chosen large enough.
More precisely, calling 
${\D_R}\equiv [v_+,v_+ +R]$, we have, for all fixed $R$
$$
\eqalign{
\GG_{\b,\g,\L}\left(A^+(R)\right)&=
\frac{1}{Z_{\b,\g,\L}}
\E_{\s_\L}\left[
e^{-\b H_{\g,\L\setminus {\D_R}}(\s_{\L\setminus {\D_R}})}
e^{-\b\left[H_{\g,{\D_R}}(\s_{{\D_R}})
+W_{\g,{\D_R}}(\s_{{\D_R}},\s_{\L \setminus {\D_R}})\right]}
\1_{\{\s\in A^+(R)\}}\right]\cr
&\leq e^{ 4 c\g^{-1}}
\frac 1{Z_{\b,\g,\D_R}} \E_{\s_{\D_R}}\left[ e^{-\b H_{\g,{\D_R}}(\s_{{\D_R}})}
\1_{\{\s\in A^+(R)\}}\right]
}
\Eq(4.42)
$$
with a $\P_\xi$-probability greater than $1-4\g^{-1}e^{-c\g^{-1}}$
for some positive constant $c$,
where we have used Lemma 4.5 to bound the interaction between
${\D_R}$ and $\L \setminus {\D_R}$. 
To estimate the last term in \eqv(4.42), we express it in terms
of block spin variables on the scale $\ell$. Using \eqv(2.9) we get 
$$
\GG_{\b,\g,{\D_R}}\left(A^+(R)\right)
\leq e^{2c \g^{-1}|{\D_R}|(4\g\ell+\g M)}
\frac{\E_{\s_{{\D_R}} }
e^{-\b\g^{-1}E_{\g,{\D_R}}^{\ell}(m_{\ell}(\s))}
\1_{\{\s\in A^+(R)\}}}
{\E_{\s_{{\D_R}}}
e^{-\b\g^{-1}E_{\g,{\D_R}}^{\ell}(m_{\ell}(\s))}}
\Eq(4.45)
$$
with a $\P_\xi$-probability greater than $1-e^{-c \g^{-1}|{\D_R}|}$

We derive first a lower bound on the denominator which  will be given 
effectively by restricting the configurations to be in the neighborhood of 
a constant profile near one of the equilibrium positions $sa(\b) e^\mu$.
We will choose without lost of generality to be $s=1, \mu=1$.
To make this precise, 
we define for any $\rho>0$ the balls   
$$
\BB_{\rho}^{(\mu,s)}\equiv\left\{m\in\R^M\, \Big|\,
\|m-m^{(\mu,s)}\|_2\leq \rho\right\}
\Eq(4.45bis)
$$
We will moreover write
$$
\BB_{\rho}\equiv\bigcup_{(\mu,s)\in\{1,\dots,M\}\times\{-1,1\}}
\BB_{\rho}^{(\mu,s)}
\Eq(4.45ter)
$$
Obviously,
$$ 
\E_{\s_{{\D_R}}}
e^{-\b\g^{-1}E_{\g,{\D_R}}^{\ell}(m_{\ell}(\s))}
\geq 
\E_{\s_{{\D_R}}}
e^{-\b\g^{-1}E_{\g,{\D_R}}^{\ell}(m_{\ell}(\s))}
\prod_{x\in \D_R}\1_{\{  
m_{\ell}(x,\s)\in\BB^{(1,1)}_{\rho}\}}
\Eq(4.46)
$$
It can easily be shown that, on the set
$\{m_{\ell}(x,\s)\in\BB_{\rho}, \forall x\in{\D_R}\}$, 
$$
-\g^{-1}E_{\g,{\D_R}}^{\ell}(m_{\ell}(\s))
\geq
\frac{\ell}{2}\sum_{x\in{\D_R}}(\|m_{\ell}(x,\s)\|_2^2 
-4\rho^2)
\Eq(4.47)
$$
from which \eqv(4.46) yields
$$
\eqalign{
\E_{\s_{{\D_R}}}
e^{-\b\g^{-1}E_{\g,{\D_R}}^{\ell}(m_{\ell}(\s))}
&\geq  
e^{-4\b \g^{-1}|{\D_R}|\rho^2}
\prod_{x\in{\D_R}}\E_{\s_x}e^{\frac{\b\ell}{2}
\|m_{\ell}(x,\s)\|_2^2}
\1_{\displaystyle{\{m_{\ell}(x,\s)\in\BB^{(1,1)}_{\rho}\}}}\cr
&=e^{-4\b \g^{-1}|{\D_R}|\rho^2}\prod_{x\in{\D_R}}
Z_{x,\b,\rho}\left(a(\b)e^1\right)
}
\Eq(4.48)
$$
 provided that $\rho$ is sufficiently 
large so that $\BB_\rho^{(1,1)}$ contains the lowest minimum of $\Phi$ 
in  the neighborhood of $a(\b)e^1$, which is the case if 
$\rho\geq c\sqrt {\frac M\ell}$, for some finite constant $c$ with a 
$\P_\xi$-probability $\geq 1- e^{-c M}$ .

Next we derive an upper bound for the
numerator of the ratio in \eqv(4.45). Using 
 the inequality $ab\leq\frac{1}{2}(a^2+b^2)$ we get
$$
-\g^{-1}E_{\g,{\D_R}}^{\ell}(m_{\ell}(\s))
\leq
\frac{\ell}{2}\sum_{x\in{\D_R}}\|m_{\ell}(x,\s)\|_2^2 
\Eq(4.50)
$$
and whence
$$
\E_{\s_{{\D_R}}}
e^{-\b\g^{-1}E_{\g,{\D_R}}^{\ell}(m_{\ell}(\s))}
\1_{\{\s\in A^+(R))\}}
\leq
\E_{\s_{{\D_R}}}
e^{\frac{\b\ell}{2}\sum_{x\in{\D_R}}\|m_{\ell}(x,\s)\|_2^2}
\1_{\{\s\in A^+(R)\}}
\Eq(4.51)
$$
Let us now recall that, by definition,
$$
A^+(R)=
\left\{\s\in\SS \,\Big|\, \forall_{u\in{\D_R}} \exists_{r\in u} :
\inf_{\mu,s}\|m^{(\mu,s)}-m_L(r,\s)\|_2>\z
\right\}
\Eq(4.52)
$$
Using that $m_L(r,\s)=\frac{\ell}{L}\sum_{x\in r}m_{\ell}(x,\s)$
 we have 
$$
\|m^{(\mu,s)}-m_L(r,\s)\|_2
\leq 
\frac{\ell}{L}\sum_{x\in r}\|m^{(\mu,s)}-m_{\ell}(x,\s)\|_2
\Eq(4.53)
$$
so that
$$
A^+(R)\subset
\left\{\s\in\SS \,\Big|\, \forall_{u\in{\D_R}} \exists_{r\in u} :
 \inf_{\mu,s}
\frac{\ell}{L}\sum_{x\in r}\|m^{(\mu,s)}-m_{\ell}(x,\s)\|_2>\z
\right\}
\Eq(4.54)
$$
We will use the following fact

\lemma{4.6}{\it Let $\left\{X_k, k=1,2,\dots,K\right\}$ be a sequence of 
real numbers satisfying $0\leq X_k\leq c$ for some $c<\infty$. 
Let  $\z<c$ and assume that
$$
\frac{1}{K}\sum_{k=1}^K X_k>\z
\Eq(4.lemma0)
$$
For $0\leq\d\leq\z$, 
define the set 
$
V_{\d,\z}\equiv \left\{k| X_k\leq \d\z\right\}
$. Then
$$
\left|\left\{1\leq k\leq K : X_k>\d\z\right\}\right|
\geq K\frac{\z(1-\d)}{c-\d\z}
\Eq(4.lemma1)
$$
}
\noindent {\bf Proof :} Set  $V_{\d,\z}^c\equiv\{1,\dots,K\}\setminus V_{\d,\z}$. Then
$$
\eqalign{
\frac{1}{K}\sum_{k=1}^K X_k
\leq
\frac{1}{K}\sum_{k\in V_{\d,\z}}X_k+\frac{1}{K}\sum_{k\in V_{\d,\z}^c}X_k
\leq 
\frac{1}{K}c|V_{\d,\z}|+\frac{1}{K}\d\z|V_{\d,\z}^c|
=
\frac{1}{K}(c-\d\z)|V_{\d,\z}|+\d\z
}
\Eq(4.lemma2)
$$
which, together with \eqv(4.lemma1) implies the bound \eqv(4.lemma2)
\endproof

Let us denote by $\VV_{\d,\z}(r)$ the set of all subsets 
$S\subset \{x\in r\}$ with 
cardinality 
$\frac{L}{\ell}\frac{\z(1-\d)}{2-\d\z}$, respectively volume
$$
|S|\geq
\g L\frac{\z(1-\d)}{2-\d\z}
\Eq(4.55)
$$
Then, since 
$\|m^{(\mu,s)}-m_{\ell}(x,\s)\|_2<2$, Lemma 4.7 implies
$$
A^+(R)\subset 
\left\{\s\in\SS \,\Big|\,\forall_{u\in{\D_R}} \exists_{r\in u} \exists_{S\in \VV_{\d,\z}(r)} : 
\forall_{ x\in S}\,,\, m_{\ell}(x,\s)\in\BB_{\d\z}^c
\right\}
\Eq(4.56)
$$
Therefore
$$
\eqalign{
&\E_{\s_{{\D_R}}}
e^{-\b\g^{-1}E_{\g,{\D_R}}^{\ell}(m_{\ell}(\s))}
\1_{\{\s\in A^+(R)\}}\cr
\leq &
\prod_{u\in{\D_R}}
\E_{\s_u}e^{\frac{\b\ell}{2}\sum_{x\in u}\|m_{\ell}(x,\s)\|_2^2}
\1_{\displaystyle{\{
\exists_{r\in u} \exists_{S\in \VV_{\d,\z}(r)} : 
\forall x\in S\,,\, m_{\ell}(x,\s)\in\BB_{\d\z}^c
\}}}
\cr
\leq &
\prod_{u\in{\D_R}}\sum_{r\in u}
\sum_{S\in \VV_{\d,\z}(r)}
\E_{\s_u}e^{\frac{\b\ell}{2}\sum_{x\in u}\|m_{\ell}(x,\s)\|_2^2}
\prod_{x\in S}
\1_{\displaystyle{\{m_{\ell}(x,\s)\in\BB_{\d\z}^c\}}}\cr
}
\Eq(4.58)
$$
Inserting this and  \eqv(4.48) into \eqv(4.45) we have 
$$
\eqalign{
&\GG_{\b,\g,{\D_R}}[\o]
\left(A^+(R)\right)\cr
&\leq e^{\g^{-1}|{\D_R}|(16\g\ell+4\g M+4\b\rho^2)}
\prod_{u\in{\D_R}}\sum_{r\in u}\sum_{S\in \VV_{\d,\z}(r)}
\prod_{x\in u\setminus S}
\frac{Z_{x,\b}}{Z_{x,\b,\rho}(a(\b)e^1)}
\,\prod_{x\in S}
\frac{Z^c_{x,\b,\d\z}}{Z_{x,\b,\rho}(a(\b)e^1)}
\cr
&\equiv  e^{\g^{-1}|{\D_R}|(16\g\ell+4\g M+4\b\rho^2)}
\prod_{u\in{\D_R}}\sum_{r\in u}\sum_{S\in \VV_{\d,\z}(r)}
T^{(1)}_S\,T^{(2)}_S
}
\Eq(4.59)
$$
where we have defined
$$
 Z^c_{x,\b,\d\z}\equiv
\E_{\s_x}e^{\frac{\b\ell}{2}\|m_{\ell}(x,\s)\|_2^2}
\1_{\displaystyle{\{m_{\ell}(x,\s)\in\BB_{\d\z}^c\}}}
\Eq(4.61)
$$

It follows from Proposition 2.3 of [BGP1] that 
$$
Z_{x,\b}\leq \exp\left(-\b\ell\left[\phi(a(\b))-c \sqrt{\sfrac M\ell}\right]
\right)
\Eq(4.62)
$$
so that using Lemma 3.1 we get that
$$
T_S^{(1)} \leq \prod_{x\in u\ba S} \exp\left(+\b\ell c\sqrt{\sfrac M\ell}
\right)
\leq e^{+\b\g^{-1}c\sqrt{\frac M\ell  }}
\Eq(4.63)
$$
with a $\P_\xi$-probability $\geq 1-(\g\ell)^{-1}e^{-c M}$
On the other hand, to bound $Z_{x,\b,\d\z}^c$, we proceed as in
[BG2] and first note that
$\|m_{\ell}(x,\s)\|_2^2\leq 2$ for all $\s$. Next, we introduce the 
lattice $\WW_{\ell,M}$ with spacing $1/\sqrt{\ell}$ in $\R^M$ and we 
denote by $\WW_{\ell,M}(2)$ the intersection of this lattice with the ball
of radius 2 in $\R^M$. We have
$$
|\WW_{\ell,M}(2)|\leq \exp
\left(M\ln\left(\frac{2\ell}{M}\right)\right)
\Eq(4.64)
$$
Now, we may cover the ball of radius 2 in $\R^M$ with balls of radii
$\hrho\equiv\sqrt{M/\ell}$ centered at the points of 
$\WW_{\ell,M}(2)$. Supposing that $\d\z>\hrho$ 
this yields,
$$
\eqalign{
Z_{x,\b,\d\z}^c
\leq &
\sum_{m\in \WW_{\ell,M}(2)}
\1_{\displaystyle{\{m\in\BB_{\d\z-\hrho}^c\}}}
Z_{x,\b,\hrho}(m)[\o]
\cr
\leq & 
\sum_{m\in \WW_{\ell,M}(2)}
\1_{\displaystyle{\{m\in\BB_{\d\z-\hrho}^c\}}}
\exp\left(-\b\ell \left(\Phi_{x,\b}(m)[\o]
-\frac{1}{2}\hrho^2\right)\right)\cr
}
\Eq(4.65)
$$
 Let us now assume that $\d\z-\hrho$ satisfies the
hypothesis of Proposition 3.2, then
$$
Z_{x,\b,\d\z}^c
\leq 
\exp\left(-\b\ell \left(
\phi(a(\b))
+\e(\d\z-\hrho)-4(\d\z-\hrho)\sqrt{\sfrac{M}{\ell}}
-\sfrac{1}{2}\hrho^2
-\sfrac{M}{\b\ell}
\ln\left(\sfrac{2\ell}{M}\right)
\right)
\right)
\Eq(4.66)
$$
with a $\P_\xi$-probability 
$\geq 1-e^{-cM}$,
where $\e(\cdot)$ is the function defined in Proposition 3.2.
We will assume that $\d\z\gg \sqrt{\frac M\ell}$. Thus

$$
\frac {Z_{x,\b,\d\z}^c}{Z_{x,\b,\rho}(a(\b)e^1)}
\leq \exp\left(-\b\ell\left[\e\left(\d\z-\hrho\right)-c\d\z\sqrt{\sfrac M\ell}
\right]\right)
\Eq(4.67)
$$
with a $\P_\xi$-probability $\geq 1 -e^{-c' M}$.
Thus the product
$T^{(1)}_ST^{(2)}_S$
defined in \eqv(4.59) is bounded by
$$
T^{(1)}_ST^{(2)}_S
\leq
\exp\left(\b\g^{-1}c\left[\sqrt{\sfrac M\ell}- \e(\z)|S|\right]\right)
\Eq(4.69)
$$
with a $\P_\xi$-probability $\geq 1 - (\g\ell)^{-1} |S|e^{-c' M}$.
Hence
$$
\eqalign{
&
\prod_{u\in{\D_R}}\sum_{r\in u}\sum_{S\in \VV_{\d,\z}(r)}
T^{(1)}_ST^{(2)}_S\cr
&\leq \prod_{u\in{\D_R}}\sum_{r\in u}\sum_{S\in \VV_{\d,\z}(r)}
\exp\left(-\b\g^{-1}c\left[|S|\e(\z)-\sqrt{\sfrac M\ell}\right]\right)\cr
&\leq
\exp\left(-\b\g^{-1}
|{\D_R}|\left[\g L\z c\e(\z)
-\g|\ln(\g L)|-\g L\frac{\ln 2}{\ell}-c\sqrt{\sfrac M\ell}
\right]\right)
}
\Eq(4.70)
$$
with a $\P_\xi$-probability 
$\geq 1 - (\g)^{-1} Re^{-c' M}$
and finally, inserting \eqv(4.70) in \eqv(4.59) we arrive at
$$
\GG_{\b,\g,{\D_R}}[\o]\
\left(A^+(R)\right)
\leq
\exp\left(-\b\g^{-1}R\left[\g L c\z\e(\z)-c'\left(
\sqrt{\sfrac M\ell}+8\g \ell +2\rho^2\right)\right]\right)
\Eq(4.71)
$$
with a $\P_\xi$-probability
$\geq 1 - (\g\ell)^{-1} Re^{-c' \ell}$,
where we have used the fact that $M\ll \ell$. \endproof

\newpage

\chap{ 5. Self averaging properties of the free energy}5

 In this chapter we study the self averaging properties of 
the free energy
of the  Hopfield-Kac model with {\it mesoscopic}
boundary conditions.

We denote the partition function on the volume $\D$ with 
boundary condition $s^-a(\b)e^{\mu^-}$ on the left of
$\D$ and $s^+a(\b)e^{\mu^+}$ on the 
right of $\D$ by
$$
{ Z}_{\D}^{(\mu^\pm,s^\pm)}\equiv
\E_{\s_{\D}}\left[ e^{- \b \left(H_{\g,\D}(\s) +
 W_{\g,\D,\partial^-\D}(\s_\D| m^{(\mu^-,s^-)})
+ W_{\g,\D,\partial^+\D}(\s_\D)| m^{(\mu^+,s^+)})
\right)}\right]
\Eq(5.1)
$$
and the corresponding free energy
$$
f_\D^{(\mu^\pm,s^\pm)}\equiv f_\D=-\frac \g{\b|\D|} \ln 
{Z}_{\D}^{(\mu^\pm,s^\pm)}
\Eq(5.2)
$$ 
To include the case of free boundary conditions, we set
$m^{(0,0)}\equiv 0$.

  We are interested in the behavior of the fluctuations of 
$f_\D^{(\mu^\pm,s^\pm)}$ around it mean value. 
 We will use the Theorem 6.6 of Talagrand [T2] that we state for
the convenience of the reader. We  denote  by 
$\M X$ a median of the random variable $X$.
Recall that a number $x$ is called the median of a random variable $X$ if
both $\P[X\geq x]\geq \frac 12$ and $\P[X\leq x]\geq \frac 12$. 

\theo {5.1}[T2]{\it Consider a real valued function $f$ 
defined on $[-1,+1]^N$. 
We assume that, for each real number $a$ the set $ \{ f\leq a \}$ is convex.
Consider a convex set $B\subset [-1,+1]^N$, and assume that for all
$x,y \in B$, $| f(x) -f(y)| \leq k \|x-y\|_2$ for some positive $k$. 
Let $X$ denote a random vector with i.i.d. components
$\{X_i\}_{1\leq i \leq N}$ taking values in $[-1,+1]$. Then
for all $t>0$, 
$$
\P\left[|f(X)-\M f(X)|\geq t\right]
\leq 4b+\frac 4{1-2b}\exp\left(-\frac {t^2}{16k^2}\right)
\Eq(5.5)
$$
where 
$b\equiv \P\left[X\not\in B\right]$ and we assume that $b<\frac 12$.
}

The main result of this chapter is the following proposition:

\proposition {5.2}{\it . 
If  $\g \ell$, $M/\ell$ and $\g M$ are small enough, then for all $t>0$,
there exists a universal numerical constant $K$ such that 
$$
\eqalign{
\P&\left[ \left| f_\D^{(\mu^\pm,s^\pm)}
-\E f_\D^{(\mu^\pm,s^\pm)}\right|
\geq t+K\sqrt{\g^{-1}|\D|} \right]
\cr
&\leq K \exp\left(- \frac {\g^{-1}}8 
 |\D|(\sqrt {1+t^2} -1)\right)
}
\Eq(5.3)
$$
}

\proof  Note first that the set $ \{ f_\D\leq a \}$ is convex.
This follows from the fact that the Hamiltonian $H_{\g,\D}$ is a convex
function of the variable $\xi$.
 The main difficulty that remains is to 
 establish that $f_\D$ is a Lipshitz function of the 
independent random variables $\xi$ with a constant $k$ that is small with 
large probability. 
To prove the Lipshitz continuity of $f_\D$ it is obviously enough to 
prove the corresponding bounds for $H_{\g,\D}(\s)$ and $W_{\g,\D,\del^\pm\D}
(\s_\D|m^{(\mu^\pm,s^\pm)})$.

\def\hxi{\hat\xi} 

Let us  first prove that $H_{\g,\D}(\s)$ is  Lipshitz in the random
variable $\xi$. Let us write $\xi\equiv \xi[\o]$ and $\hat\xi\equiv \xi[\o']$. 
Denoting by $\xi^{\mu}\s$ the coordinatewise product of the two vectors 
$\xi^{\mu}$ and
$\s$ and $J_{\g}(i-j)$ the 
symmetric $\g^{-1} |\D|\times \g^{-1}|\D|$  matrix with $i,j$ 
entries, we have
 
$$
|H_{\g,\D}[\o](\s)-H_{\g,\D}[\o'](\s) |=
\left|\sum_{\mu=1}^M 
\left( \left[\xi^\mu\s-\hxi^\mu\s\right],J_{\g}
\left[\xi^\mu\s+\hxi^\mu\s\right]\right)\right|
\Eq(5.6)
$$
Since  $J_{\g}$  is a symmetric and positive definite matrix, its 
square root $J^{1/2}_{\g}$ exists. Thus using the Schwarz inequality we may 
write
$$
\eqalign{
&\left|\sum_{\mu=1}^M 
\left( [\xi^\mu\s-\hxi^\mu\s],J_{\g}[\xi^\mu\s+\hxi^\mu\s]\right)\right|\leq\cr
&\sum_{\mu=1}\| J_{\g}^{1/2}[\xi^\mu\s-\hxi^\mu\s]\|_2 
\| J_{\g}^{1/2}[\xi^\mu\s+\hxi^\mu\s]\|_2 \cr
&\leq {\cal J^+}{\cal J^-}\cr
}
\Eq(5.7)
$$
where
$$
{\cal J^+}\equiv
 \left(\sum_{\mu=1}^M ( [\xi^\mu\s+\hxi^\mu\s],
J_{\g}[\xi^\mu\s+\hxi^\mu\s])\right)^{1/2}
\Eq(5.8)
$$
and
$$
{\cal J^-}\equiv \left(\sum_{\mu=1}^M 
( [\xi^\mu\s-\hxi^\mu\s],J_{\g}[\xi^\mu\s-\hxi^\mu\s])\right)^{1/2}
\leq \|\xi -\hxi\|_2
\Eq(5.9)
$$
The last inequality in \eqv(5.9) follows since $\|J_\g\|\leq 1 $.

 On the other hand, by convexity
$$  
\eqalign{
\left({\cal J^+}\right)^2&\leq 2 \sum_{\mu=1}^M (\xi^\mu\s J_\g \xi^\mu\s)+
2 \sum_{\mu=1}^M (\hxi^\mu\s J_\g \hxi^\mu\s)\cr
&= 2 H_{\g,\D}[\o](\s)+2H_{\g,\D}[\o'](\s)\cr
}\Eq(5.12)
$$
Collecting, we get
$$
|H_{\g,\D}[\o](\s)-H_{\g,\D}[\o'](\s) |\leq \sqrt {2}\|\xi - \hxi\|_2
\left(  H_{\g,\D}[\o](\s)+H_{\g,\D}[\o'](\s)\right)^{1/2}
\Eq(5.13)
$$
This means that as in [T2], we are in a situation where the upper bound
for the Lipshitz norm of $H_{\g,\D}[\o](\s)$ is not uniformly
bounded. However  the estimates of  Section 2,  allow
us to give reasonable  estimates on the probability distribution of 
this Lipshitz norm.  
Recalling \eqv(2.9) we have
$$
\P\left[\sup_{\s\in \SS_{\D}}|\D H_{\g,\D}(\s)|\geq \g^{-1} |\D|
(16(1+c))\g \ell + 4 \g M )\right] \leq 16 e^{-c \g^{-1}|\D|}
\Eq(5.14)
$$
Therefore, using \eqv(2.1) we get 
$$
\eqalign{
&\P\left[\sup_{\s\in \SS_{\D}}| H_{\g,\D}(\s)|\geq \g^{-1} |\D|
(C + (16(1+c))\g \ell + 4\g M) \right]\cr
& \leq 16 e^{-C \g^{-1}|\D|}
+\P\left[ \sup_{\s\in \SS_{\D}}| \g^{-1} E^{\ell}_{\g,\D}(m_{\ell}(\s))|
\geq C \g^{-1}\D \right]
}
\Eq(5.15)
$$
To estimate this last probability, we notice that by convexity
$$
2 (m_{\ell}(x,\s),m_{\ell}(y,\s)) \leq  \| m_{\ell}(x,\s)\|_2^2 
+ \| m_{\ell}(y,\s)\|_2^2
\Eq(5.16)
$$
Therefore
$$
\eqalign{
|\g^{-1} E^{\ell}_{\g,\D}(m_{\ell}(\s))|&=
 1/2 \left|\sum_{x,y \in \D} J_{\g \ell}(x-y) 
(m_{\ell}(x,\s),m_{\ell}(y,\s))\right|\cr
&\leq  \ell/2 \sum_{x \in \D} \| m_{\ell}(x,\s)\|_2^2\cr
}\Eq(5.17)
$$
Now we have
$$
\eqalign{
&\P\left[ \sup_{\s\in \SS_{\D}}  \ell \sum_{x \in \D} \| m_{\ell}(x,\s)\|_2^2
\geq 2C \g^{-1}|\D| \right]  \cr
& \leq 2^{\g^{-1} |\D|} \P\left[ \ell \sum_{x \in \D} \| m_{\ell}(x,\s)\|_2^2
\geq 2C \g^{-1}|\D| \right] \cr
&\leq 2^{\g^{-1} |\D|} \inf_{0\leq t< 1/2} e^{-2 C \g^{-1}|\D| t} 
\prod_{x\in \D}\prod_{\mu=1}^M
\E e^{t\ell \left(\frac 1\ell  \sum_{i\in x}  \xi_i^{\mu}\s_i \right)^2}\cr
}\Eq(5.18)
$$
Using the well known inequality [BG1]
$$
\E \exp \left( t\ell \left(\frac 1\ell \sum_{i\in x} \xi_i^{\mu}\s_i 
\right)^2\right)
\leq \frac 1{\sqrt {1-2t}}
\Eq(5.19)
$$
and choosing $t=1/4$, the r.h.s of \eqv(5.18) is bounded from above by
$$
 \exp \left(- \g^{-1} |\D| \left( \frac C2 - (1+ M/2\ell)\ln 2\right) \right)
\Eq(5.20)
$$
Collecting, we get
$$
\P\left[ \sup_{\s\in \SS_{\D}}  \ell \sum_{x \in \D} \| m_{\ell}(x,\s)\|_2^2
\geq  \g^{-1}|\D| 2\left( c + (1+ M/2\ell)\ln 2\right) \right] \leq
e^{-c\g^{-1} |\D|}
\Eq(5.21)
$$ 
which implies, if $\g \ell$,  $\g M$ and $M/\ell$ are small enough, that 
 
$$
\P\left[\sup_{\s\in \SS_{\D}}| H_{\g,\D}(\s)|\geq \g^{-1} |\D|
(4c+1) \right]
\leq  17 e^{-c \g^{-1} |\D|}
\Eq(5.22)
$$
which is the estimate we wanted.

To treat the boundary terms, {\it c.f} \eqv(4.9), let us call
$W_{\g,\D}^-[\o]$ (respectively $W_{\g,\D}^+[\o]$)  the terms corresponding 
to interactions with the left ( respectively right)
part  of the boundary $\partial \D$.  We estimate first the Lipshitz norm of
$W_{\g,\D}^-[\o]$, the one of $W_{\g,\D}^+[\o]$ being completely identical.
$$
\eqalign{
&|W_{\g,\D}^-[\o](\s_\D,m^{(\mu^-,s^-)})-
W_{\g,\D}^-[\o'](\s_\D,m^{(\mu^-,s^-)}) | \cr
&\leq  a(\b)\left| \sum_{i\in \D}\s_i(\xi_i^{\mu^-}-\hxi_i^{\mu^-})
\left(\sum_{j\in\partial^-\D} J_\g(i-j)\right)\right|\cr
&\leq a(\b)\left( \sum_{i\in \D}(\xi_i^{\mu^-}-\hxi_i^{\mu^-})^2\right)^{1/2}
\left( \sum_{i\in \D}\left(\sum_{j\in\d^-\D} J_\g(i-j)\right)^2\right)^{1/2}\cr
&\leq \g^{1/2} a(\b)\|\xi -\hxi\|_2^2 \cr
&\leq \g^{1/2}\|\xi -\hxi\|_2^2
}\Eq(5.23)
$$
where we have used the Schwarz inequality and 
$$
\sum_{i\in \D}\left(\sum_{j\in\partial^-\D} J_\g(i-j)\right)^2 \leq  \g^{-1}
\Eq(5.24)
$$
  Therefore if we denote by
$$
\O_B \equiv \left\{ \xi \in [-1,+1]^{\g^{-1}\D M}\big|
 \sup_{\s \in {\cal S}_\D}
|H_{\g,\D}(\s) | \leq \g^{-1} |\D|
(4c +1)\right\}
\Eq(5.25)
$$
Using \eqv(5.5), \eqv(5.22), \eqv(5.23) and some easy computations, we get
$$
\P\left[ \left| f_\D^{(\mu^\pm,s^\pm)}
-\M f_\D^{(\mu^\pm,s^\pm)}\right|
\geq t\g^{-1}|\D| \right]\leq 68 e^{-c \g^{-1} |\D|} + 
68 e^{-\frac {t^2}{16(4c +2)}\g^{-1} |\D|}
\Eq(5.26)
$$
Choosing $c$  such that $c= \frac {t^2}{16(4c +2)}$
we get
$$
\P\left[ \left| f_\D^{(\mu^\pm,s^\pm)}
-\M f_\D^{(\mu^\pm,s^\pm)}\right|
\geq t\g^{-1}|\D| \right]\leq 136 \exp\left(- \frac {\g^{-1}}8 
 |\D|(\sqrt {1+t^2} -1)\right)
\Eq(5.27)
$$
Finally, a simple calculation shows that \eqv(5.27)  implies that 
$$
|\M f_\D^{(\mu^\pm,s^\pm)}-
\E f_\D^{(\mu^\pm,s^\pm)}|\leq 544\sqrt{\g^{-1}|\D|}
\Eq(5.28)
$$
and this implies the claim of Proposition 5.2.
\endproof

We will mainly use the Proposition 5.2 in the following form

\corollary {5.3}{\it If $|\D| \leq \g^{-1} g(\g)$ for some $g(\g)$ with
$g(\g)\downarrow 0$ and $\g^{-1} g(\g)>c$, for all $\g$ small enough
then there exists a set $\O_g$ with
$\P[\O_g] \geq 1-Ke^{-c(g(\g))^{-1/2}}$ for some positive constants $c$
and $K$, such that for all $\o \in \O_g$

$$
\left| \ln Z_{\D}^{(\mu^\pm,s^\pm)} 
-\E\left[\ln Z_{\D}^{(\mu^\pm,s^\pm)}\right]\right| 
\leq \b\g^{-1} (g(\g))^{1/4}
\Eq(5.29)
$$
}

\proof The Corollary  follows from Proposition 5.2 by choosing 
$t= \g^{1/2}|\D|^{-1/2}(g(\g))^{-1/4}$ \endproof

\newpage
\chap{6. Localization of the Gibbs measures II: The block-scale}6

\line{\bf 6.1. Finite  volume, free boundary conditions\hfill}

Instead of dealing with the  measures $\GG_{\b,\g,\L}^{(\mu_\pm,s_\pm)}[\o]$ 
immediately, we will first consider the simpler case of Gibbs measures in 
a finite volume $\L\equiv [v_-,v_+]$ of order $|\L|=o(\g^{-1})$ with {\it free}
(Dirichlet) boundary conditions. This will be considerably simpler
and the result will actually be needed as a basic input in order to deal with 
the full problem. 
On the other hand, the result may be seen as interesting in its own right
and exhibits, to a large extent, the main relevant features of the model. 
This may indeed satisfy many readers  who may not wish to follow the 
additional 
technicalities. With this in mind, we give a more detailed exposition
of this case.  

Our basic result here will be that the free boundary conditions measure 
in volumes small compared to $\g^{-1}$ are concentrated on ``constant
profiles'' with very large probability. More precisely, we have

\theo{6.1} {\it Assume that $\g |\L|\downarrow 0$, 
 $\b$ large enough ($\b>1$) and
$\g M(\g)\downarrow 0$. Then we can find $\g^{-1}\gg\hat L\gg 1$ and 
$\hat\z\downarrow 0$, such that on a subset $\O_\L\subset \O$ with 
$\P(\O_\L^c)\leq e^{-c g^{-1/2}(\g)}$
where $g(\g)\downarrow 0$ and $\g^{-1} g(\g)>c$,
 we have that for all $\o\in \O_\L$
$$ 
\GG_{\b,\g,\L}[\o]\left(\exists_{u\in \L}
\eta_{\hat \z,\hat L}(u,\s)=0\right)
\leq e^{-\hat L h(\hat\z)}
\Eq(61.1)
$$
and 
$$
\GG_{\b,\g,\L}[\o]\left(\exists_{u\in\L}
\eta_{\hat\z,\hat L}(u,\s)\neq  \eta_{\hat \z,\hat L}(u+1,\s)\right)
\leq e^{-\hat L h(\hat\z)}
\Eq(61.2)
$$
where $h(\z)=c\b \z\e(\z)$ and $\e(\z)$ is defined in \eqv(B.28).
}

The proof of this theorem relies on a large deviation type estimate for 
events that take place on a scale much smaller than the size of $\L$.
We will consider  events $F$ that are in the cylinder algebra
with base $I=[u_-,u_+]\subset \L$, where $|I|\ll 1/(\g\ell)$ is 
very small compared to $\L$ and that  in addition are
measurable with respect to the sigma-algebra
generated by the variables $\{m_\ell(\s,x)\}_{x\in I}$.
Let us define the functions
$
U_\D^{s^\pm,\mu^\pm}
$ and $\FF_{\D,\b,\rho}^{s^\pm,\mu^\pm}$ by
$$
\eqalign{
U_\D^{s^\pm,\mu^\pm}(m_\ell)&\equiv 
\g\ell\sum_{x,y\in   \D} J_{\g\ell}(x-y)\frac{\|m_\ell(x)-
m_\ell(y)\|_2^2}4\cr
&+\g\ell  \sum_{x\in  \D,y\in \del \D}
J_{\g\ell}(x-y)\frac{\|m_\ell(x)-m^{(\mu^\pm,  s^\pm)}\|_2^2}2
}
\Eq(61.14)
$$
and
$$
\FF^{(  \mu^\pm,  s^\pm)}_{ \D , \b ,\rho}(m_\ell)\equiv
U^{(  \mu^\pm,  s^\pm)}_{ \D}(m_\ell)
+\g\ell\sum_{x\in   \D} f_{x,\b,\rho} (m_\ell(x))
\Eq(61.15)
$$
where
$$
f_{x,\b,\rho}(m_\ell(x))\equiv-\frac 1{\b\ell}\ln \E_\s
e^{\frac {\b\ell}2\|m_\ell(\s,x)\|_2^2}\1_{\{\|m_\ell(\s,x)-m_\ell(x)\|_2\leq 
\rho\}}
\Eq(61.16)
$$
For any $\d>0$
define the $\d$-covering $F_\d$ of $F$ as
    $F_\d\equiv \{ \s |  \exists_{\s'\in F} :\forall_{x\in I}\|m_\ell(\s,x)
-m_\ell(\s',x)\|_2<\d\}$. 

With these notations we 
 have the following large deviation estimates:

\theo {6.2} {\it Let $F$ and $F_\d$ be as defined above.
Assume that $|\L|\leq g(\g)\g^{-1}$
where $g(\g) $ satisfies the hypothesis of Corollary 5.3. 
Then there exist $\ell,L,\zeta, R$ all depending on $\g$
and a set $\O_\L\subset \O$ with $\P[\O_\L^c]\leq 
Ke^{-c(g(\g))^{-1/2}}+e^{-cR/\g}$ such that for all $\o\in \O_\L$
$$
\eqalign{
&\frac \g\b\ln\GG_{\b,\g,\L}[\o](F)\cr
&\leq -\inf_{\mu^\pm,s^\pm ,\pm(w_\pm-u_\pm)\leq R\,\,}
\left[\inf_{m_\ell\in F}\FF_{[w_-,w_+],\b,\g}^{(\mu^\pm, s^\pm)}(m_\ell)
-\inf_{m_\ell} \FF_{[w_-,w_+],\b,\g}^{(1,1,1,1)}(m_\ell)\right]
+er(\ell,L,M,\zeta, R)
}
\Eq(61.201)
$$
and for any $\d>0$, for $\g$ small enough
$$
\eqalign{
&\frac \g\b\ln\GG_{\b,\g,\L}[\o](F_\d)\cr
&\geq -\inf_{\mu^\pm,s^\pm ,\pm(w_\pm-u_pm)\leq R\,\,}
\left[\inf_{m_\ell\in F}\FF_{[w_-,w_+],\b,\g}^{(\mu^\pm, s^\pm)}(m_\ell)
-\inf_{m_\ell} \FF_{[w_-,w_+],\b,\g}^{(1,1,1,1)}(m_\ell)\right]
-er(\ell,L,M,\zeta, R)
}
\Eq(61.202)
$$
where $er(\ell,L,M,\hat\zeta, R)$ is a function of $\a\equiv \g M$
that tends to zero as $\a\downarrow 0$.
}

\proof
Relative to the interval $I$ we introduce again the partition $S$ 
from Section 4.  While we will use again the estimate \eqv(4.10) we treat the
terms corresponding to $S_R$ somewhat differently. 
Let us introduce the constrained partition functions 
$$
Z_{\b,\g,\L}[\o](F)\equiv\GG_{\b,\g,\L}[\o](F) Z_{\b,\g,\L}[\o]
\Eq(61.3)
$$
Just as in Proposition 4.1 we have that 
$$
\eqalign{
Z_{\b,\g,\L}(F\cap \AA(\mu^\pm,s^\pm,w_\pm)) 
&\leq Z_{\b,\g,\L^-}(\{\eta(w_-,\s)=s^- e^{\mu^-}\})
Z_{\b,\g,\D}^{(\mu^\pm,s^\pm)}(F)Z_{\b,\g,\L^+}(\{\eta(w_+,\s)=s^+ e^{\mu^+}\})
\cr
&\times
e^{8\g^{-1}(\zeta+2\g L)}
}
\Eq(61.4)
$$
and
$$
\eqalign{
Z_{\b,\g,\L}(F\cap \AA(\mu^\pm,s^\pm,w_\pm))
&\geq Z_{\b,\g,\L^-}(\{\eta(w_-,\s)=s^- e^{\mu^-}\})
Z_{\b,\g,\D}^{(\mu^\pm,s^\pm)}(F)Z_{\b,\g,\L^+}(\{\eta(w_+,\s)=s^+ e^{\mu^+}\})
\cr
&\times 
e^{-8\g^{-1}(\zeta+2\g L)}
}
\Eq(61.5)
$$
where $\D=[w_-+\frac{1}{2},w_+-\frac{1}{2}]$ and $\L^\pm $ are the two 
connected components of the complement of $\D$ in $\L$.
Using the trivial observation that 
$$
Z_{\b,\g,\L}\geq Z_{\b,\g,\L}(\AA(\mu^\pm=1,s^\pm=1,w_\pm))
\Eq(61.6)
$$
this combines to 
$$
\eqalign{
\GG_{\b,\g,\L}(F\cap \AA(\mu^\pm,s^\pm,w_\pm))&\leq
\frac {Z_{\b,\g,\D}^{(\mu^\pm,s^\pm)}(F)}
      { Z_{\b,\g,\D}^{(1,1,1,1)}}\cr
&\times
\frac { Z_{\b,\g,\L^-}(\{\eta(w_-,\s)=s^- e^{\mu^-}\})}
      { Z_{\b,\g,\L^-}(\{\eta(w_-,\s)=e^1\})}
\frac { Z_{\b,\g,\L^+}(\{\eta(w_+,\s)=s^+ e^{\mu^+}\})}
      { Z_{\b,\g,\L^+}(\{\eta(w_+,s)=e^1\})}\cr
&\times e^{16\g^{-1}(\zeta +2\g L)}
}
\Eq(61.7)
$$
The point is now that the ratios of partition functions on 
$\L^\pm$ are in fact ``close'' to one.  Indeed we have

\lemma {6.3}{\it Let $\L=[w_--\frac{1}{2},w_++\frac{1}{2}]$ with 
$|\L|\leq \g^{-1} g(\g)$, where $g(\g)\downarrow 0$
and $g(\g)/\g\geq c>0$. Then 
$$
\left|\ln Z_{\b,\g,\L}(\{\eta(w_-,\s)=s^- e^{\mu^-}\})
 -  \ln  Z_{\b,\g,\L}(\{\eta(w_-,\s)=e^1\})\right|
\leq \b\g^{-1}\left[(g(\g))^{1/4}+10\z +48\g L\right] 
\Eq(61.8)
$$
with probability greater than $1- e^{-c\g^{-1}}-Ke^{-c(g(\g))^{-1/2}}$.
}

\proof Introducing a carefully chosen zero and using the 
triangle inequality, we see that
$$
\eqalign{
&\left|\ln Z_{\b,\g,\L}(\{\eta(w_-,\s)=s^- e^{\mu^-}\})
 -  \ln  Z_{\b,\g,\L}(\{\eta(w_-,\s)=e^1\})\right|\cr
&\leq \left|\ln Z_{\b,\g,\L}(\{\eta(w_-,\s)=s^- e^{\mu^-}\})
-\ln Z_{\b,\g,\L\ba w_-}^{(0,0,\mu^-,s^-)}
+\ln Z_{\b,\g,\L\ba w_-}^{(0,0,1,1)}
- \ln  Z_{\b,\g,\L}(\{\eta(w_-,\s)=e^1\})\right|\cr
&+\left|\ln Z_{\b,\g,\L\ba w_-}^{(0,0,\mu^-,s^-) }
-\E \ln Z_{\b,\g,\L\ba w_-}^{(0,0,\mu^-,s^-)}\right|\cr
&+\left|\E \ln Z_{\b,\g,\L\ba w_-}^{(0,0,\mu^-,s^-)}
-\E \ln Z_{\b,\g,\L\ba w_-}^{(0,0,1,1)}\right|\cr
&+\left|\E\ln Z_{\b,\g,\L\ba w_-}^{(0,0,1,1)}
-\ln Z_{\b,\g,\L\ba w_-}^{(0,0,1,1)}\right|\cr
}
\Eq(61.9)
$$
The third term on the right hand side of \eqv(61.9) is zero by 
symmetry, while the second  and fourth are bounded by Corollary 5.3
by $\g^{-1}(g(\g))^{-1/4}$ with probability at least 
$1- e^{-c\g^{-1}}-Ke^{-c(g(\g))^{-1/2}}$. To bound the first term  we proceed
as in the proof of Proposition 4.1, part i, that is we use the same 
decomposition as in \eqv(4.4) and \eqv(4.41). This gives that
$$
\eqalign{
\ln Z_{\b,\g,\L}(\{\eta(w_-,\s)=s^- e^{\mu^-}\})
-\ln Z_{\b,\g,\L\ba w_-}^{(0,0,\mu^-,s^-)}
&= \ln Z_{w_-,\b,\g}(\{\eta(w_-,\s)=s^- e^{\mu^-}\})\cr &+
O\left(4\g^{-1}\left(\z +2\g L\right)\right)
}
\Eq(61.10)
$$
The constraint partition function on the block $w_-$ is easily dealt with. 
First, we note that by \eqv(2.9) with probability 
greater than $1-\exp(-c\g^{-1})$ we can replace
the Hamiltonian by its blocked version on scale $L$ 
at the expense of an error of order
$\g^{-1}(16 \g L)$. Then we can repeat the steps \eqv(4.46) to \eqv(4.48)
and use Lemma 3.1
to get that with the same probability,
$$
\ln Z_{w_-,\b,\g}(\{\eta(w_-,\s)=s^- e^{\mu^-}\})
\geq -\b \g^{-1}\left[\phi(a(\b))+\z^2+\sfrac{\ln 2}{\ln L}\right]
- \b \g^{-1}(16 \g L) 
\Eq(61.11)
$$
provided $\z\geq 2\sqrt{\sfrac ML}$. Using \eqv(4.50) and the large deviation 
bound \eqv(B.012), we also get
$$
\ln Z_{w_-,\b,\g}(\{\eta(w_-,\s)=s^- e^{\mu^-}\})
\leq -\b \g^{-1}\left[\phi(a(\b))-\sfrac 12 \z^2\right]
+\b \g^{-1}(16 \g L)
\Eq(61.12)
$$
The same bounds hold of course for the the term with $(s^-,\mu^-)$
replaced by $(1,1)$, so that we get an upper bound
$$
\b \g^{-1}\left[48 \g L+ 8\z +\frac 32\z^2\right]
\Eq(61.13)
$$
for the first term on the right of (61.9). Putting all things together, 
we arrive at the assertion of the lemma.\endproof

Lemma 6.3 asserts that to leading order, only the first ratio of
partition functions is relevant in \eqv(61.7). On the other hand, since 
by Proposition 4.1, part (ii), we 
only need to consider the case $|\D|\leq R$, we can use the block approximation
on scale $\ell$ for those, committing an error of order $\b\g^{-1}( R\g\ell)$
only. We will make this precise in the next lemma.

\lemma {6.4} {\it For any $(s^\pm, \mu^\pm, w_\pm)$ and $I\subset\D\subset\L$
and any 
$F$ that is measurable with respect to the sigma algebra 
generated by $\{m_\ell(\s,x)\}_{x\in I}$
$$
\eqalign{
\sfrac\g\b\ln \frac{Z_{\b,\g,\D}^{(\mu^\pm,s^\pm)}(F)}
                   {Z_{\b,\g,\D}^{(1,1,1,1)}}
\leq -&\inf_{m_\ell\in F}\FF^{( \mu^\pm,  s^\pm)}_{ \D,\b,\rho}
(m_\ell)
+\inf_{m_\ell}\FF^{(1,1,1,1)}_{ \D,\b,\rho}(m_\ell)\cr
&+c' \left(|\D| \g\ell
+|\D| \g M|\ln\sfrac {2\ell}M|+|\D|\sfrac {M}{\ell}\right)
}
\Eq(61.17)
$$
and $\forall{\d>0}$ for sufficiently small $\g$
$$
\eqalign{
\sfrac\g\b\ln \frac{Z_{\b,\g,\D}^{(\mu^\pm,s^\pm)}(F_\d)}
                   {Z_{\b,\g,\D}^{(1,1,1,1)}}
\geq -&\inf_{m_\ell\in F}\FF^{( \mu^\pm,  s^\pm)}_{ \D,\b,\rho}
(m_\ell)
+\inf_{m_\ell}\FF^{(1,1,1,1)}_{ \D,\b,\rho}(m_\ell)\cr
&+c' \left(|\D| \g\ell
+|\D| \g M|\ln\sfrac {2\ell}M| +|\D|\sfrac {M}{\ell}\right)
}
\Eq(61.171)
$$
with probability greater than $1-e^{-c |\D|/\g}$.
}

\proof Using Lemma 2.1, we see that 
$$
\eqalign{
 Z_{\b,\g,\D}^{(\mu^\pm,s^\pm)}(F)
 &\leq \E_\s\1_{\{m_\ell(\s)\in F\}}e^{-\b\g^{-1}
\left[E^\ell_{\g,\D}(m_\ell(\s))
+E^{\ell,L}_{\g,\D}\left(m_\ell(\s_\D),m^{(\mu^\pm,s^\pm)}\right)\right]}
\cr 
&\times e^{\b\g^{-1}40 |\D|\g\ell}
}
\Eq(61.18)
$$
and
$$
\eqalign{
 Z_{\b,\g,\D}^{(\mu^\pm,s^\pm)}(F)
 &\geq \E_\s\1_{\{m_\ell(\s)\in F\}}e^{-\b\g^{-1}
\left[E^\ell_{\g,\D}(m_\ell(\s))
+E^{\ell,L}_{\g,\D}\left(m_\ell(\s_\D),m^{(\mu^\pm,s^\pm)}\right)\right]}
\cr 
&\times e^{-\b\g^{-1}40 |\D|\g\ell}
}
\Eq(61.181)
$$
Now
$$
\eqalign{
&E^\ell_{  \D}\left(m_\ell(\s_{  \D})\right)+
E^{ \ell,L}_{  \D,\del \D}\left(m_\ell(\s_{  \D})|
m^{( \mu^\pm,  s^\pm)}\right)\cr
&=E^\ell_{  \D}\left(m_\ell(\s_{  \D})\right)+
E^{ \ell,L}_{  \D,\del \D}\left(m_\ell(\s_{  \D})|
m^{( \mu^\pm,  s^\pm)}\right)+\g \ell\sum_{x\in  \D}
\frac {\|m_\ell(\s,x)\|_2^2}2+\g \ell\sum_{x\in \del  \D}\frac {[a(\b)]^2}2\cr
&-\g \ell\sum_{x\in  \D}
\frac {\|m_\ell(\s,x)\|_2^2}2 -\g \ell\sum_{x\in \del  \D}\frac {[a(\b)]^2}2
\cr
&=-\frac 12\g\ell\sum_{(x,y)\in  \D\times  \D}
J_{\g\ell}(x-y)\left(m_\ell(\s,x),m_\ell(\s,y)\right)
-\g\ell  \sum_{x\in  \D,y\in \del \D}J_{\g\ell}(x-y)
\left(m_\ell(x,\s) ,m^{( \mu^\pm,  s^\pm)}\right)  
\cr
&+\g\ell\sum_{x\in   \D}\frac 12 \left(m_\ell(x,\s),m_\ell(x,\s)\right)
 +\g\ell \sum_{x\in \del \D} \frac 12 \left(m^{( \mu^\pm,  s^\pm)},
m^{( \mu^\pm,  s^\pm)}\right)\cr
&-\g \ell\sum_{x\in  \D}
\frac {\|m_\ell(\s,x)\|_2^2}2 -\g \ell\sum_{x\in \del  \D}\frac {[a(\b)]^2}2
\cr
}
\Eq(61.19)
$$
On the other hand
$$
\eqalign{
&\g\ell\sum_{x,y\in   \D} J_{\g\ell}(x-y)\frac{\|m_\ell(\s,x)-
m_\ell(\s,y)\|_2^2}4+\g\ell  \sum_{x\in  \D,y\in \del \D}
J_{\g\ell}(x-y)\frac{\|m_\ell(\s,x)-m^{( \mu^\pm,  s^\pm)}\|_2^2}2\cr
&=-\g\ell\sum_{x,y\in   \D} J_{\g\ell}(x-y)\frac 12
\left(m_\ell(\s,x),m_\ell(\s,y)\right)
-\g\ell  \sum_{x\in  \D,y\in \del \D}J_{\g\ell}(x-y)\frac 12
\left(m_\ell(\s,x),m^{( \mu^\pm,  s^\pm)}\right)\cr
&+\g\ell\sum_{x,y\in   \D} J_{\g\ell}(x-y)\frac 12 \|m_\ell(\s,x)\|_2^2
+\g\ell \sum_{x\in  \D,y\in \del \D}J_{\g\ell}(x-y)
\left(\frac 12 \|m_\ell(\s,x)\|_2^2+ \frac 12[a(\b)]^2\right)\cr
&=-\g\ell\sum_{x,y\in   \D} J_{\g\ell}(x-y)\frac 12
\left(m_\ell(\s,x),m_\ell(\s,y)\right)
-\g\ell  \sum_{x\in  \D,y\in \del \D}J_{\g\ell}(x-y)\frac 12
\left(m_\ell(\s,x),m^{( \mu^\pm,  s^\pm)}\right)\cr
&+\g\ell\sum_{x\in   \D}\frac 12 \|m_\ell(\s,x)\|_2^2
+\g\ell \sum_{x\in  \D,y\in \del \D}J_{\g\ell}(x-y)\frac 12[a(\b)]^2
}
\Eq(61.20)
$$
Comparing \eqv(61.19) and \eqv(61.20) we find that

$$
\eqalign{
&E^\ell_{  \D}\left(m_\ell(\s_{  \D})\right)+
E^{ \ell,L}_{  \D,\del \D}\left(m_\ell(\s_{  \D})|
 m^{( \mu^\pm,  s^\pm)}\right)+\g \ell\sum_{x\in  \D}
\frac {\|m_\ell(\s,x)\|_2^2}2+\g \ell\sum_{x\in \del  \D}\frac {[a(\b)]^2}2\cr
&=\g\ell\sum_{x,y\in   \D} J_{\g\ell}(x-y)\frac{\|m_\ell(\s,x)-
m_\ell(\s,y)\|_2^2}4+\g\ell  \sum_{x\in  \D,y\in \del \D}
J_{\g\ell}(x-y)\frac{\|m_\ell(\s,x)-m^{( \mu^\pm,  s^\pm)}\|_2^2}2\cr
&-\g\ell \sum_{x\in  \D,y\in \del \D}J_{\g\ell}(x-y)\frac 12[a(\b)]^2
\cr
&\equiv U^{  \mu^\pm,  s^\pm}_{ \D}
\left(m_\ell(\s_{\D})\right) -C(| \D|,\b)
}
\Eq(61.21)
$$
where $C(| \D|,\b)$ is an irrelevant $\s$-independent constant 
that will drop out of all relevant formulas and may henceforth be ignored.
For suitably chosen $\rho$ we introduce a lattice $\WW_{M,\rho}$ in 
$\R^M$ with 
spacing $\rho/\sqrt M$. Then for any domain $D\subset\R^M$, 
the balls of radius $\rho$ centered at the points of  $\WW_{M,\rho}\cap D$
cover $D$. For reasons that should be clear from 
Section 3, 
we choose $\rho=2\sqrt{\frac M\ell}$. With probability greater than 
$1-\exp(-c\ell)$,
$f_{x,\b,\rho}(m_\ell(x))=\infty$
if $\|m\|_2^2>2$, while the number of lattice points within the 
ball of radius $2$ are bounded by $\exp\left(M \ln \frac {2\ell}M\right)$.
But this implies that 
$$
\eqalign{
&\ln \left(\E_{\s_{ \D}}\1_{\{m_\ell(\s)\in F\}}
e^{-\b\g^{-1} \left[E^\ell_{  \D}(m_\ell(\s_{\D})
+
E^{ \ell,L}_{  \D,\del \D}\left(m_\ell(\s_{  \D})|
m^{( \mu^\pm,  s^\pm)}\right)\right]}\right)\cr
&\leq-\g^{-1}\b\inf_{m_\ell\in F} 
\left[
\FF^{(  \mu^\pm,  s^\pm)}_{ \D,\b,\rho}(m_\ell)-
C(| \D|,\b)\right]
+| \D|\left( M|\ln\sfrac {2\ell}M|+2\frac {M}{\ell}\right) 
}
\Eq(61.22)
$$
and also, if $\d>2\sqrt{\frac M\ell}$,
$$
\eqalign{
&\ln \left(\E_{\s_{ \D}}\1_{\{m_\ell(\s)\in F_\d\}}
e^{-\b\g^{-1} \left[E^\ell_{  \D}(m_\ell(\s_{\D})
+
E^{ \ell,L}_{  \D,\del \D}\left(m_\ell(\s_{  \D})|
m^{( \mu^\pm,  s^\pm)}\right)\right]}\right)\cr
&\geq-\g^{-1}\b\inf_{m_\ell\in F} 
\left[
\FF^{(  \mu^\pm,  s^\pm)}_{ \D,\b,\rho}(m_\ell)-
C(| \D|,\b)\right]
-| \D|2\frac {M}{\ell} 
}
\Eq(61.221)
$$
Treating the denominator in the first line of \eqv(61.7) in the same
way and 
putting everything together concludes the proof of the lemma.\endproof

An immediate corollary of Lemma 6.4 is 

\lemma {6.5} {\it For any $(s^\pm, \mu^\pm, w_\pm)$, $|\L|\leq \g^{-1} g(\g)$
 and any 
$F$ that is measurable with respect to the sigma algebra 
generated by $\{m_\ell(\s,x)\}_{x\in I}$, 
$$
\eqalign{
\sfrac\g\b\ln \GG_{\b,\g,\L}(F\cap \tilde 
\AA( \mu^\pm,s^\pm, w_\pm))
\leq -&\inf_{m_\ell\in F}\FF^{( \mu^\pm,  s^\pm)}_{ \D,\b,\rho}
(m_\ell) 
+\inf_{m_\ell}\FF^{(1,1,1,1)}_{ \D,\b,\rho}(m_\ell)\cr
&+c' \left(\g L+(g(\g))^{1/4} + \zeta +|\D| \g\ell
+|\D| \g M|\ln\sfrac {2\ell}M|+|\D|\sfrac {M}{\ell}\right)
}
\Eq(61.172)
$$
with probability greater than $1-Ke^{-c(g(\g))^{-1/2}}-2e^{-c /\g}$
for some finite positive numerical constants $c,c',K$. 
}
 
\proof This is an immediate consequence of \eqv(61.7) and Lemmata 6.3 and 6.4.
\endproof

We are now set to prove the upper bound in Theorem 6.2. Using the 
notation of Section 4 we have that
$$
\eqalign{
\ln\GG_{\b,\g,\L}(F)&\leq \ln \left(\GG_{\b,\g,\L}(F\cap S_R)+
\GG_{\b,\g,\L}(F\cap S_R^c)\right)\cr
&=\ln\GG_{\b,\g,\L}(F\cap S_R)+\ln\left(1+\frac{\GG_{\b,\g,\L}(F\cap S_R^c)}
{\GG_{\b,\g,\L}(F\cap S_R)}\right)\cr
&\leq 4M^2 2R \sup_{\mu^\pm,s^\pm,\pm(w_\pm-u_\pm)\leq R\,\,}\ln\GG_{\b,\g,\L}
(F\cap\AA(\mu^\pm,s^\pm,w_\pm)) +\ln \left(1+\frac{\exp\left(-c_2\b LR\zeta
\e(\zeta)\right)}
{\GG_{\b,\g,\L}(F\cap S_R)}\right) \cr
}
\Eq(61.173)
$$
where we used \eqv(4.10). We see that the last term can be made irrelevantly
small by choosing $R$ sufficiently large. In fact, since we will  consider
events $F$ those probability  will be at least of
order
$\exp(-\g^{-1}\b C)$, it will suffice to choose
$$
 R\gg \frac 1{\g L\zeta\e(\zeta)}
\Eq(61.2211)
$$
 On the other hand,
 in order for the error terms in \eqv(61.17) to go to zero,
we must assure that (note that $|\D|=|I|+2R$ is of order $R$)
$R(\g\ell+\frac M\ell)$ tends to zero. With $\a\equiv \g M$, this means
$$
R\left(\g\ell + \sfrac \a{\g\ell}\right)\downarrow 0
\Eq(61.222)
$$
From this we see that $\ell$ should be chosen as $\g\ell =\sqrt\a$
while $R$ must satisfy
$R\sqrt\a\downarrow 0$. \eqv(61.2211) and \eqv(61.222) impose conditions
on $L$ and $\zeta$, namely that
$$
\sfrac{\sqrt\a}{\g L\z\e(\z)}\downarrow 0
\Eq(61.223)
$$
Of course we also need that $\zeta \downarrow 0$ and $\g L\downarrow 0$,
but clearly these constraints can be satisfied provided that $\a\downarrow 0$
as $\g\downarrow 0$. Thus the upper bound of Theorem 6.2 follows.

To prove the lower bound, we will actually need to make use of the upper bound.
To do so, we need more explicit control of the functional $\FF$, i.e.
 we have to use the explicit 
bounds on $f_{x,\b,\rho}(m_\ell(x))$ in terms of the function $\Phi$ 
from Lemma 3.1. 

\lemma {6.6} {\it The functional $ \FF$ defined in \eqv(61.15) satisfies
$$
\FF_{\D,\b,\rho}^{(\mu^\pm,s^\pm)}(m_\ell)
\geq U_\D^{(\mu^\pm,s^\pm)}(m_\ell)+\g\ell\sum_{x\in \D}\Phi_{x,\b}(m_\ell(x))
-\frac 12 |\D|\rho^2
\Eq(61.23)
$$
and
$$
\inf_{m_\ell}\FF_{\D,\b,\rho}^{(1,1,1,1)}(m_\ell)
\leq |\D| \phi_\b(a(\b))+|\D| \frac{\ln 2}{\ell\b}
\Eq(61.24)
$$
where $\phi_\b(a)\equiv \frac {a^2}2-\b^{-1}\ln \cosh(\b a)$.
}

\proof Eq.\eqv(61.23) follows straightforward from \eqv(B.012). To get 
\eqv(61.24), just note that $U$ is non-negative and is equal to zero 
for any constant $m_\ell$, while from Lemma 3.1 it follows that 
$$
\eqalign{
\inf_{m_\ell(x)}f_{x,\b,\rho}(m_\ell(x)) &\leq 
\inf_{m_\ell(x)}\Phi_{x,\b}(m_\ell(x))+\frac{\ln 2}{\ell\b}\cr
&\leq\Phi_{x,\b}(m^{(1,1)})+\frac{\ln 2}{\ell\b}\cr
&= \phi_\b(a(\b))+\frac{\ln 2}{\ell\b}
}
 \Eq(61.25)
$$
\endproof

This concludes the derivation of the upper bound. We now turn to the
corresponding lower bound.  
What is needed for this is an upper bound on the partition function that 
would be comparable to the lower bound \eqv(61.6).
Now 
$$
\eqalign{
Z_{\b,\g,\L}&=\sum_{( \mu^\pm,  s^\pm)}
\E_\s e^{-\b H_\L(\s_\L) }
\1_{\{\eta(  w_\pm,\s)=  s^\pm e^{  \mu^\pm}\}}
 \frac {Z_{\b,\g,\L}}
{ \sum_{( \mu^\pm,  s^\pm)} \E_\s e^{-\b H_\L(\s_\L)}
\1_{\{\eta(  w_\pm,\s)=  s^\pm e^{  \mu^\pm}\}}
}\cr
&=\sum_{( \mu^\pm,  s^\pm)}
\E_\s e^{-\b H_\L(\s_\L) }
\1_{\{\eta(  w_\pm,\s)=  s^\pm e^{  \mu^\pm}\}}
\frac {Z_{\b,\g,\L}}
{\E_\s e^{-\b H_\L(\s_\L)}
\left(1-\1_{\{\eta(  w_\pm,\s)=0\}}\right)
}
\cr
&=\sum_{( \mu^\pm,  s^\pm)}
\E_\s e^{-\b H_\L(\s_\L)}
\1_{\{\eta(  w_\pm,\s)=  s^\pm e^{  \mu^\pm}\}}
\left[1-\GG_{\b,\g,\L}\left(\{\eta(  w_\pm,\s)=0\}\right)
\right]^{-1}
}
\Eq(61.26)
$$
This is almost the same form as the one  we want, except for the last factor.
The point is now that we want to use our upper bound from Theorem 6.2 to
show that  $\GG_{\b,\g,\L}\left(\{\eta(  w_\pm,\s)=0\}\right)$ is 
small, e.g. smaller than $1/2$. so that this entire factor 
is negligible on our scale. Remembering our estimate \eqv(4.10),
one may expect an estimate of the order $\exp(-c_2\b L \z\e(\z))$,
up to the usual errors. Unfortunately, these errors are of order
$\exp(\pm \b\g^{-1}(\zeta+\g L))$ and thus may offset completely the 
principle term. A way out of this apparent dilemma is given by our
remaining freedom of choice in the parameters $\zeta$ and $L$; that is 
to say, to obtain the lower bound, we will use a $\hat\zeta$ and a $\hat L$
in such that 
first they still satisfy the requirement \eqv(61.223) while second
$c_2 \hat L \hat\z(\e(\hat\z)\gg \g^{-1}\z +L $. 
This is clearly possible. With this in mind we get

\lemma {6.7} {\it  With the same probability as in Lemma 6.5,
$$
\eqalign{
&\frac \g\b\ln \GG_{\b,\g,\L}\left(\{\eta_{\hat\z,\hat L}(  w_\pm,\s)=0\}\right)
\cr
&\leq -\g \hat L \hat\z\frac {1-\d}{2-\d\hat\z}\e(\d\hat\z)
+ c'\left(\g L+(g(\g))^{1/4} + \zeta +R \g\ell
+R \g M|\ln\sfrac {2\ell}M|+R\sfrac {M}{\ell}\right)
}
\Eq(61.27)
$$
}

\proof The proof of this Lemma is very similar to the proof of (ii) of 
Proposition 4.1, except that in addition we use the upper bound of
Lemma 6.5 to reduce the error terms. We will skip the details of the 
proof.\endproof 

Choosing $\hat L$ and $\hat \z$ appropriately, we can thus achieve that
$\left[1-\GG_{\b,\g,\L}\left(\{\eta(  w_\pm,\s)=0\}\right)
\right]^{-1}\leq 2$ so that 
$$
\eqalign{
Z_{\b,\g,\L}
\leq &2\sum_{( \mu^\pm,  s^\pm)}
\E_\s e^{-\b H_\L(\s_\L)}
\1_{\{\eta(  w_\pm,\s)=  s^\pm e^{  \mu^\pm}\}}\cr
\leq &2(2M)^2 \sup_{\mu^\pm,s^\pm}
Z_{\b,\g,\L_-}(\{\eta(w_-\s)=s^-e^{\mu^-}\})
Z_{\b,\g,\D}^{(\mu^\pm,s^\pm)}
Z_{\b,\g,\L_+}(\{\eta(w_+\s)=s^+e^{\mu^+}\})\cr
&e^{+8\g^{-1}\b 
(\hat\z +2\g \hat L)}
}
\Eq(61.28)
$$
(we will drop henceforth the distinction between $\hat L$ and $L$
and $\hat \zeta$ and $\z$). 
The first and third  factor in the last line are, by Lemma 6.3,
independent of $\mu^\pm, s^\pm$, up to the usual errors. The second partition 
function is maximal for $(\mu^+,s^+)=(\mu^-,s^-)$, (this will be shown 
later).   
Thus with probability greater than $1-e^{-c \g^{-1}}-Ke^{-c(g(\g))^{-1/2}}$
$$
\GG_{\b,\g,\L}(F\cap \AA(\mu^\pm,s^\pm,w_\pm))\geq 
\frac {Z_{\b,\g,\D}^{(\mu^\pm,s^\pm)}(F)}
      { Z_{\b,\g,\D}^{(1,1,1,1)}}
e^{-c'\b\g^{-1}(\z+\g L+(g(\g))^{1/4})}
\Eq(61.29)
$$
for some numerical constant $c,c'$. Using the second assertion of Lemma 6.4
allows us to conclude the poof of Theorem 6.2. \endproof\endproof  

We are now ready to prove Theorem 6.1:

\proofof{Theorem 6.1} Notice first that the first assertion \eqv(61.1)
follows 
immediately from 
Lemma 6.7. Just note that 
$$
\GG_{\b,\g,\L}[\o]\left(\exists_{u\in \L}
\eta_{\hat\zeta,\hat L}(u,\s)=0\right)
\leq \sum_{u\in \L}\GG_{\b,\g,\L}[\o]\left( \{\eta_{\hat\z,\hat L}(u,s)=0
\}\right)\leq |\L| e^{-c\b\hat L\hat\z(\e(\hat\z))}
\Eq(61.30)
$$
for suitably chosen $\hat L, \hat z$.
To prove \eqv(61.2), note that we need only consider the case where
both $\eta(u,\s)$ and $\eta(u+1,\s)$ are non-zero. This follows then 
simply from the upper bound of Theorem 6.2 and the lower bound
$$
\inf_{\mu^\pm,s^\pm}
\inf_{m_\ell: \eta(u, m_\ell)\neq\eta(u+1, m_\ell)\neq 0}
U_\D^{(\mu^\pm,s^\pm)} (m_\ell)
\geq \sfrac 14\g\ell \sum_{x\in u}\sum_{y\in u+1}J_{\g\ell}(x-y)   
\|m_\ell(x)-m_\ell(y)\|_2^2
\Eq(61.31)
$$
Using convexity, we see that
$$
\eqalign{
&\g\ell \sum_{x\in u}\sum_{y\in u+1}J_{\g\ell}(x-y)   
\|m_\ell(x)-m_\ell(y)\|_2^2\cr
&\geq (\g\ell)^2 \sum_{{r\in u,s\in u+1}\atop{|r-s|\leq (\g\hat L)^{-1}-2}}
\sum_{x\in r}\sum_{y\in s}\|m_\ell(x)-m_\ell(y)\|_2^2\cr
&\geq (\g \hat L)^2 \sum_{{r\in u,s\in u+1}\atop{|r-s|\leq (\g\hat L)^{-1}-2}}
\left\|\sfrac{\ell}{\hat L}\sum_{x\in r}m_\ell(x)-
\sfrac{\ell}{\hat L}\sum_{y\in s}m_\ell(y)
\right\|_2^2\cr
&= (\g \hat L)^2 \sum_{{r\in u,s\in u+1}\atop{|r-s|\leq (\g\hat L)^{-1}-2}}
\left\|m_{\hat L}(r)-m_{\hat L}(s)\right\|_2^2
}
\Eq(61.32)
$$
Inserting this inequality into \eqv(61.31)
gives immediately that
$$
\eqalign{
\inf_{\mu^\pm,s^\pm}
\inf_{m_\ell: \eta(u, m_\ell)\neq\eta(u+1, m_\ell)\neq 0}
U_\D^{(\mu^\pm,s^\pm)} (m_\ell)&
\geq \sfrac 14  \sum_{{r\in u,s\in u+1}\atop{|r-s|\leq (\g\hat L)^{-1}-2}}
\left((a(\b))^2-2 a(\b)\hat \z\right)\cr
&\geq\sfrac 18 (1-2\g\hat L)^2 \left((a(\b))^2-2 a(\b)\hat \z\right)
}
\Eq(61.33)
$$
From here the proof of \eqv(61.2) is obvious. \endproof\endproof

This concludes our analysis of the free boundary condition measure in 
volumes of order $o(\g^{-1})$. We have seen that this measures are concentrated
on constant profiles on some scale $\hat L\ll \g^{-1}$ (microscopic scale).
In the next subsection we will  analyse the measures 
with fixed equilibrium boundary conditions.

\vskip 1cm
\line{\bf 6.2 Finite volume, fixed symmetric boundary conditions\hfill}

To proceed in order of increasing difficulty, we consider first the 
case where the boundary conditions are the same on both sides of the 
box $\L$. Since these are compatible with one of the preferred 
constant profiles of the free boundary conditions measures and since 
the size of the box $\L$ we consider is so small that by our self-averaging 
results we know that the random fluctuations do not favour one of the constant
values by a factor on the scale $\exp(\b\g^{-1})$, we expect that the optimal 
profile will be the constant profile compatible with the boundary 
conditions. Indeed, we will prove

\theo {6.8}   {\it Assume that $|\L|\leq g(\g)\g^{-1}$
where $g(\g) $ satisfies the hypothesis of Corollary 5.3. 
Then there exist $\ell,L,\zeta, R$ all depending on $\g$
and a set $\O_\L\subset \O$ with $\P[\O_\L^c]\leq 
Ke^{-c(g(\g))^{-1/2}}+e^{-cR/\g}$ such that for all $\o\in \O_\L$
$$
\eqalign{
&\frac \g\b\ln\GG_{\b,\g,\L}^{(\mu,s,\mu,s)}[\o](F)\cr
&\leq -\inf_{\pm(w_\pm-u_\pm)\leq R\,\,}
\left[\inf_{m_\ell\in F}\FF_{[w_-,w_+],\b,\g}^{(\mu,s,\mu,s)}(m_\ell)
-\inf_{m_\ell} \FF_{[w_-,w_+],\b,\g}^{(1,1,1,1)}(m_\ell)\right]
+er(\ell,L,M,\zeta, R)
}
\Eq(62.1)
$$
and for any $\d>0$, for $\g$ small enough
$$
\eqalign{
&\frac \g\b\ln\GG_{\b,\g,\L}^{(\mu,s,\mu,s)}[\o](F_\d)\cr
&\geq -\inf_{\pm(w_\pm-u_\pm)\leq R\,\,}
\left[\inf_{m_\ell\in F}\FF_{[w_-,w_+],\b,\g}^{(\mu,s,\mu,s)}(m_\ell)
-\inf_{m_\ell} \FF_{[w_-,w_+],\b,\g}^{(1,1,1,1)}(m_\ell)\right]
-er(\ell,L,M,\zeta, R)
}
\Eq(62.2)
$$
where $er(\ell,L,M,\zeta, R)$ is a function of $\a\equiv \g M$
that tends to zero as $\a\downarrow 0$.
}

An immediate corollary of Theorem 6.8 is the analog of Theorem 6.1 
for the measures $ \GG_{\b,\g,\L}^{(\mu,s,\mu,s)}[\o]$:

\theo{6.9} {\it Assume that $\g |\L|\downarrow 0$, 
 $\b$ large enough ($\b>1$) and
$\g M(\g)\downarrow 0$. Then we can find $\g^{-1}\gg\hat L\gg 1$ and 
$\hat\z\downarrow 0$, such that on a subset $\O_\L\subset \O$ with 
$\P(\O_\L^c)\leq e^{-c g^{-1/2}(\g)}$
where $g(\g)\downarrow 0$ and $\g^{-1} g(\g)>c$,
 we have that for all $\o\in \O_\L$
$$ 
\GG_{\b,\g,\L}^{(\mu,s,\mu,s)}[\o]\left(\exists_{u\in \L}
\eta_{\hat \z,\hat L}(u,\s)\neq se^\mu\right)
\leq e^{-\hat L g(\hat\z)}
\Eq(62.1bis)
$$
where $h(\z)=c\b \z\e(\z)$ and $\e(\z)$ is defined in \eqv(B.28).
}

\remark Eq. \eqv(62.1bis) implies that with $\P$-probability one 
$$
\lim_{\g\downarrow 0}\GG_{\b,\g,\L}^{(\mu,s,\mu,s)}[\o]\left(\forall_{u\in \L}
\eta_{\hat \z,\hat L}(u,\s)=se^\mu\right)=1
\Eq(62.1ter)
$$

{\bf \noindent Proof of Theorem 6.8:} 
%
Many of the technical steps in this proof are similar to those of 
the preceeding subsection, and we will stress only the new features here.
Let us fix without restriction of generality $(\mu,s)=(1,1)$.  
We consider again the upper bound first.
Proceeding as in \eqv(61.1), the first major  difference is that \eqv(61.7) is
replaced by 
$$
\GG_{\b,\g,\L}^{(1,1,1,1)}(F\cap \AA(\mu^\pm,s^\pm,w_\pm))\leq
\frac { Z_{\b,\g,\L^-\setminus w_-}^{(1,1,\mu^-,s^-)}}
      { Z_{\b,\g,\L^-\setminus w_-}^{(1,1,1,1)}}
\frac {Z_{\D,\b,\g}^{(\mu^\pm,s^\pm)}(F)}
      { Z_{\D,\b,\g}^{(1,1,1,1)}}
\frac { Z_{\b,\g,\L^+\setminus w_+}^{(\mu^+,s^+,1,1)}}
      { Z_{\b,\g,\L^+\setminus w_+}^{(1,1,1,1)}}
\times e^{c\g^{-1}(\zeta +\g L)}
\Eq(62.3)
$$
where we have also used \eqv(61.10) through \eqv(61.12) to replace 
partition functions with boundary condition on one side and constraint on the
 other by partition functions with two-sided boundary conditions. While in 
the free boundary condition case, by symmetry, the ratios of partition 
functions on $\L^\pm$ were seen to be negligible, we will show here that they
favour $(\mu^\pm, s^\pm) =(1,1)$. To make this precise,
define for any box $\L\equiv [\l_-,\l_+]$ with 
$|\L|=o(\g^{-1})$,
$$
P_{\b,\g,\L}^{(\tilde\mu,\tilde s,\mu,s)}\equiv 
\frac{Z_{\b,\g,\L}^{(\tilde\mu,\tilde s,\mu,s)}}{Z_{\b,\g,\L}^{(1,1,1,1)}}
\Eq(62.4)
$$
In the case of symmetric boundary conditions,
Corollary 5.3 provides the following estimates 
$$
e^{-c \b\g^{-1} (g(\g))^{1/4}}
\leq
 P_{\b,\g,\L}^{(\mu,s,\mu,s)} 
\leq e^{c \b\g^{-1} (g(\g))^{1/4}}
\Eq(62.04bis)
$$
All we need are thus
estimates on the quantity $P_\L^{(\tilde\mu,\tilde s,\mu,s)}$
for $(\tilde\mu, \tilde s)\neq (\mu,s)$. Without loss of generality we may 
consider the case $(\tilde\mu,\tilde s,\mu,s)=(1,1,2,1)$ only. As shown in the
forthcoming lemma, the quantity
$$
P_0\equiv\sup_{{[w_-,w_+]\subset\L\cup\partial\L}\atop{|w_--w_+|< 2R}}
 \frac{Z_{\b,\g,[w_-+1,w_+-1]}^{(1,1,2,1)}}
{Z_{\b,\g,[w_-+1,w_+-1]}^{(1,1,1,1)}}
\Eq(62.15)
$$
with $R$ chosen as in \eqv(61.222),
will prove to be of special relevance in estimating 
$P_{\b,\g,\L}^{(1,1,2,1)}$. 
It has in a reasonable sense the interpretation of the probability of having 
a ``jump''. Note that
the  logarithm of $P_0$ is self-averaging so that,
up to the usual error terms, by Corollary 5.3,  
the random quantity  can be replaced by the following deterministic one
$$
\bar P_0
\equiv\sup_{{[w_-,w_+]\subset\L\cup\partial\L}\atop{|w_--w_+|< 2R}}
\exp\left(\E\ln Z_{\b,\g,[w_-+1,w_+-1]}^{(1,1,2,1)}-
\E\ln Z_{\b,\g,[w_-+1,w_+-1]}^{(1,1,1,1)}\right)
\Eq(62.16)
$$
With this notations we have the

\lemma {6.10} {\it
Assume that $R$ satisfies \eqv(61.2211) and that $|\L|<\g^{-1}g(\g)$ where
$g(\g)$ is chosen as in Corollary 5.3. Then, there exists 
$\ell,L,\hat L,\z,\hat\z$ 
all depending on $\g$ such that,
with a probability greater than $1-e^{-c'\g^{-1}R}-Ke^{-c''(g(\g))^{-1/2}}$, 
where $K, c, c'$ and $c''$ are strictly positive numerical constants,
$$
P_{\b,\g,\L}^{(1,1,2,1)}\leq
\bar P_0 e^{\g^{-1}er'(\ell,L,M,\z,R)}
\Eq(62.4bis)
$$
and, if in addition $|\L|>R$, for $\g$ small enough,
$$
P_{\b,\g,\L}^{(1,1,2,1)}\geq
\bar P_0 e^{-\g^{-1}er'(\ell,\hat L,M,\hat\z,R)}
\Eq(62.4ter)
$$
where, $er'(\ell,\hat L,M,\z,R)$ is a function of $\a$ that tends to zero as 
$\a\downarrow 0$.
}

\remark Lemma 6.10 states  a very crucial result that can be paraphrased as 
follows: If the boundary conditions over a volume $\L$ with 
$|\L|=o(\g^{-1})$  require a ``jump'', than this jump takes place 
somewhere in the volume over a region smaller than $2R$; in particular, 
and this will become evident in the proof, there will occur one single 
``jump''.  Note that we cannot determine the precise location of this 
jump. The optimal position will be determined by the randomness.  
   
The proof of Lemma 6.10. relies on the important fact that, 
as stated in the next lemma,  the quantity $P_0$ 
is exponentially small.

\lemma {6.11} {\it With the notations of Lemma 6.4 we have:
\item{i)}
With a probability greater than $1-e^{-c M}$, for some constant $c>0$,
$$
P_0 \geq 
e^{-\frac 12 \b\g^{-1}a^2(\beta)}
e^{-c\b\g^{-1}\left(R\g\ell+R\g M |\ln\frac {2\ell}M|+R\frac M\ell
+2R\frac{\ln 2}{\ell}\right)}
\Eq(62.17)
$$
\item{ii)}
There exists $\tilde\z_0>0$ depending on $\b$ such 
that for all $\tilde\z_0\geq \tilde\z\geq 2a(\b)\sqrt{\sfrac M\ell}$,
with a probability greater than $1-e^{-c' M}$, for some constant $c'>0$,
$$
P_0 \leq 
e^{-\b\g^{-1}\sqrt{\e(\tilde\z)}\left(\sqrt{12((a(\b))^2 -4\tilde\z^2)}-3
\sqrt{\e(\tilde\z)}\right)}
e^{c\b\g^{-1}\left(R\g\ell+R\g M |\ln\frac {2\ell}M|+R\frac M\ell\right)}
\Eq(62.17bis)
$$
}

We will assume in the sequel that the parameters $\ell, L, M$ and $R$ satisfy 
the set of conditions \eqv(61.2211) to \eqv(61.223) from Section 6.1.
It is then clear that the parameter $\tilde\z$ in part ii) of Lemma 6.11
can always be chosen in such a way that the exponential decrease of the first 
term in the r.h.s. of \eqv(62.17bis) compensates the increase of the second 
one.
We will postpone the proof of Lemma 6.11 to the end of this subsection.

\proofof{lemma 6.10} 
Without loss of generality we will, for convenience, consider only sets $\L$ 
of the form $\L\equiv[\l^--\frac{1}{2},\l^++\frac{1}{2}]$ where $\l^\pm$ are 
assumed to be integers. 
We start with the proof of the upper bound \eqv(62.4bis). Let us  
define the set
$$
B\equiv\left\{\s\,: \forall_{u\in\L} \eta(u,\s)\in\{0,e^1,e^2\}\right\}
\Eq(62.5)
$$
We further define  
$$
u_1(\s)\equiv \cases{ \sup \left\{u\in[\l_--\frac{1}{2},\l_++\frac{1}{2}]
\mid \eta(u,\s)=e^1\right\}&,
if such $u$ exists\cr
   \l_--1&, otherwise}
\Eq(62.6)
$$
$$
u_2(\s)\equiv \cases{ \inf \left\{u\in (u_1(\s),\l_++\frac{1}{2}]\mid 
\eta(u,\s)=e^2
\right\}&,
if such $u$ exists\cr
   \l_++1&, otherwise}
\Eq(62.7)
$$
and we set 
$$
B(u_1,u_2)\equiv \left\{\s\in B\mid u_1(\s)=u_1,u_2(\s)=u_2\right\} 
\Eq(62.8)
$$
A piece of profile between locations $u_1(\s)$ and $u_2(\s)$ will be called
a ``jump'' between equilibrium (1,1) and $(2,1)$.
For $R$ chosen as in \eqv(61.222), we will set moreover
$$
C\equiv \bigcup_{{\l_--1\leq u_1<u_2\leq \l_++1}\atop{|u_2-u_1|<2R}}
B(u_1,u_2)
\Eq(62.9)
$$
and
$$
D\equiv \bigcup_{{\l_--1\leq u_1<u_2\leq \l_++1}\atop{|u_2-u_1|\geq 2R}}
B(u_1,u_2)
\Eq(62.10) 
$$
Obviously,
$$
Z_{\b,\g,\L}^{(1,1,2,1)}=Z_{\b,\g,\L}^{(1,1,2,1)}(B)
+Z_{\b,\g,\L}^{(1,1,2,1)}(B^c)
\Eq(62.11)
$$
and 
$$
Z_{\b,\g,\L}^{(1,1,2,1)}(B)=Z_{\b,\g,\L}^{(1,1,2,1)}(C)
+Z_{\b,\g,\L}^{(1,1,2,1)}(D)
\Eq(62.11bis)
$$
Now, on the one hand, we have
$$
\eqalign{ 
\frac{Z_{\b,\g,\L}^{(1,1,2,1)}(D)}{Z_{\b,\g,\L}^{(1,1,1,1)}}
\leq &\sum_{{\l_--1\leq u_1<u_2\leq \l_++1}\atop{|u_2-u_1|\geq 2R}}
\frac{Z_{\b,\g,\L}^{(1,1,2,1)}\left(\forall_{u_1<u<u_2} \eta(u,\s)=0\right)}
{Z_{\b,\g,\L}^{(1,1,1,1)}} \cr
\leq &\sum_{{\l_--1\leq u_1<u_2\leq \l_++1}\atop{|u_2-u_1|\geq 2R}}
\frac{Z_{\b,\g,[u_1,u_2]}\left(\forall_{u_1<u<u_2} \eta(u,\s)=0\right)}
{Z_{\b,\g,[u_1,u_2]}^{(1,1,1,1)}}
e^{+c\b\g^{-1}\left(\g L+\z+(g(\g))^{1/4}\right)}\cr
}
\Eq(62.12)
$$
where we have proceeded by complete analogy with the proof of the upper bound 
of Theorem 6.2 (see \eqv(61.3)-\eqv(61.7) and Lemma 6.3) to chop out
the partition functions in $[\l_--\frac{1}{2}, u_1]$ and 
$[u_2, \l_++\frac{1}{2}]$, and where we have 
dropped the boundary conditions of $Z_{\b,\g,[u_1, u_2]}$ 
in the numerator of the
last line. This holds with a  probability greater than 
$1-Ke^{-c(g(\g))^{-1/2}}$. Up to some minor 
modifications, it then follows from the proof of
Proposition 4.1, part ii), that,
with a probability greater than $1-e^{-c' M}-Ke^{-c(g(\g))^{-1/2}}$, 
$$
\frac{Z_{\b,\g,\L}^{(1,1,2,1)}(D)}{Z_{\b,\g,\L}^{(1,1,1,1)}}
\leq |\L|^2 e^{-\b L R \z\e(\z)}
e^{c\b\g^{-1}\left(\g L+\z+(g(\g))^{1/4}\right)}
\Eq(62.13)
$$
On the other hand, 
we have also that, with a probability greater than $1-Ke^{-c(g(\g))^{-1/2}}$,
$$
\eqalign{
\frac{Z_{\b,\g,\L}^{(1,1,2,1)}(C)}{Z_{\b,\g,\L}^{(1,1,1,1)}}
&\leq
\sum_{{\l_--1\leq u_1<u_2\leq \l_++1}\atop{|u_2-u_1|< 2R}}
  \frac{Z_{\b,\g,[u_1+1,u_2-1]}^{(1,1,2,1)}
\left(\forall_{u_1<u<u_2}\eta(u,\s)=0\right)}
{Z_{\b,\g,[u_1+1,u_2-1]}^{(1,1,1,1)}}
\cr
&\times e^{c\b\g^{-1} (\g L+\z +(g(\g))^{1/4})}
\cr
&\leq 2R|\L|P_0 e^{c\b\g^{-1}\left(\g L+\z+(g(\g))^{1/4}\right)}
\cr
}
\Eq(62.14)
$$
From now on we will abstain from specifying the probability with which our
various estimates hold; this will straightforwardly follow from the different
results called into play.
Now, using the lower bound \eqv(62.17) of Lemma 6.11 and 
recalling that $R$ is chosen large enough  to satisfy the
constraint \eqv(61.2211), we see that the r.h.s of \eqv(62.13) is negligibly 
small compared with that of \eqv(62.14). Combining this with \eqv(62.11) and 
using Corollary 5.3 we then arrive at
$$
\frac{Z_{\b,\g,\L}^{(1,1,2,1)}(B)}{Z_{\b,\g,\L}^{(1,1,1,1)}}
\leq \bar P_0 e^{c'\b\g^{-1} (\g L+\z +(g(\g))^{1/4})}
\Eq(62.18)
$$

We are therefore left to consider the constrained partition function 
$Z_{\b,\g,\L}^{(1,1,2,1)}(B^c)$. 
By definition, for any $\s\in B^c$, there must exist $u\in \L$ such that 
$\eta(u,\s)=se^\mu$, with $(s,\mu)\not\in \{ (1,1),(1,2)\}$.   
This means that we can define the four random locations
$$
u_1^+(\s)= \sup \left\{u\in[\l_--\sfrac{1}{2},\l_++\sfrac{1}{2}]
\mid \eta(u,\s)\not\in\{0, e^2\}\right\}
\Eq(62.19)
$$
$$
u_2^+(\s)\equiv \cases{ \inf \left\{u\in[u_1^+(\s),\l_++\sfrac{1}{2}]
\mid \eta(u,\s)=e^2
\right\}&,
if such $u$ exists and\cr
   \l_++1&, otherwise}
\Eq(62.20)
$$
and
$$
u_1^-(\s)\equiv \inf \left\{u\in[\l_--\sfrac{1}{2},\l_++\sfrac{1}{2}]
\mid \eta(u,\s)\not\in\{0, e^1\}
\right\}
\Eq(62.21)
$$
$$
u_2^-(\s)\equiv \cases{ \sup \left\{u\in [\l_--\sfrac{1}{2},u_2^-(\s)]\mid 
\eta(u,\s)=e^1\right\}&,
if such $u$ exists and\cr
   \l_--1&, otherwise}
\Eq(62.22)
$$
and can be sure that, for all $\s\in B^c$, $u_1^-(\s)\leq u_1^+(\s)$.
In other words, any configuration in $B^c$ contains two ``jumps''. The
following two events, $B^-(u_2^-)$ and $B^-(u_2^+)$, describe, respectively the
leftmost and rightmost of these jumps.
For $u_2^->\l_--1$ we set
$$
B^-(u_2^-)\equiv
\left\{\s\Big| \forall_{\l^--1<u<u_2^-}
\eta(u,\s)\in\{0, e^1\},\,
\eta(u_2^-,\s)=e^1,\,
\forall_{u_2^-<u<u_1^-}\eta(u,\s)=0\right\}
\Eq(62.221)
$$
while
$$
B^-(\l_--1)\equiv
\Bigl\{\s\Big| \forall_{\l^--1<u<u_1}\,\eta(u,\s)=0\Bigr\}
\Eq(62.221bis)
$$
Similarly we set, for $u_2^+<\l_++1$, 
$$
B^+(u_2^+)\equiv\Bigl\{\s\Big|\forall_{u_2^+<u<\l^+-1}
\eta(u,\s)\in\{0, e^2\},\, 
\eta(u_2^+,\s)=e^2\,
\forall_{u_+^1<u<u_2^+}\eta(u,\s)=0\Bigr\}
\Eq(62.222)
$$
and
$$
B^+(\l_++1)\equiv\Bigl\{\s\Big|
\forall_{u_+^1+1<u<\l^+-1}\eta(u,\s)=0\Bigr\}
\Eq(62.222bis)
$$
Proceeding in the (by now) usual way, we see from here that
$$
\eqalign{
\frac{Z_{\b,\g,\L}^{(1,1,2,1)}(B^c)}{Z_{\b,\g,\L}^{(1,1,1,1)}}
&\leq 
\sup_{\l_--1\leq u_2^-<u_1^-\leq u_1^+<u_2^+\leq \l_++1}
\sup_{{(\mu^-,s^-)\neq (1,1)}\atop
      {(\mu^+,s^+)\neq (1,2)}}
\Biggl\{
\frac {Z_{[u_1^-+\frac 12,u_1^+-\frac 12],\b,\g}^{(\mu^-,s^-,\mu^+,s^+)}}
      {Z_{\b,\g,[u_1^-+\frac 12,u_1^+-\frac 12]}^{(1,1,1,1)}}\cr
&\times
\frac {Z_{\b,\g,[\l_--\frac 12,u_1^--\frac 12]}^{(1,1,\mu^-,s^-)}(B^-(u_2^-))}
      {Z_{\b,\g,[\l_--\frac 12,u_1^--\frac 12]}^{(1,1,1,1)}}
\frac {Z_{\b,\g,[u_1^++\frac 12,\l_++\frac 12]}^{(\mu^+,s^+,2,1)}(B^+(u_2^+))}
     {Z_{\b,\g,[u_2^++\sfrac 12,\l_+]}^{(1,1,1,1)}}
\Biggr\}
\cr
&\times |\L|^4M^2 e^{c\b\g^{-1} (\g L+\z +(g(\g))^{1/4})}\cr
}
\Eq(62.23)
$$
Clearly, each of the two terms in the second line of \eqv(62.23)
is bounded as in \eqv(62.18), so that,
up to the error term, we get the relation\note{Observe that this inequality 
shows in particular that the probability of having more than one jump 
is bounded by the square of the probability of having one jump.}
$$
\frac{Z_{\b,\g,\L}^{(1,1,2,1)}(B^c)}{Z_{\b,\g,\L}^{(1,1,1,1)}}
\leq (\bar P_0)^2\sup_{\l_--1\leq u_2^-<u_1^-\leq u_1^+<u_2^+\leq \l_++1}
\sup_{{(\mu^-,s^-)\neq (1,1)}\atop
      {(\mu^+,s^+)\neq (1,2)}} 
P_{\b,\g,[u_1^-+\frac 12,u_1^++\frac 12]}^{(\mu^-,s^-,\mu^+,s^+)}
\Eq(62.24)
$$
and combining this with \eqv(62.11) and \eqv(62.18) we arrive, still up to 
the error term, at
$$
P_\L^{(1,1,2,1)}\leq \bar P_0 + 
(\bar P_0)^2\sup_{\l_--1\leq u_2^-<u_1^-\leq u_1^+<u_2^+\leq \l_++1}
\sup_{{(\mu^-,s^-)\neq (1,1)}\atop
      {(\mu^+,s^+)\neq (1,2)}} 
P_{\b,\g,[u_1^-+\frac 12,u_1^++\frac 12]}^{(\mu^-,s^-,\mu^+,s^+)}
\Eq(62.25)
$$
We immediately see from this recursion that the supremum over the 
$\mu^\pm,s^\pm$ will be realized for $(\mu^-,s^-)=(\mu^+,s^+)$.
But putting the estimates \eqv(62.04bis) together with the upper bound 
\eqv(62.17bis) of Lemma 6.11  and Corollary 5.3 we get that, for small enough 
$\g$, 
$\bar P_0P_{\b,\g,[u_1^-+\frac 12,u_1^++\frac 12]}^{(\mu,s,\mu,s)} <<1$.
From this the upper bound \eqv(62.4bis) is readily obtained.
\endproof 

We now turn to the proof of the lower bound \eqv(62.4ter). 
First note that for any $[w_-,w_+]\subset \L$,
$$
\eqalign{
Z_{\b,\g,\L}^{(1,1,2,1)}
&\geq Z_{\b,\g,\L}^{(1,1,2,1)}(\AA(1,1,2,1,w_\pm))\cr
&\geq Z_{\b,\g,\L^-}^{(1,1,0,0)}(\{\eta(w_-,\s)=e^1\})
Z_{\D,\b,\g}^{(1,1,2,1)}
Z_{\b,\g,\L^+}^{(0,0,2,1)}(\{\eta(w_+,\s)=e^2\})e^{-8\g^{-1}(\z+2\g L)}
}
\Eq(62.28)
$$
where we obtained the second inequality by proceeding just as in \eqv(61.4).
The difficulty thus lies in establishing a corresponding upper bound 
for the partition function $Z_{\b,\g,\L}^{(1,1,1,1)}$. But this can be done 
by, basically, repeating the proof of the lower bound of Theorem 6.2. I.e., we 
first use the decomposition \eqv(61.26) to write
$$
Z_{\b,\g,\L}^{(1,1,1,1)}=\sum_{\mu^\pm,s^\pm}\E_{\s}
e^{H_{\g,\L}(\s_\L)+W_{\g,\L}(\s_\L, m^{(\mu^\pm,s^\pm)})}
\1_{\{\eta(w_\pm,\s^\pm)=s^\pm e^{\mu^\pm}\}}
\left[1-\GG_{\b,\g,\L}^{(1,1,1,1)}(\{\eta(w_\pm,\s^\pm)=0\})\right]^{-1}
\Eq(62.29)
$$
anticipating that 
$\GG_{\b,\g,\L}^{(1,1,1,1)}(\{\eta(w_\pm,\s^\pm)=0\})$ can be shown to be very 
small. We will prove that Lemma 6.7 still holds when the Gibbs measure 
with free boundary conditions in \eqv(61.27) is replaced by
$\GG_{\b,\g,\L}^{(1,1,1,1)}$. Assuming for the moment that this is true
we get, as in \eqv(61.28) and with the same choices of the parameters
$\hat\z$ and $\hat L$, 
$$
\eqalign{
Z_{\b,\g,\L}^{(1,1,1,1)}\leq &2(2M)^2\sup_{\mu^\pm,s^\pm}
Z_{\b,\g,\L^-}^{(1,1,0,0)}(\{\eta(w_-,\s)=s^- e^{\mu^-}\})
Z_{\D,\b,\g}^{(\mu^\pm,s^\pm)}
Z_{\b,\g,\L^+}^{(0,0,1,1)}(\{\eta(w_+,\s)=s^+ e^{\mu^+}\})\cr
&e^{-8\g^{-1}(\hat\z+2\g\hat L)}\cr
}
\Eq(62.30)
$$
Next, proceeding as in \eqv(62.3) 
to replace constrained partition functions with 
free boundary condition on one side, by partition functions with two-sided 
boundary conditions, we have
$$
\eqalign{
&\frac{Z_{\b,\g,\L^-}^{(1,1,0,0)}(\{\eta(w_-,\s)=s^- e^{\mu^-}\})}
     {Z_{\b,\g,\L^-}^{(1,1,0,0)}(\{\eta(w_-,\s)=e^1\})}
\frac{Z_{\D,\b,\g}^{(\mu^\pm,s^\pm)}}
     {Z_{\D,\b,\g}^{(1,1,1,1)}}
\frac{Z_{,\b,\g,\L^+}^{(0,0,1,1)}(\{\eta(w_+,\s)=s^+ e^{\mu^+}\})}
     {Z_{\b,\g,\L^+}^{(0,0,1,1)}(\{\eta(w_+,\s)=e^1\})}
\cr
\leq &P_{\b,\g,\L^-\setminus w_-}^{(1,1,\mu^-,s^-)}
      P_{\D,\b,\g}^{(\mu^\pm,s^\pm)}
      P_{\b,\g,\L^+\setminus w_+}^{(\mu^+,s^+,1,1)}
e^{c\g^{-1}(\z+2\g L)}\cr
\leq &e^{c'\g^{-1}(\z+2\g L+(g(\g)^{1/4}))}\cr
}
\Eq(62.31)
$$
where, to obtain the last line, we used \eqv(62.04bis) to treat terms with 
symmetric boundary conditions while we used, in the case of 
asymmetric boundary conditions, the upper bound \eqv(62.4bis) of Lemma 6.10 
together with Lemma 6.11 and Corollary 5.3.
From this and \eqv(62.30) it follows that, up to the error term,
$$
Z_{\b,\g,\L}^{(1,1,1,1)}\leq 2(2M)^2\sup_{\mu^\pm,s^\pm}
Z_{\b,\g,\L^-}^{(1,1,0,0)}(\{\eta(w_-,\s)=e^1\})
Z_{\D,\b,\g}^{(1,1,1,1)}
Z_{\b,\g,,\L^+}^{(0,0,1,1)}(\{\eta(w_+,\s)=e^1\})
\Eq(62.32)
$$
Following a procedure with which the reader is now well acquainted,
\eqv(62.28) together with \eqv(62.32) easily yields
\eqv(62.4ter) by choosing $w_-$ and $w_+$
as those which satisfy the constrained supremum problem in \eqv(62.15). 
Of course we must ask that $|\D|>R$ to ensure that this choice is always 
possible.

To complete the proof of \eqv(62.4ter) it remains to show that Lemma 6.7 
holds when replacing $\GG_{\b,\g,\L}$ by $\GG_{\b,\g,\L}^{(1,1,1,1)}$. 
To do so, all we need is to prove the analogous of Lemma 6.5 for the measure 
$\GG_{\b,\g,\L}^{(1,1,1,1)}$. But this is an immediate consequence of 
\eqv(62.3). Indeed, we can for the present purpose be content with the bounds 
from Corollary 5.3. to estimate (although roughly) the first and third factor 
in the r.h.s. of \eqv(62.3), while using Lemma 6.4 to treat the middle term.
This concludes the proof of Lemma 6.10.\endproof

We are now ready to continue the proof of the upper bound of Theorem 6.8.
Remember that we were left in \eqv(62.3) to estimate the ratios of partition
functions in $\L^\pm$. In the case of asymmetric boundary 
conditions i.e., $(\mu^\pm,s^\pm)\neq(1,1)$, Lemma 6.10 enables us to
replace these quantities by the corresponding ratios in boxes of length 
at least $R$.
More precisely, consider two boxes $\L'$ and $\L$ such that
$\L'\subset\L$ and $R<|\L'|<|\L|<\g^{-1}g(\g)$ where $g(\g)$ 
is chosen as in Corollary 5.3. Then, Lemma 6.10 implies, 
for any $(\tilde\mu,\tilde s)\neq (\mu,s)$, that
$$
P_{\b,\g,\L}^{(\tilde\mu,\tilde s,\mu,s)}
\leq P_{\b,\g,\L'}^{(\tilde\mu,\tilde s,\mu,s)}
e^{+\g^{-1}(er'(\ell,L,M,\z,R)+er'(\ell,L,M,\hat\z,R))}
\Eq(62.4corol)
$$
Therefore, defining the boxes
$\widetilde\L^-\equiv [w_-+\frac 12-R,w_-+\frac 12]$ and 
$\widetilde\L^+\equiv [w_+-\frac 12,w_+-\frac 12+R]$, adjacent to $\D$
on its left, respectively right hand side, we have, up the the error terms,
$$
\GG_{\b,\g,\L}^{(1,1,1,1)}(F\cap \AA(\mu^\pm,s^\pm,w_\pm))
\leq P_{\widetilde\L^-\setminus w_-,\b,\g}^{(1,1,\mu^-,s^-)}
      \frac{Z_{\D,\b,\g}^{(\mu^\pm,s^\pm)}(F)}
     {Z_{\D,\b,\g}^{(1,1,1,1)}}
      P_{\widetilde\L^+\setminus w_+,\b,\g}^{(\mu^+,s^+,1,1)}
\Eq(62.33)
$$
By \eqv(62.04bis) a relation of the form \eqv(62.32), and hence
\eqv(62.33), trivially holds in the case of symmetric boundary conditions.
From here we can easily reconstruct the ratio of partition functions
in $\widetilde\D\equiv\widetilde\L^-\cup\D\cup\widetilde\L^+$ with 
$(1,1,1,1)$-boundary conditions. I.e., proceeding much along the line of the 
proof of the upper bound of Lemma 6.10
(using in particular \eqv(62.32)) we obtain, up to the usual error term,
$$
\eqalign{
\GG_{\b,\g,\L}^{(1,1,1,1)}(F\cap \AA(\mu^\pm,s^\pm,w_\pm))
&\leq
      \frac {Z_{\widetilde\D,\b,\g}^{(1,1,1,1)}
                (F\cap\{\eta(w_\pm,\s)=s^\pm e^{\mu^\pm}\})}
      { Z_{\widetilde\D,\b,\g}^{(1,1,1,1)}}\cr
&\leq \frac {Z_{\widetilde\D,\b,\g}^{(1,1,1,1)}(F)}
      { Z_{\widetilde\D,\b,\g}^{(1,1,1,1)}}\cr
}
\Eq(62.34)
$$
The upper bound of Theorem 6.8 then follows from \eqv(62.34) and Lemma 6.4
just as the upper bound of Theorem 6.2 follows from Lemma 6.5. 

At this point, the proof of the lower bound \eqv(62.4ter) is a simple 
matter. In full
generality, for arbitrary $\D\equiv [w_-+\frac{1}{2},w_+-\frac{1}{2}]$ and
any $(\mu,s)$,
$$
\GG_{\b,\g,\L}^{(\mu,s,\mu,s)}(F)
\geq\frac{Z_{\b,\g,\L}^{(\mu,s,\mu,s)}(F\cap \AA(\mu^\pm=\mu,s^\pm=s,w_\pm))}
     {Z_{\b,\g,\L}^{(\mu,s,\mu,s)}}
\Eq(62.341)
$$
Proceeding as in \eqv(61.4) to bound the numerator in \eqv(62.34) and
using \eqv(62.32) to treat the denominator we get
$$
\eqalign{
&\frac{Z_{\b,\g,\L}^{(\mu,s,\mu,s)}(F\cap \AA(\mu^\pm=\mu,s^\pm=s,w_\pm))}
     {Z_{\b,\g,\L}^{(\mu,s,\mu,s)}}\cr
&\approx
\frac{Z_{\b,\g,\L^-}^{(\mu,s,0,0)}(\{\eta(w_-,\s)=s e^{\mu}\})}
     {Z_{\b,\g,\L^-}^{(\mu,s,0,0)}(\{\eta(w_-,\s)=e^1\})}
\frac{Z_{\D,\b,\g}^{(\mu,s,\mu,s)}(F)}
     {Z_{\D,\b,\g}^{(\mu,s,\mu,s)}}
\frac{Z_{\b,\g,\L^+}^{(0,0,\mu,s)}(\{\eta(w_+,\s)=s e^{\mu}\})}
     {Z_{\b,\g,\L^+}^{(0,0,\mu,s)}(\{\eta(w_+,\s)=e^1\})}\cr
}
\Eq(62.342)
$$
Again, we recognise in the first and last factor above the 
quantities $P_{\b,\g,\L}^{(\mu,s,\mu,s)}$ for which we have the estimates 
\eqv(62.04bis). Thus, up to the usual error term,
$$
\GG_{\b,\g,\L}^{(\mu,s,\mu,s)}(F)
\geq \frac{Z_{\D,\b,\g}^{(\mu,s,\mu,s)}(F)}
     {Z_{\D,\b,\g}^{(\mu,s,\mu,s)}}
\Eq(62.343)
$$
Now, \eqv(62.343) and Lemma 6.4
yield \eqv(62.2) by choosing $w_\pm$ as the solutions of the
variational problem in \eqv(62.1).
This completes the proof of Theorem 6.8.
\endproof

We finally are left to give the proof of Lemma 6.11

\proofof{lemma 6.11}
To prove the lower bound \eqv(62.17), just note that for any 
event $F$ satisfying the assumptions of Lemma 6.4, and, making use of the 
lower bound \eqv(61.171), we have.
$$
\eqalign{
P_0& \geq 
\exp\left(-\b\g^{-1}\left[\inf_{m_\ell\in F}\FF^{(1,1,2,1)}_{[\l_-,\l_+]}
(m_\ell)-\inf_{m_\ell}\FF_{[\l_-,\l_+]}^{(1,1,1,1)}(m_\ell) \right]\right)\cr
&\times e^{-c\b\g^{-1}\left(R\g\ell+R\g M |\ln\frac {2\ell}M|
+R\frac M\ell\right)}
}
\Eq(62.lemme691)
$$
Now, choosing the event $F$ as 
$$
F\equiv\bigl\{\{m_\ell(x)\}_{x\in I}
\big| m_\ell(x)=a(\b)e^1 \forall x\leq 0, m_\ell(x)=a(\b)e^2\forall x> 0\bigr\}
\Eq(62.lemme692)
$$
it easily follows from the definition \eqv(61.15) of $\FF$ together with
the estimates of Lemma 3.1 and Proposition 3.2 that, under their respective 
assumptions,
$$
\eqalign{
\inf_{m_\ell\in F}\FF^{(1,1,2,1)}_{[\l_-,\l_+]}
(m_\ell)-\inf_{m_\ell}\FF_{[\l_-,\l_+]}^{(1,1,1,1)}(m_\ell)
&\leq \inf_{m_\ell\in F} U^{(1,1,2,1)}_{[\l_-,\l_+]}(m_\ell)
+2R\frac{\ln 2}{\b\ell}+R\frac{M}{\ell}\cr
&\leq  \frac{a^2(\b)}{2}
+2R\frac{\ln 2}{\b\ell}+R\frac{M}{\ell}\cr
}
\Eq(62.lemme693)
$$
from which \eqv(62.17) follows.
To prove the lower bound \eqv(62.17bis) we make use of the bound \eqv(61.17) 
of Lemma 4 to write
$$
\eqalign{
P_0& \leq
\sup_{{\l_--1\leq u_1<u_2\leq \l_++1}\atop{|u_2-u_1|< 2R}}
\exp\left(-\b\g^{-1}\left[\inf_{m_\ell}\FF^{(1,1,2,1)}_{[u_1+1,u_2-1]}
(m_\ell)-\inf_{m_\ell}\FF_{[u_1+1,u_2-1]}^{(1,1,1,1)}(m_\ell) \right]\right)\cr
&\times e^{c\b\g^{-1}\left(R\g\ell+R\g M |\ln\frac {2\ell}M|
+R\frac M\ell\right)}\cr
}
\Eq(62.lemme694)
$$
\eqv(62.17bis) is then an immediate consequence of \eqv(62.lemme694)
together with the following proposition

\proposition {6.12} {\it There exists $\tilde\z_0>0$ depending on $\b$ such 
that for all $\tilde\z_0\geq \tilde\z\geq 2a(\b)\sqrt{\sfrac M\ell}$ 
and for all boxes $\D$
$$
\inf_{m_\ell}\FF^{(1,1,2,1)}_{\D}
(m_\ell)-\inf_{m_\ell}\FF_{\D}^{(1,1,1,1)}(m_\ell)
\geq\sqrt{\e(\tilde\z)}\left(\sqrt{12((a(\b))^2 -4\tilde\z^2)}
-3\sqrt{\e(\tilde\z)}\right)
\Eq(62.prop610)
$$
with probability greater than $1-e^{-c M}$.
}

The proof of Proposition 6.12, which is somewhat technical, will
be the object of Section 7. With this, the proof of Lemma 6.11 is
concluded. \endproof

\vskip 1cm

\newpage

\line{\bf 6.3 Finite volume, fixed asymmetric boundary conditions\hfill}
In this last subsection we consider the case where the boundary conditions
to the right and to the left of the box $\L$ are distinct. We would expect 
here that the optimal profile will be the ``jump'' profile compatible with 
these conditions. We will prove

\theo  {6.13}   {\it Assume that $|\L|\leq g(\g)\g^{-1}$
where $g(\g) $ satisfies the hypothesis of Corollary 5.3. 
Then there exist $\ell,L,\zeta, R$ all depending on $\g$
and a set $\O_\L\subset \O$ with $\P[\O_\L^c]\leq 
Ke^{-c(g(\g))^{-1/2}}+e^{-c/\g}$ such that for all $\o\in \O_\L$, for any
$(\tilde\mu,\tilde s)\neq (\mu,s)$, 
$$
\eqalign{
&\frac \g\b\ln\GG_{\b,\g,\L}^{(\tilde\mu,\tilde s,\mu,s)}[\o](F)\cr
&\leq -\inf_{\pm(w_\pm-u_\pm)\leq R\,\,}
\left[
\inf_{m_\ell\in F}\FF_{[w_-,w_+]}^{(\tilde\mu,\tilde s,\mu,s)}(m_\ell)
-\inf_{m_\ell} \FF_{[w_-,w_+]}^{(1,1,2,1)}(m_\ell)\right]
+er(\ell,L,M,\zeta, R)
}
\Eq(62.35)
$$
and for any $\d>0$, for $\g$ small enough,
$$
\eqalign{
&\frac \g\b\ln\GG_{\b,\g,\L}^{(\tilde\mu,\tilde s,\mu,s)}[\o](F_\d)\cr
&\geq -\inf_{\pm(w_\pm-u_\pm)\leq R\,\,}
\left[
\inf_{m_\ell\in F}\FF_{[w_-,w_+]}^{(\tilde \mu,\tilde s,\mu,s)}(m_\ell)
-\inf_{m_\ell} \FF_{[w_-,w_+]}^{(1,1,2,1)}(m_\ell)\right]
-er(\ell,L,M,\zeta, R)
}
\Eq(62.36)
$$
where $er(\ell,L,M,\hat\zeta, R)$ is a function of $\a\equiv \g M$
that tends to zero as $\a\downarrow 0$.
}

\proof The proof of Theorem 6.13  presents no additional technical 
difficulties compared to that of Theorem 6.8. We shall thus be very brief and 
restrict ourselves to detail the only subtle step. This one enters in 
the proof 
of the upper bound for the quantity 
$\GG_{\b,\g,\L}^{(\tilde\mu,\tilde s,\mu,s)}(F\cap \AA(\mu^\pm,s^\pm,w_\pm))$. 
From now on we will place ourselves on the subset of the probability space on 
which our various estimates from Section 6.1 and 6.2 hold. 
Without loss of generality we may only consider the case 
$(\tilde\mu,\tilde s,\mu,s)=(1,1,2,1)$. It is a simple matter to establish that
$$
\GG_{\b,\g,\L}^{(1,1,2,1)}(F\cap \AA(\mu^\pm,s^\pm,w_\pm))\leq
\frac { Z_{\b,\g,\L^-\setminus w_-}^{(1,1,\mu^-,s^-)}}
      { Z_{\b,\g,\L^-\setminus w_-}^{(1,1,1,1)}}
\frac {Z_{\D,\b,\g}^{(\mu^\pm,s^\pm)}(F)}
      { Z_{\D,\b,\g}^{(1,1,2,1)}}
\frac { Z_{\L^+\setminus w_+,\b,\g}^{(\mu^+,s^+,2,1)}}
      { Z_{\L^+\setminus w_+\b,\g}^{(2,1,2,1)}}
\times e^{c\g^{-1}(\zeta +\g L)}
\Eq(62.37)
$$
Just as in \eqv(62.33) we replace the ratios of partition functions
in boxes $\L^\pm$ above by the corresponding ratios in boxes 
$\widetilde\L^\pm$ of length $R$. Thus, up to negligible errors,
$$
\GG_{\b,\g,\L}^{(1,1,2,1)}(F\cap \AA(\mu^\pm,s^\pm,w_\pm))\leq
P_{\widetilde\L^-\setminus w_-,\b,\g}^{(1,1,\mu^-,s^-)}
\frac {Z_{\D,\b,\g}^{(\mu^\pm,s^\pm)}(F)}
      { Z_{\D,\b,\g}^{(1,1,2,1)}}
P_{\widetilde\L^+\setminus w_+,\b,\g}^{(\mu^+,s^+,2,1)}
\Eq(62.38)
$$
From this we want to reconstruct the Gibbs measure in
$\widetilde\D\equiv\widetilde\L^-\cup\D\cup\widetilde\L^+$ with 
$(1,1,2,1)$-boundary conditions. Treating the numerator just as in 
\eqv(62.34), all we need is to show that, still up to negligible errors,
$$
Z_{\widetilde\L^-,\b,\g}^{(1,1,0,0)}(\{\eta(w_-,\s)=e^1\})
Z_{\D,\b,\g}^{(1,1,2,1)}
Z_{\widetilde\L^+,\b,\g}^{(0,0,2,1)}(\{\eta(w_-,\s)=e^2\})
\geq
Z_{\widetilde\D,\b,\g}^{(1,1,2,1)}
\Eq(62.39)
$$
To prove this we start by piecing together the first and second term in the 
l.h.s. of \eqv(62.39).  Using in turn the lower bound \eqv(62.4ter) and the 
upper bound \eqv(62.4bis),
$$
\eqalign{
Z_{\widetilde\L^-,\b,\g}^{(1,1,0,0)}(\{\eta(w_-,\s)=e^1\})
Z_{\D,\b,\g}^{(1,1,2,1)}
&=Z_{\widetilde\L^-,\b,\g}^{(1,1,0,0)}(\{\eta(w_-,\s)=e^1\})
  P_{\D,\b,\g}^{(1,1,2,1)}
  Z_{\D,\b,\g}^{(1,1,1,1)}\cr
&\geq Z_{\widetilde\L^-\cup\D,\b,\g}^{(1,1,1,1)}\bar P_0
     e^{-\g^{-1}er'(\ell,L,M,\hat\z,R)}\cr
&\geq Z_{\widetilde\L^-\cup\D,\b,\g}^{(1,1,2,1)}
      e^{-\g^{-1}(er'(\ell,L,M,\hat\z,R)+er'(\ell,L,M,\z,R))}\cr
}
\Eq(62.40)
$$
where we have in addition used \eqv(62.32) in the first inequality.
In the same manner
$$
Z_{\widetilde\L^-\cup\D,\b,\g}^{(1,1,2,1)}
Z_{\widetilde\L^+,\b,\g}^{(0,0,2,1)}(\{\eta(w_-,\s)=e^2\})
\geq 
Z_{\widetilde\D,\b,\g}^{(1,1,2,1)}
      e^{-\g^{-1}(er'(\ell,L,M,\hat\z,R)+er'(\ell,L,M,\z,R))}
\Eq(62.41)
$$
Combining \eqv(62.40) and \eqv(62.41) thus gives \eqv(62.39) and making use of
the later with \eqv(62.38) yields, up to the usual error term,
$$
\GG_{\b,\g,\L}^{(1,1,2,1)}(F\cap \AA(\mu^\pm,s^\pm,w_\pm))\leq
\frac {Z_{\widetilde\D,\b,\g}^{(1,1,2,1)}(F)}
      { Z_{\widetilde\D,\b,\g}^{(1,1,2,1)}}
\Eq(62.42)
$$
The upper bound \eqv(62.35) of Theorem 6.13 now simply follows from 
\eqv(62.42) just as the upper bound of Theorem 6.8 follows from \eqv(62.34).
The proof of the lower bound is a mere repetition of that of the lower of 
Theorem 6.8: simply substitute the boundary conditions $(1,1,1,1)$ by 
$(1,1,2,1)$ and use \eqv(62.39) instead of \eqv(62.31).
This concludes the proof of Theorem 6.13.\endproof.

Finally, we want to give a characterization of the typical profile 
in the case of asymmetric boundary conditions. 
The relevant estimates and notations for this have been introduced already in 
the proof of Lemma 6.10. 

Let us define the events
$$
E_{1,\L}^{(\mu,s,\mu',s')}\equiv\left\{ 
\s\,\Big| \,\exists_{{u_0\leq u_1\in \L}\atop{u_1-u_0\leq 2R}}
\forall_{\l_-\leq u<U_0}
\eta_{\hat\z,\hat L}(u,\s)=se^\mu \,\wedge \forall_{u_0<v\leq \l_+}
\eta_{\hat\z,\hat L}(u,\s)=s'e^{\mu'}
\right\}
\Eq(62.42bis)
$$

\theo {6.14}  {\it Assume that $\g |\L|\downarrow 0$, 
 $\b$ large enough ($\b>1$) and
$\g M(\g)\downarrow 0$. Then we can find $\g^{-1}\gg\hat L\gg 1$ and 
$\hat\z\downarrow 0$, such that on a subset $\O_\L\subset \O$ with 
$\P(\O_\L^c)\leq e^{-\g^{-1} f(\z')}$ we have that for all $\o\in \O_\L$
$$ 
\GG_{\b,\g,\L}^{(\mu,s,\mu',s')}[\o]\left(E_{1,\L}^{(\mu,s,\mu',s')}
 \right)\geq 1-2R e^{-\hat L c(\hat \z)}
\Eq(62.43)
$$
}

\remark This theorem implies that for any volume $\L$ 
such that $\g |\L|\downarrow 0$, we have $\P$-almost surely,
$$
\lim_{\g\downarrow 0} \GG_{\b,\g,\L}^{(\mu,s,\mu',s')}[\o]
\left(E_{1,\L}^{(\mu,s,\mu',s')} \right) =1
\Eq(62.44)
$$
(Here one may, to avoid complications with the 
``almost sure'' statement due to the uncountability of the number of 
possible sequences $\g_n$, assume for simplicity 
that $\lim_{\g\downarrow 0}$
is understood to be taken along some fixed discrete sequence, e.g. $\g_n=
1/n$. To show that the convergence holds also with probability 
one for {\it all} sequences tending to zero, one can use a continuity 
result as given inLemma 2.3 of [BGP2]).  

\proof The proof of this Theorem follows from Lemma 6.10 and its proof. 
We leave the details to the reader. \endproof

We are now ready to state a precise version of the main result 
announced in the introduction. We define the events
$$
E_{0,\L}^{(\mu,s)}\equiv \left\{\s\,\Big|\,\forall_{u\in \L}
\eta_{\hat\zeta,\hat L}(u,\s)=s e^\mu\right\}
\Eq(62.45)
$$
and set 
$$
E_{0,\L}\equiv \cup_{(\mu,s)}E_{0,\L}^{(\mu,s)}
\Eq(62.46)
$$
$$
E_{1,\L}\equiv \cup_{(\mu,s)\neq(\mu',s')} E_{1,\L}^{(\mu,s,\mu',s')}
\Eq(62.46bis)
$$
This this notation we have

\theo{6.15} {\it For any macroscopic box $V$ such that 
$\lim_{\g\downarrow 0}\g|V\|=0$, $\P$-almost surely,
$$
\lim_{\g\downarrow 0}\lim_{\L\uparrow\Z}  \GG_{\b,\g,\L}[\o]\left(
E_{0,V}\cup E_{1,V}\right)=1
\Eq(62.47)
$$
}

\proof  This theorem follows immediately Corollary 4.2 and the Theorems
6.9 and 6.14. The remark after \eqv(62.44) also applies here.\endproof

\newpage

\chap{7. Conclusions and conjectures}7

In the preceeding sections we have labored hard to prove that typical profiles
in the one dimensional Kac-Hopfield model are constant on a scale 
of the order $o(\g^{-1}$. The careful reader will have noticed the 
conspicuous absence of any argument that would proof that they are non-constant
on any larger scale. The reason for the absence of such an argument lies in
Section 5. There, we prove upper bounds on the fluctuations of the quantities 
$f_\D^{(\mu^\pm,s^\pm)}$ that imply that they are typically not larger than
$\sqrt {\g^{-1}|\D|}$.  What is not shown, and what would be needed, is that 
these fluctuations are actually of that size, and in particular that for
$\mu\neq \mu'$, $f_\D^{(\mu,s,\mu,s)}-f_\D^{(\mu',s,\mu',s)}$ 
typically differ by a random amount of that order. We certainly believe that 
this is true, but rigorous proofs of such statements are notoriously 
difficult to obtain and many problems in the theory of disordered systems 
are unsolved for very similar reasons. To our knowledge the only known method
in this direction is the work of Aizenman and Wehr [AW] that yields, however, 
no good quantitative results for finite volume objects. In fact, it appears that 
even the uniqueness of the Gibbs state in two dimensions (which 
one should expect to be provable with this method) cannot be 
shown using their approach (just as, and for similar reasons,
is the case in the two dimensional spin glass). A general method that would 
allow to get lower bounds on fluctuations corresponding to Theorem 5.1 
is thus still a great desideratum.

A natural question that poses itself is of course ``What about
dimensions greater than one?''. Here, again, conjectures come easy, but at 
some of them may be provable. First, as mentioned, we would expect that 
in dimension $d=2$ we still have a unique Gibbs state. This is motivated 
by the fact that at least  the block-approximation  looks very much like 
a multi-state random field model, for which this result would follow 
from Aizenman-Wehr. But as for a proof, see above.....
The same argument suggests, on the basis of the results of Bricmont-Kupiainen
[BrKu] and Bovier-K\"ulske [BK] that in dimension $d\geq 3$ we will 
have many Gibbs states, at least one for every pattern and its mirror image.
We would expect that this can actually proven, although 
technically this would be quite hard. To our surprise, it turns out 
that such a result is not even known in the ferromagnetic Kac model
(see Cassandro, Marra and Presutti [CMP] for a conjecture), and techniques to
take into account the the weak but long range interaction in proofs
of phase-stability have still to be worked out. However, this 
problem appears to be solvable. This entire line of research is very 
interesting and will be pursued in forthcoming publications.

\newpage

\chap{Appendix A: Proof of Proposition 6.12}8

In this section we prove  a lower bound on the infimum of the  free 
energy functional
 over all the profiles that form an interface between a ``phase''
 where the local overlaps
are close to   $a(\b)s^-e^{\mu^-}$ and another one where they are close
 $a(\b)s^+e^{\mu^+}$. In the case of the ferromagnetic Kac model, the shape 
of the interface was described  in [COP] chap. 6. In the case of the
Kac-Hopfield model due to our
restricted knowledge of the Hopfield model  free energy with fixed overlaps, 
we cannot perform such a detailed  analysis. 

Instead of working with the full free energy functional $\FF$ defined in 
\eqv(61.15) we will replace it by a lower bound (that is also 
suitably normalized to have its minimal value equal to zero) defined as
follows:
Given a macroscopic volume $\wt \D$ that could be chosen without lost of 
generality
to be  $[1, u_3]$  we denote by
$$
\wt {\cal F}_{\wt\D}^{\mu^\pm,s^\pm}
\equiv
V_{\wt\D}
^{\mu^\pm,s^\pm}+ \g\ell \sum_{x \in\sminn\wt\D}\Phi^{T}(m_{\ell}(x))
\Eq(7.1)
$$
where
$$\eqalign{
V_{\wt\D}^{\mu^\pm,s^\pm}&\equiv 
 \g\ell\sum_{x,y\in \sminn\wt\D} J_{\g\ell}(x-y)\frac{\|m_\ell(x)-
m_\ell(y)\|_2^2}4 \cr
&+\g\ell  \sum_{x\in\sminn\wt\D,y\in \del\wt\D}
J_{\g\ell}(x-y)\frac{\|m_\ell(x)-m^{
(\mu^\pm,s^\pm)}\|_2^2}2\cr
}
\Eq(7.2)
$$
and for any $\z \geq 2 a(\b)(\frac M \ell )^{1/2}$ ( cf Proposition 3.1), 
$$
\Phi^{T}(m_{\ell}(x))\equiv \cases{ 0, &if 
 $\exists_{\mu,s}\|m_\ell(x)-a(\b)se^{\mu}\|_2\leq \z$;\cr
\e(\z),&otherwise.\cr}
\Eq(7.3)
$$
 The set of profiles that form an interface between the
$(s^-,\mu)$ and the $(s^+,\mu^+)$ within the volume $\tilde \D$ is denoted by
$$
{\cal T}^{\pm}(\wt\D)
\equiv \{ m_{\ell}(x), x \in {\wt \D}\,\big |\,\, m_{\ell}(x)= a(\b)
{s}^- e^{\mu^-}, \forall_x \leq 0, m_{\ell}(x)=a(\b)s^+e^{\mu^+}, 
\forall_x\geq y_3 \}
\Eq(7.4)
$$
 where $ y_3\equiv\sup \{y| y\in u_3\}$

Proposition 6.12 then follows immediately from 

\proposition {7.1}{\it  There exists a $\z_0\equiv\xi(\b, M, \ell)$ such that
for all $\z$,  $\z_0\geq \z\geq 2a(\b)(\frac M \ell)^{1/2}$, 
we have}
 
$$
\inf_{m_{\ell} \in {\cal T}^{\pm}(\wt\D)}
\wt {\cal F}_{\wt\D}^{\mu^\pm,s^\pm}\geq
\sqrt{\e(\z)}\left(\sqrt{12((a(\b))^2 -4\z^2)}-3\sqrt{\e(\z)}\right)
\Eq(7.5)
$$
\proof

 For any given  profile in ${\cal T}^{\pm}(\wt\D)$, we denote by
$$
y_1\equiv \sup\{x\geq 0\big|
 \|m_{\ell}(x-1)-a(\b)s^{-}e^{\mu^-}\|_2\leq \z\} 
\Eq(7.6)
$$
the last exit of the $\z$ neighborhood of the $(s^-,{\mu^-})$ phase
and
$$
y_2=\inf\{x\geq y_1,\|m_{\ell}(x)-a(\b)s^{+}e^{\mu^+}\|_2\leq \z\} 
\Eq(7.7)
$$
the first entrance in the $\z$ neighborhood of the $(s^+,\mu^+)$ phase 
{\it after} $y_1$. 
Notice that by definition of ${\cal T}^{\pm}(\wt\D)$,
$y_1$ and $y_2$ exist  
and satisfy $0\leq y_1 \leq y_2 \leq  y_3$.
We defined also the overlap increments:
$$
D(x)\equiv m_{\ell}(x)-m_{\ell}(x-1)
\Eq(7.8)
$$
We write for   $1=y_0\leq x_1\leq y_1 $

$$
m_{\ell}(x_1)-a(\b)s^- e^{\mu^-}=\sum_{x=1}^{x_1} D(x)
\Eq(7.9)
$$
and for  $i=1,2$ and all $y_{i}\leq x_{i+1}\leq y_{i+1}$
$$\eqalign{
m_{\ell}(y_i)-m_{\ell}(x_i)&=\sum_{x=x_i +1}^{y_i} D(x)\cr
m_{\ell}(x_{i+1})-m_{\ell}(y_i)&=\sum_{x=y_i +1}^{x_{i+1}}D(x)
}
\Eq(7.10)
$$
at last
$$
a(\b)s^{+}e^{\mu^+}-
m_{\ell}(x_3)=\sum_{x=x_3 +1}^{y_3} D(x)
\Eq(7.11)
$$
We define now the quantity
$$
{\cal L}\equiv \sum_{i=1}^3 
\sum_{x_i=y_{i-1}+1}^{y_i} 
\left\|\sum_{x=y_{i-1}+1}^{x_i} D(x)\right\|_2^2+ 
\left\|\sum_{x=x_i +1}^{y_i} D(x)\right\|_2^2
\Eq(7.12)
$$
we first show that ${\cal L}$ can be bounded from above 
in term of $V_{\wt\D}^{\mu^\pm,s^\pm}$. Then we will bound from
below ${\cal L}$ by solving elementary variational problems. Putting 
those two bounds together will give a lower bound for 
$V_{\wt\D}^{\mu^\pm,s^\pm}$.
 
Let us start with the upper bound.
We first perform for each value of  $b \in 1,\dots, [1/\g \ell]$ a
block summation with blocks of length $b$, the location of the leftmost part
of the first block being a point $z \in 1,\dots, b$.  
Explicitly, calling
$$
D(u,b,z)\equiv \sum_{x=(u-1)b+z+1}^{ub+z} D(x)
\Eq(7.13)
$$
we write
$$
\sum_{x=1}^{x_1} D(x)= \sum_{x=1}^{z} D(x) + 
\sum_{u=1}^{\left[\frac{x_1 -z}b\right]}D(u,b,z) + D_+(x_1,b,z)
\Eq(7.14)
$$
where
$$
D_+(x_1,b,z)= \sum_{x=\left[\frac {x_1 -z}b\right]b+z+1}^{x_1} D(x)
\Eq(7.15)
$$
 The second sum in the r.h.s of \eqv(7.14) will be called the bulk term,
while the first sum and $D_+$ will be called  boundary terms.
We have  also
$$
\sum_{x=x_i +1}^{y_i} D(x)= D_{-}(x_i,b,z) + 
\sum_{u=\left[ \frac{x_i +1 -z}{b}\right]+2}
^{\left[ \frac {y_i -z}b\right]} D(u,b,z) + D_+(y_i,b,z)
\Eq(7.16)
$$
with the `boundary' terms
$$
D_-(x_i,b,z)= \sum_{x=x_i +1}^{\left[\frac{x_i +1 -z}b \right]b +b +z+1} D(x)
\Eq(7.17)
$$
and 
$$
 D_+(y_i,b,z)=\sum_{x=\left[\frac{y_i -z}b\right]b +z+1}^{y_i} D(x)
\Eq(7.18)
$$
We have also
$$
\sum_{x=y_{i-1}+1}^{x_i} D(x)= D_{-}(y_{i-1},b,z) + 
\sum_{u=\left[ \frac {y_{i-1}+1-z}b\right]+2}
^{\left[\frac{x_i-z}b\right]} D(u,b,z) +D_+(x_i,b,z)
\Eq(7.19)
$$
here the `boundary' terms are:
$$ 
D_{-}(y_{i-1},b,z)\equiv
\sum_{x=y_{i-1}+1}^{\left[\frac {y_{i-1}+1-z}b \right]b +b+z+1}
D(x)
\Eq(7.20)
$$
and
$$
 D_+(x_i,b,z)\equiv \sum_{x=\left[\frac {x_i -z}{b} \right] b +z+1}^{x_i} D(x)
\Eq(7.21)
$$

For a given $b$ and $z$, the  Schwarz Inequality implies
$$
\eqalign{
&\left\|\sum_{x=y_{i-1}+1}^{x_i} D(x)\right\|_2^2
\leq\cr
&\left(\frac {x_i -y_{i-1} +1} b +2\right)
\left( \|D_{-}(y_{i-1},b,z)\|_2^2+ 
\sum_{u=\left[\frac {y_{i-1}+1-z}{b}\right]+2 }^{\left[\frac {x_i -z}b\right]}
 \|D(u,b,z)\|_2^2
+\|D_+(x_i,b,z)\|_2^2\right)\cr
}\Eq(7.22)
$$

We want to take the mean of the two sides of  \eqv(7.22) over all the
possible choices of block 
lengths $b$ in $ 1,\dots,[(\g\ell)^{-1}$ and $z$ in $1,\dots,b$. 
To do this we  use a weighted mean for the block lengths and
an uniform mean for the $z\in 1,\dots b$. We  use  
$$
\eqalign{
\sum_{b=1}^{\left[(\g \ell)^{-1}\right]} b^2=
& \frac {\left[(\g \ell)^{-1}\right]
 (\left[(\g \ell)^{-1}\right] +1)( 2\left[(\g \ell)^{-1}\right] +1)}6\cr
&= (\left[(\g \ell)^{-1}\right])^3 \frac 13 \left( 1+ O(\g \ell)\right)
}
\Eq(7.23)
$$
to define a weighted mean on $ 1,\dots,[(\g\ell)^{-1}$. Performing explicitly
these weighted means gives
$$
\eqalign{
&\left\|\sum_{x=y_{i-1}+1}^{x_i} D(x)\right\|_2^2 \leq\cr
&\frac 3{\left[(\g \ell)^{-1}\right]^3(1+O(\g \ell))}
\left(x_i - y_{i-1} +1 + \frac 2 {\g\ell}\right)\cr
&\sum_{b=1}^{\left[(\g \ell)^{-1}\right]} \sum_{z=1}^b 
\left( \|D_{-}(y_{i-1},b,z)\|_2^2+ 
\sum_{u=\left[\frac {y_{i-1}+1-z}{b}\right]+2 }^{\left[\frac {x_i -z}b\right]}
\| D(u,b,z)\|_2^2
+\|D_+(x_i,b,z)\|_2^2\right)
}\Eq(7.24)
$$
and by the very same argument
$$
\eqalign{
&\left\|\sum_{x=x_{i}+1}^{y_i} D(x)\right\|_2^2 \leq\cr
&\frac 3{\left[(\g \ell)^{-1}\right]^3(1+O(\g \ell))}
\left(y_i - x_{i} +1 + \frac 2 {\g\ell}\right)\cr
&\sum_{b=1}^{\left[(\g \ell)^{-1}\right]} \sum_{z=1}^b 
\left( \|D_{-}(x_{i},b,z)\|_2^2+ 
\sum_{u=\left[\frac {x_{i}-z}{b}\right]+2 }^{\left[\frac {y_i -z}b\right]}
\| D(u,b,z)\|_2^2
+\>D_+(y_i,b,z)\|_2^2\right)
}\Eq(7.25)
$$
Collecting the `bulk' terms in \eqv(7.24) and \eqv(7.25) to bound $\cal L$,
it is not difficult to check that 
$$\eqalign{
& \frac 3{\left[(\g \ell)^{-1}\right]^3 (1+O(\g \ell))}
\sum_{x_i=y_{i-1}}^{y_i}
\sum_{b=1}^{\left[(\g \ell)^{-1}\right]} \sum_{z=1}^b \cr
&\left(x_i - y_{i-1} +1 + \frac 2{\g\ell}\right)
\sum_{u=\left[\frac {y_{i-1}+1-z}{b}\right]+2 }^{\left[\frac {x_i -z}b\right]}
 \|D(u,b,z)\|_2^2+
\left(y_i - x_{i} +1 + \frac 2{\g\ell}\right)
\sum_{u=\left[\frac {x_{i}-z}{b}\right]+2 }^{\left[\frac {y_i -z}b\right]} 
\|D(u,b,z)\|_2^2\cr
&\leq 3 \left[\g\ell\right]^3 \left( 1+ O(\g \ell)\right)
\left(y_i -y_{i-1}\right)\left(y_i -y_{i-1} + \frac 2{\g \ell}\right)
\sum_{b=1}^{\left[(\g \ell)^{-1}\right]} \sum_{z=1}^b
\sum_{u=\left[\frac {y_{i-1}+1-z}{b}\right]}^{\left[\frac {y_i -z}b\right]} 
\|D(u,b,z)\|_2^2\cr
&\leq 3 \left(1+ O(\g \ell)\right) \left(\g \ell (y_i -y_{i-1})\right)
\left(\g \ell (y_i -y_{i-1}) +2 \right) \sum_{y_{i-1} \leq x,y \leq y_i} 
J_{\g \ell}(x-y)
\left\|m_\ell (x) -m_\ell (y)\right\|^2_2\cr
}
\Eq(7.26)
$$

It remains to consider the "boundary'' terms, putting together the 
terms $ D_{+}(x_i,b,z)$ and $D_{-}(x_i,b,z)$, it is not too difficult
to check that 

$$\eqalign{
& \frac 3{\left[(\g \ell)^{-1}\right]^3 (1+O(\g \ell))}
\sum_{x_i=y_{i-1}}^{y_i}
\sum_{b=1}^{\left[(\g \ell)^{-1}\right]} \sum_{z=1}^b \cr
&\left(x_i - y_{i-1} +1 + \frac 2{\g\ell}\right)
\left\| \sum_{x=\left[\frac {x_i -z}{b} \right] b +z+1}^{x_i} 
D(x) \right\|_2^2 +
\left( y_i -x_i +1 +\frac 2{\g\ell}\right) 
\left\|\sum_{x=x_i +1}^{\left[\frac{x_i +1 -z}b \right]b +b +z+1} 
D(x) \right\|_2^2  \cr
&\leq 3 \left( (y_i -y_{i-1} +\frac 2{\g\ell})\g \ell \right)
\sum_{y_{i-1} \leq x,y \leq y_i} J_{\g \ell}(x-y)
\left\|m_\ell (x) -m_\ell (y)\right\|^2_2\cr
}
\Eq(7.27)
$$
Therefore we get
$$
{\cal L} \leq 4\sum_{i=1}^3 \g\ell ( y_i- y_{i-1})( \g\ell(y_i -y_{i-1})+3)
\sum_{y_{i-1} \leq x,y \leq y_i} J_{\g \ell}(x-y)
\left\|m_\ell (x) -m_\ell (y)\right\|^2_2
\Eq(7.28)
$$
which is the upper bound we wanted.

 Now we want to bound from below $\cal L$. Notice first that by 
solving explicitly the variational problem we have that for all 
$m_1, m_2 \in \R^M$
$$\eqalign{
&\inf_{m_{\ell}(x)} \{ \|m_{\ell}(x)-m_1\|^2_2
+\|m_{\ell}(x)-m_2\|^2_2\}\cr
&\geq \frac 12 \|m_1 -m_2\|^2_2
}\Eq(7.29)
$$
using \eqv(7.6) and \eqv(7.7) and convexity, we get  
$$\eqalign{
{\cal L}&\geq (y_1-y_0)\frac {\z^2}2 + \frac 12 (y_2 -y_1) 
\|m_\ell(y_1)-m_\ell(y_2)\|^2_2
+(y_3 -y_2) \frac {\z^2}2\cr
&\geq \frac 12 (y_2 -y_1) \left((a(\b))^2-4\z^2\right)
}\Eq(7.30)
$$
 On the other hand, {\it c.f} \eqv(7.3), we have
$$
\g\ell \sum_{x \in\sminn\wt\D}\Phi^{T}(m_{\ell}(x)) \geq 
\g \ell ( y_2 -y_1) \e(\z)
\Eq(7.31)
$$
therefore, introducing the macroscopic variables $u_i= \g \ell y_i$ 
we get
$$\eqalign{
\wt {\cal F}_{\wt\D}^{\mu^\pm,s^\pm}&\geq
(u_2 -u_1)\e(\z) +\frac {12}{u_2-u_1 +3} \left((a(\b))^2-4\z^2\right)\cr
&\geq \sqrt{\e(\z)}\left( \sqrt{ 12\left((a(\b))^2-4\z^2\right)}
-3\sqrt{\e(\z)}\right)\cr
}\Eq(7.32)
$$
where the last step follows from  the explicit computation of the infimum over all 
possible values of $u_2 -u_1$.\endproof

\frenchspacing
\chap{References}4
\item{[AGS]} D.J. Amit, H. Gutfreund and H.
Sompolinsky, ``Statistical mechanics of neural networks near saturation'',
Ann. Phys. {\bf 173}, 30-67 (1987).
\item{[ALR]}  M. Aizenman, J.L. Lebowitz, and D. Ruelle, ``Some rigorous
 results on the Sherrington-Kirkpatrick spin glass model''. Commun. Math.
Phys. {\bf 112}, 3-20 (1987).
 \item{[AW]}  M. Aizenman, and J. Wehr, ``Rounding effects on
quenched randomness on first-order phase transitions'',
Commun. Math. Phys. {\bf 130}, 489 (1990).
\item{[BG1]} A. Bovier and V. Gayrard, ``Rigorous results on the
thermodynamics of the dilute Hopfield model'', J. Stat. Phys. {\bf 69},
597-627 (1993).
\item{[BG2]} A. Bovier and V. Gayrard, ``An almost sure large deviation 
principle for the Hopfield model'', to appear in  Ann. Probab, (1996).
\item{[BG3]} A. Bovier and V. Gayrard, ``The retrieval phase of the 
Hopfield model, A rigorous analysis of the overlap distribution'',
submitted to Prob. Theor. Rel. Fields (1995).
\item{[BGP1]} A. Bovier, V. Gayrard, and P. Picco, ``Gibbs states
of the Hopfield model in the regime of perfect  memory'',
Prob. Theor. Rel. Fields {\bf 100}, 329-363 (1994).
\item{[BGP2]} A. Bovier, V. Gayrard, and P. Picco,
``Large deviation principles for the Hopfield model and the
Kac-Hopfield model'', Prob. Theor. Rel. Fields {\bf 101}, 511-546 (1995).
\item {[BGP3]} A. Bovier, V. Gayrard, and P. Picco,
``Gibbs states of the Hopfield model with extensively many patterns'',
 J. Stat. Phys. {\bf 79}, 395-414 (1995).
\item{[BF]} A. Bovier and J. Fr\"ohlich, ``A heuristic theory of the
spin glass phase'', J. Stat.Phys. {\bf 44}, 347-391 (1986).
\item{[BK]}  A. Bovier and Ch. K\"ulske, A rigorous renormalization group
method for interfaces in random media,
Rev. Math. Phys. {\bf 6}, 413-496 (1994).
\item{[BrKu]} J. Bricmont, and A. Kupiainen, ``Phase transition in the
3d random field Ising model'', Commun. Math. Phys. {\bf
116}, 539-572 (1988).
\item{[Lu]} J.M. Luttinger,  ``Exactly Soluble Spin-Glass Model'',
 Phys.Rev. Lett. {\bf 37}, 778-782 (1976).
\item{[CMP]} M. Cassandro, R. Marra, and E. Presutti, 
``Corrections to the critical temperature in 2d Ising systems with 
Kac potentials'', J. Stat. Phys. {\bf 78}, 1131-1138 (1995).
\item{[CN]} F. Comets and J. Neveu, ``The Sherrington-Kirkpatrick model
of spin glasses and stochastic calculus, the high temperature case, 
Commun. Math. Phys. {\bf 166}, 549-564 (1995).
\item{[COP]} M. Cassandro, E. Orlandi, and E. Presutti,
``Interfaces and typical Gibbs configurations for one-dimensional
Kac potentials'', Prob. Theor. Rel. Fields {\bf 96}, 57-96 (1993).
\item{[DOPT]} A. De Masi, E. Orlandi, E. Presutti, and L. Triolo,
``Glauber evolution with Kac potentials, I. Mesoscopic and macroscopic limits,
interface dynamics'', Nonlinearity {\bf 7}, 633-696 (1994); 
``II. Spinodal decomposition'', to appear. 
\item {[EA]} Edwards, P.W. Anderson, ``Theory of spin glasses'', 
J. Phys. {\bf F 5}, 965-974 (1975).
\item{[vE]} A.C.D. van Enter, ``Stiffness exponent, number of pure states,
and Almeida-Thouless line in spin glasses'', J. Stat. Phys. {\bf 60}, 275-279
 (1990).
\item{[FH]}  D.S. Fisher and D.A. Huse, ``Pure phases in spin glasses'', 
J. Phys. {\bf A 20}, L997-L1003 (1987); ``Absence of many states in magnetic 
spin glasses'', J. Phys.  {\bf A 20}, L1005-L1010 (1987).
\item{[FP1]} L.A. Pastur and A.L. Figotin, ``Exactly soluble model
of a spin glass'', Sov. J. Low Temp. Phys. {\bf 3(6)}, 378-383
(1977).
\item{[FP2]} L.A. Pastur and A.L. Figotin, ``On the theory of
disordered spin systems'', Theor. Math. Phys. {\bf 35}, 403-414
(1978).
\item{[FP2]} L.A. Pastur and A.L. Figotin, ``Infinite range limit for a class
 of disordered spin systems'', Theor. Math. Phys. {\bf 51}, 564-569 (1982).
\item{[FZ]} J. Fr\"ohlich and B. Zegarlinski, ``Some comments on the 
Sherrington-Kirkpatrick model of spin glasses'', Commun. Math. Phys. {\bf 
112}, 553-566 (1987). 
\item{[Ge]} S. Geman, ``A limit theorem for the norms of random matrices'',
Ann. Probab. {\bf 8}, 252-261 (1980).
\item{[Ho]} J.J. Hopfield, ``Neural networks and physical systems
with emergent collective computational abilities'', Proc. Natl.
Acad. Sci. USA {\bf 79}, 2554-2558 (1982).
\item{[K]} H. Koch, ``A free energy bound for the Hopfield
model'', J. Phys. A  {\bf A 26}, L353-L355 (1993).
\item{[KUH]} M. Kac, G. Uhlenbeck, and P.C. Hemmer, ``On the van
der Waals theory of vapour-liquid equilibrium. I. Discussion of a
one-dimensional model'' J. Math. Phys. {\bf 4}, 216-228 (1963);
``II. Discussion of the distribution functions''
J. Math. Phys. {\bf 4}, 229-247 (1963);
``III. Discussion of the critical region'',
J. Math. Phys. {\bf 5}, 60-74 (1964).
\item{[LP]} J. Lebowitz and O. Penrose, ``Rigorous treatment of
the Van der Waals Maxwell theory of the liquid-vapour
transition'', J. Math. Phys. {\bf 7}, 98-113 (1966)
\item{[Ma]} D.C. Mattis, ``Solvable spin system with random interactions'', 
Phys. Lett.
{\bf 56A}, 421-422 (1976).
\item{[MPR]} E. Marinari, G. Parisi, and F. Ritort, ``On the 3D Ising 
spin glass'', J. Phys.  {\bf A 27}, 2687-2708.
\item{[MPV]} M. M\'ezard, G. Parisi, and M.A. Virasoro, 
``Spin-glass theory
and beyond'', { World Scientific}, Singapore (1988).
\item {[NS]} Ch.M. Newman and D.L. Stein, ``Non-mean-field behaviour in
realistic spin glasses'', preprint cond-mat/9508006 (1995).
\item{[PS]} L. Pastur and M. Shcherbina, ``Absence of self-averaging
of the order parameter in the Sher\-ring\-ton-Kirkpatrick model'', 
J. Stat. Phys. {\bf 62 }, 1-19 (1991).
\item{[SK]} D. Sherrington and S. Kirkpatrick, ``Solvable model of a
spin glass'', { Phys. Rev. Lett.}
{\bf 35}, 1792-1796 (1972).
\item{[ST]} M. Shcherbina and B. Tirozzi, ``The free energy for a class
of Hopfield models'', J. Stat. Phys. {\bf 72}, 113-125 (1992).
\item{[T1]} M. Talagrand, ``Concentration of measure and isoperimetric
inequalities in product space'', preprint (1994). 
\item{[T2]} M. Talagrand, ``A new look at independence'', preprint (1995).

\end